\documentstyle[12pt]{article}

\newcommand{\x}{{\mbox{\bf x}}}
\newcommand{\I}{{\mbox{I}}}
\newcommand{\2}{{\mbox{II}}}
\newcommand{\3}{{\mbox{III}}}
\newcommand{\4}{{\mbox{IV}}}
\newcommand{\1}{\bf 1}
\newcommand{\T}{\cal T}
\newcommand{\LXI}{{\cal L}_{\xi}}
\newcommand{\R}{{\hat {\cal R}}}
\newcommand{\tr}{{\mbox{tr}}}
\newcommand{\J}{\stackrel{\bullet}{J}}
\newcommand{\q}{{\mbox{q}}}
\makeatletter
\@addtoreset{equation}{section}
\makeatother
\begin{document}

{\renewcommand{\theequation}{1.\arabic{equation}}

\begin{center}
{\LARGE\bf Particle creation \\
 in the  effective action method}
\end{center}
\begin{center}
{\bf A.G. Mirzabekian}$ ^{\mbox{\small a}}$ {\bf and G.A. Vilkovisky}$ ^{
\mbox{\small a,b}}$
\end{center}
\vspace{2cm} 
$^{\mbox{\small a}}$Lebedev 
Research Center in Physics, Leninsky Prospect 53, Moscow 117924,
Russia \\
$^{\mbox{\small b}}$Lebedev Physics Institute, Academy of Sciences of Russia, Leninsky
Prospect 53, Moscow 117924, Russia

\begin{abstract}
The effect of particle creation by nonstationary external fields is considered 
as a radiation effect in the expectation-value spacetime. The energy of created
massless particles is calculated as the vacuum contribution in the 
energy-momentum tensor of the expectation value of the metric. The calculation
is carried out for an arbitrary quantum field coupled to all 
external fields entering
the general second-order equation. The result is obtained as a functional of 
the external fields. The paper gives a systematic 
derivation of this result on the basis of the 
nonlocal effective action. Although the derivation is quite involved
and touches on many aspects of the theory, the result itself is remarkably
simple. It brings the quantum problem of particle creation to the level of
complexity of the classical radiation problem. For external fields like the 
electromagnetic or gravitational field there appears a quantity, the radiation 
moment, that governs both the classical radiation of waves and the quantum 
particle production. The vacuum radiation of an electrically charged source is 
considered as an example. The research is aimed at the problem of 
backreaction of the vacuum radiation.
\end{abstract}

\newpage
\begin{center}
{\Large\bf Contents}
\end{center}
\begin{enumerate}
\item {\Large Introduction}
\begin{itemize}
\item Introduction proper
\item Outline of the contents
\item The model
\item The limits ${\cal I}^+$ and $i^+$
\item Vacuum radiation in the expectation-value spacetime
\end{itemize}
\item {\Large The vacuum energy-momentum tensor}
\begin{itemize}
\item The effective action
\item Obtaining $T^{\mu\nu}_{\mbox{\scriptsize vac}}$ at ${\cal I}^+$
\item The result for $T^{\mu\nu}_{\mbox{\scriptsize vac}}$ at ${\cal I}^+$
\item The problem of quantum noise
\end{itemize}
\item {\Large Retarded kernels of the nonlocal operators}
\begin{itemize}
\item The retarded resolvent
\item The kernels of $1/\Box$ , $\log(-\Box)$ , and of the vertex operators
\item The kernels of the vertex operators superposed with $1/\Box$
\end{itemize}
\item {\Large The asymptotic behaviours at ${\cal I}^+$}
\begin{itemize}
\item Null hyperplanes
\item The $1/\Box$ and $\log(-\Box)$ at ${\cal I}^+$
\item The vertex operators at ${\cal I}^+$
\item The conserved charges
\item The vector and cross vertices at ${\cal I}^+$
\item The gravitational vertex at ${\cal I}^+$
\item The trees at ${\cal I}^+$
\end{itemize}
\item {\Large The early-time behaviours}
\begin{itemize}
\item Geometrical two-point functions in presence of the Killing vector
\item Retarded kernels in presence of the Killing vector
\item Causality of the vacuum radiation
\item Convergence of the vertex operators
\end{itemize}
\item {\Large The late-time behaviours}
\begin{itemize}
\item The retarded Green function in the future of ${\cal I}^+$
\item Formula for the energy of vacuum radiation
\item The vertices and trees in the future of ${\cal I}^+$
\item The trees at $i^+$ and ${\cal T}^+$
\end{itemize}
\item {\Large The late-time behaviours (continued)}
\begin{itemize}
\item Spacelike hyperplanes
\item The radiation moments and conserved charges
\item The vertex operators at $i^+$
\item The non-scalar vertices at $i^+$
\item The vertices at ${\cal T}^+$
\end{itemize}
\item {\Large Creation of particles and radiation of waves}
\begin{itemize}
\item The energy of the vacuum particle production
\item Positivity
\item Radiation of waves
\end{itemize}
\item {\Large Specializations and examples}
\begin{itemize}
\item Spherical symmetry
\item Electrically charged shell expanding in the self field
\item Particle production by a spherically symmetric electromagnetic field
\item Radiation of the nonrelativistic shell
\item Radiation of the ultrarelativistic shell
\item Validity of the approximations
\end{itemize}
\end{enumerate}
{\Large Concluding remarks}\\
{\Large Acknowledgments}\\
{\Large Appendix A.}\parbox[t]{10cm}{\it IDENTITIES FOR THE
VERTEX OPERATORS.\\ \vspace{2.5mm}
 USE OF THE JACOBI IDENTITIES.}\\ 
{\Large Appendix B.}\parbox[t]{11cm}{\it ALTERNATIVE APPROACH
TO THE NON-SCALAR \\ \vspace{2.5mm} VERTICES.}\\ 
{\Large References}

\newpage

\begin{center}
\section{\bf    Introduction} 
\end{center}
\subsection*{\bf Introduction proper.}

$$ $$ 

In the present paper we consider the problem of creation of particles
from the vacuum by nonstationary external fields. This problem has been much 
discussed in the literature (see , e.g., [1-3]) but the present approach
and the result obtained are new. Namely, we calculate the energy of massless 
particles created in  external fields of arbitrary configuration and obtain
the result as a functional of these fields. The calculation is done for a set of
massless quantum fields coupled to all external fields entering the general
second-order equation. The approach used is expectation-value theory
and the method of the gravitational effective action.

The main result of the present work is briefly reported in [4]. This
final result is simple but its derivation is not because the effect of
particle creation is nonlocal and sits in the cubic terms of the effective
action. This is an effect of the one-loop vertices. A use of the
 nonlocal effective action in the context of expectation values is almost
 unknown. Therefore, we present here a systematic derivation showing
 the techniques involved.

 The result that obtains is remarkable since it brings the problem of the
 vacuum particle production to the level of complexity of the classical
 radiation problem. The strengths of external fields are expressed through
 their physical sources which are next integrated over certain spacelike
 hypersurfaces orthogonal to the geodesics. These integrals ( we call them
 radiation moments ) are direct generalizations of the moments of classical
 radiation theory. For external fields like the electromagnetic and 
 gravitational fields, the radiation moments govern both the classical 
 radiation of waves and the quantum particle production. As an example
 we consider the vacuum radiation produced by an electrically charged shell
 expanding in the self field.

 The work is motivated by the fact that the problem with external fields is
 physically incomplete. For a restoration of the energy conservation law it
 should be regarded as a part of a dynamical problem for the field's expectation
 values in an initial quantum state. An outstanding example is the 
 gravitational collapse problem [5]. In the expectation-value problem one
 first calculates the currents as functionals of the mean fields, and next 
 makes these fields subject to the self-consistent equations. At the stage
 of the calculation of the currents the expectation-value fields appear as 
 external fields.

 The setting of the problem with external fields in an asymptotically flat 
 spacetime assumes that these fields become stationary in the remote past
 and future in which case there exist the standard in-  and out- vacuum
 states for the quantum field [1]. Because the external field is nonstationary
 in the intermediate domain, the out-vacuum is generally a many-particle
 in-state, i.e. the external field creates particles from the in-vacuum.
 The method usually applied for the calculation of this effect consists in
 representing the quantum field as a sum over modes and obtaining the
 Bogoliubov transformation that relates the basis functions of the 
 in- and out- modes.

 The method of mode decomposition is normally used for an explicit solving
 of the equations in a given external field but one can also build a perturbation
 theory by solving for the basis functions iteratively in the external-field 
 strength. This brings one to the loop diagrams of expectation-value theory,
 or the Schwinger-Keldysh diagrams [6-14]. As shown in [14], the 
 Schwinger-Keldysh diagrams for the in-vacuum state are related through a  
 certain set of rules to the Euclidean or Feynman effective action. This makes
 it possible to get away from both the mode decomposition and Schwinger-Keldysh 
 diagrams, and carry out the calculation by the method of effective action.

 The vacuum energy-momentum tensor is obtained by varying the loops of the
 effective action with respect to the metric ( and next applying
 the rules of Ref.[14]). Therefore, irrespectively of the nature of the 
 external field in question, one needs the effective action for the quantum
 field coupled to an external gravitational field ( in addition to the
 external field in question). Since, for the effect of the vacuum patricle 
 production, the lowest nonvanishing order is second order in the field
 strength, one needs the terms in the effective action that are quadratic
 in the field strength in question and linear in the gravitational field 
 strength. The effect is thus contained in the one-loop triangular diagrams
 with at least one external gravitational line.

 The covariant perturbation theory for the effective action is built in 
 [14-18] where the one-loop vertices are calculated for all couplings of the
 quantum field whose small disturbances are propagated by the general
 second-order equation. These results make the starting point for the 
 present work.

 \subsection*{\bf Outline of the contents.}

 $$  $$

 In the present section, after a brief outline of the contents of the paper, 
 we introduce the field model for which the calculation will be carried out,
 and the limitations under which the result will be obtained. We introduce
 also some notions pertaining to various limits at the asymptotically flat 
 infinity, and relate the vacuum particle production to the radiation in
 the expectation-value spacetime.

 Sec.2 reviews the structure of the effective action and the procedure
 of obtaining from it the expectation-value current which in the present
 case is the energy-momentum tensor for the mean metric. The energy-momentum 
 tensor contains  a contribution of the in-vacuum,
 $T^{\mu\nu}_{\mbox{\scriptsize vac}}$ , which is the subject of study 
 in the paper. For the radiation problem,
 $T^{\mu\nu}_{\mbox{\scriptsize vac}}$ is needed only at the future null 
 infinity $({\cal I}^+)$, and there are rules which facilitate obtaining
 $T^{\mu\nu}_{\mbox{\scriptsize vac}}\Bigl|_{{\cal I}^+}$ greatly.
 These rules are also reviewed including a theorem which relates the
 limit of ${\cal I}^+$ for the kernel of an operator function to a certain
 limit for the function itself. This presentation is based on an earlier 
 work [14-23] with the exception of the final result for 
 $T^{\mu\nu}_{\mbox{\scriptsize vac}}\Bigl|_{{\cal I}^+}$
 which is obtained by a drastic simplification of the results in Ref.[18].
 The technique of this simplification is outlined in Appendix A.

 Sec.2 concludes with a discussion of the problem of averaging the
 quantum noise which is a sign-indefinite contribution in
 $T^{\mu\nu}_{\mbox{\scriptsize vac}}\Bigl|_{{\cal I}^+}$
 due to the quantum uncertainty. In two and four dimensions the mechanisms
 of this averaging are completely different since, in four dimensions,
 $T^{\mu\nu}_{\mbox{\scriptsize vac}}\Bigl|_{{\cal I}^+}$
has the form of a total derivative in retarded time. The energy radiated
by the vacuum for the whole history is thus determined by the limits of late
time and early time, and, for this energy to be nonvanishing, certain 
nonlocal functions of external fields should exhibit a growth at late time.
This offers a problem whose solution is the principal achievement of the 
present work.

The result for
$T^{\mu\nu}_{\mbox{\scriptsize vac}}\Bigl|_{{\cal I}^+}$
is obtained in Sec.2 in the form of a superposition of nonlocal operators
acting on the sources of external fields. All nonlocal operators in
$T^{\mu\nu}_{\mbox{\scriptsize vac}}$
are defined by their spectral forms in which the resolvent is the 
retarded Green function [14].
A derivation of the kernels of these operators is the subject of Sec.3.
The resolvent is taken in the approximation in which it is determined
by the geometrical two-point functions which, in their turn, are built
on the basis of causal geodesics of the expectation-value 
(or background) spacetime. Therefore, all kernels  in
$T^{\mu\nu}_{\mbox{\scriptsize vac}}$
inherit this geodetic structure and the retardation property.
The geometrical objects involved are the light cone of a point, and
the hyperboloid of equal (timelike) geodetic distance from a point.
Subsequently, two other objects are derived from these: null and
spacelike hyperplanes.

The behaviours of all kernels at 
${\cal I}^+$ 
are obtained in Sec.4. In particular, it is shown that, at the
limit of 
${\cal I}^+$,
the kernel of the operator 
$\log(-\Box)$
which stands for the external line in
$T^{\mu\nu}_{\mbox{\scriptsize vac}}\Bigl|_{{\cal I}^+}$
boils down to the kernel of 
$1/\Box$ .
Here, there first appear the radiation moments but only their ultrarelativistic
limiting cases defined as integrals over the null hyperplanes.
Upon the calculation of the behaviours of the vertex functions at
${\cal I}^+$,
there emerges a problem.
It turns out that, in the case of the external vector field (and only
in this case), the superposition of kernels in
$T^{\mu\nu}_{\mbox{\scriptsize vac}}$
fails to converge at 
${\cal I}^+$.
The problem removes if the vector field is exposed  to a special limitation
that it contains no outgoing wave. The result in the paper is obtained
under this limitation.

As mentioned above, the total energy of  vacuum radiation is determined only
by the  limits of late time and early time. The purpose of Sec.5 is a 
proof that the limit of early time makes no contribution. The proof
uses three facts: i) the retardation of all kernels in
$T^{\mu\nu}_{\mbox{\scriptsize vac}}$ ,
ii) the presence of a time derivative in the kernel of 
$\log(-\Box)$ ,
and iii) the assumed stationarity of the external fields in the past.
Technically, it involves the properties of the geometric two-point functions
and retarded kernels in presence of the Killing vector. The main
result of this section is formulated as an assertion about the causality
of the vacuum radiation. Sec.5 discusses also the question of convergence
of the massless operators
$1/\Box^n$ for $n>1.$
This question emerges in connection with some of the vertex operators in
$T^{\mu\nu}_{\mbox{\scriptsize vac}}$.

In Secs.6 and 7 we come to the heart of the matter: the behaviours at late
time. Most of Sec.6 is devoted to the analysis of the behaviour of the
function
$\Bigl(1/\Box\Bigr)X$
in the future of 
${\cal I}^+$.
The result is that the dominant contribution to this behaviour comes from
$X$
at the limit 
$i^+$
which here is defined as the limit of infinite proper time along the timelike
geodesics that reach the future asymptotically flat infinity. It follows from
this result that the radiation energy is given by an integral over the
energies of the particles at $i^+$ and is nonvanishing only if the vertex 
functions have an appropriate {\it growth} at $i^+$.

The behaviours of the vertex functions at 
$i^+$
is the chief thing. They are obtained in Sec.7. Here, there appear the full 
radiation moments defined as integrals over the spacelike hyperplanes.
Much work in Sec.7 is connected with the  vector and tensor vertices.
An alternative way of  handling these vertices (which also requires much
work) is considered in Appendix B.

The final result for the energy of vacuum radiation is presented in
Sec.8, and it is gratifying to see its positivity. The positivity is
based on two facts: i) the conservation of the currents  of external fields, 
and ii) the  self-adjointness of the equation of the quantum field. The discussion 
of the radiation moments is  completed in Sec.8 by reviewing their role
in classical radiation theory and establishing their relation to the textbook
multipole moments.
After that, it becomes visual that the  quantum problem of particle
creation is made almost the same thing as the classical problem of
radiation of waves.

In Sec.9 the result for the vacuum radiation is specialized to spherical
symmetry and to the  external electromagnetic field. The  longitudinal
projections of the radiation  moments are calculated. The vacuum radiation
produced by  an electrically charged spherical shell is considered as an 
example. The shell is assumed expanding in the self field from the state
of maximum contraction to infinity, and the loss of its energy for the whole
 time of expansion is calculated in both the  nonrelativistic and
 ultrarelativistic cases. Without accounting for the vacuum backreaction,
the radiation of the ultrarelativistic shell violates the energy conservation
law. The limits of validity of the technique based
 on the nonlocal expansion of the effective action are illustrated by 
 specializing to the problem of particle creation in an external
 electric field.

 The concluding remarks are devoted to a discussion of the limitations
 imposed on the  external fields. In particular, the emergence of  the 
 limitation on the vector field signals that  the theory contains
 another effect: the vacuum screening or amplification of the
 electromagnetic waves emitted by a source. This effect is equivalent
 to an observable renormalization of all multipole moments and is 
 analogous to the effect of the vacuum gravitational waves [23].

 \subsection*{\bf The model.}

  $$ $$

 We shall consider the vacuum of a multicomponent quantum field
 $\varphi = \varphi^A$
 coupled to external fields through the following equation for the small
 disturbances $\delta \varphi$:
 \begin{equation}
 {\hat {H}}\delta\varphi \equiv {H}^A_{\;B}\delta\varphi^B = 0
 \end{equation}
 with
 \begin{equation}
 {\hat H} = g^{\mu\nu}\nabla_{\mu}\nabla_{\nu}{\hat 1} + \Bigl(
 {\hat P} - \frac{1}{6}R\;{\hat 1}\Bigr)\;\;\;.
 \end{equation}
 Here the hat over a symbol indicates that this symbol is a matrix in the
 space of field components:
 ${\hat 1} = \delta^A_{\;\;B}$ , ${\hat P} = P^A_{\;\;B}$ , etc.
 The matrix trace will be denoted $\tr$. 
 In (1.2), ${\hat P}$ is an arbitrary matrix potential,
 $R$ is the Ricci scalar
 \footnote{
 We use the conventions for which 
 $[\nabla_{\mu},\nabla_{\nu}]X^{\alpha} = R^{\alpha}_{\;\beta\mu\nu}X^{\beta},
 \;R_{\alpha\beta} = R^{\mu}_{\;\alpha\mu\beta},\; 
 R= g^{\alpha\beta}R_{\alpha\beta}.$}
 of the metric $g_{\mu\nu}$ , and $\nabla_{\mu}$ is the covariant derivative 
 with respect to any connection defining the commutator
\begin{equation}
[\nabla_{\mu},\nabla_{\nu}]\delta \varphi = 
{\hat {\cal R}}_{\mu\nu} \delta\varphi \equiv {\cal R}^A_{\;B\mu\nu}
\delta\varphi^B\;.
\end{equation}
The full set of external fields is thus a metric, a connection, and a 
potential, and the respective set of field strengths is
 \begin{equation}
 \Re = \Bigl( R^{\alpha}_{\;\beta\mu\nu}, {\hat {\cal R}}_{\mu\nu}, 
 {\hat P}\Bigr)\;\;
 \end{equation}
where $R^{\alpha}_{\;\beta\mu\nu}$ is the Riemann tensor of the metric
in (1.2), and ${\hat {\cal R}}_{\mu\nu}$ is the commutator curvature in
(1.3).

By solving the Jacobi and Bianchi identities
\begin{eqnarray}
\nabla_{\gamma}{\hat {\cal R}}_{\alpha\beta} +
 \nabla_{\beta}{\hat {\cal R}}_{\gamma\alpha} +
 \nabla_{\alpha}{\hat {\cal R}}_{\beta\gamma} = 0\;\;,\\
 \nabla_{\sigma}R_{\mu\nu\beta}{}^{\alpha} +
\nabla_{\nu}R_{\sigma\mu\beta}{}^{\alpha} +
\nabla_{\mu}R_{\nu\sigma\beta}{}^{\alpha} = 0
\end{eqnarray}
with the vacuum initial conditions
\footnote{
i.e. with the condition of absence of the incoming waves.},
the commutator and Riemann curvatures can be expressed (nonlocally)
through their contractions
\begin{equation}
{\hat J}^{\mu} \equiv \nabla_{\nu}{\hat {\cal R}}^{\mu\nu}\quad ,\quad
J^{\mu\nu} \equiv R^{\mu\nu} - \frac{1}{2} g^{\mu\nu} R
\end{equation}
(see Refs. [15-20,23] and Appendix A below). The set of quantities
\begin{equation}
J = ( J^{\mu\nu}, {\hat J}^{\mu}, {\hat P} )
\end{equation}
represents the physical sources of external fields (1.4).
Specifically, the commutator curvature 
${\hat {\cal R}}_{\mu\nu}$ is a generalization of the Maxwell tensor,
and 
${\hat J}^{\mu}$ in (1.7) is a counterpart of the electromagnetic current.
The expressions for the field strengths in terms of the currents are
obtained iteratively but the conservation of the vector and tensor currents
holds exactly:
\begin{equation}
\nabla_{\mu}{\hat J}^{\mu} = 0\quad ,\quad \nabla_{\mu}J^{\mu\nu} = 0\quad .
\end{equation}
Specifically, for the commutator curvature one has [24]
\begin{equation}
[\nabla_{\mu},\nabla_{\nu}]{\hat {\cal R}}^{\alpha\beta} = 
{\hat {\cal R}}_{\mu\nu}{\hat {\cal R}}^{\alpha\beta} - 
{\hat {\cal R}}^{\alpha\beta}{\hat {\cal R}}_{\mu\nu} +
R^{\alpha}_{\;\gamma\mu\nu}{\hat {\cal R}}^{\gamma\beta} - 
R^{\beta}_{\;\gamma\mu\nu}{\hat {\cal R}}^{\gamma\alpha}
\end{equation}
whence
\begin{equation}
\nabla_{\mu}\nabla_{\nu}{\hat {\cal R}}^{\mu\nu} = 
\frac{1}{2}\;[\nabla_{\mu},\nabla_{\nu}]{\hat {\cal R}}^{\mu\nu} = 0\;\;.
\end{equation}

The vacuum energy of the quantum field $\varphi$ is calculated below
as a functional of the external-field strengths (1.4) to the lowest  
nonvanishing order in the number of loops and the power of $\Re.$
Since solving the identities (1.5),(1.6) with the vacuum initial
conditions is a part of the calculational procedure in covariant
perturbation theory [15-18], the result will be expressed through
the sources (1.8). At intermediate stages the potential term in (1.2)
\begin{equation}
{\hat Q} \equiv {\hat P} - \frac{1}{6} R\; {\hat 1}
\end{equation}
will as a whole figure in the capacity of a source but the final result
will be expressed in terms of ${\hat P}$ for the reason explained in
Sec.8.

The calculation in the paper is carried out under a number of limitations on
the external fields whose significance is discussed in conclusion. One
limitation is already predetermined: we consider only the fields of sources,
i.e. neither the gravitational field nor the field represented by
the commutator curvature will contain an incoming wave. It will be assumed 
that the sources of external fields have their supports in a  spacetime
tube with compact spatial sections and a timelike boundary. 
Their domain of nonstationarity will be assumed compact in both space and 
time. 
It will be assumed that the metric has no singularities and horizons since
here we consider only the case of finite particle production \footnote{
i.e. finite energy production. The {\it number} of created massless 
particles may be infinite.}. Finally, in the case of the
commutator curvature there will be one futher limitation, namely that
the vector source in (1.8) does not radiate classically. In the specific
case where ${\hat {\cal R}}_{\mu\nu}$ is the Maxwell tensor, this limitation
means that the external electromagnetic field contains no outgoing wave.
No such limitation emerges in the case of the gravitational field.

\subsection*{\bf The limits ${\cal I}^+$ and $i^+$ .}

$$ $$

An important role in what follows is played by the limits along the null
and timelike geodesics  traced towards the future. By the assumption above,
 all causal geodesics reach the future null infinity or the future timelike
 infinity (see, e.g., [25]). Here we introduce the notations and
 reference equations pertaining to these limits that will next be
 used throughout the paper.

 When dealing with the null geodesics, it is useful to build a Bondi-Sachs 
 type [26,27] frame by choosing an arbitrary timelike geodesic (referred to
 as the central geodesic) and drawing the family of the future
 light cones with vertices on this geodesic (Fig.1). Let
 \begin{equation}
 u(x) = \mbox{const.}\quad , \quad (\nabla u)^2 \equiv 0
 \end{equation}
 (with $\nabla u$ past directed) be the equation of this family,
 $4\pi r^2(x)$ be the area of a 2~-~dimensional section of a given cone
 in the induced metric,
 and $\phi (x)$ be a set of two coordinates labelling the null generators
 of a given cone. The $\phi (x)$ takes values on a 2-sphere and satisfies the
 orthogonality condition $(\nabla \phi, \nabla u) \equiv 0$.
 Then $u$ is the retarded time, and $r$ is the luminosity distance along the 
 light rays that cross the central geodesic (the radial light rays).

 One property of the Bondi-Sachs frame used in the paper is the fact that, 
 if the point ${\bar x}$ is in the causal past of the point $x$, then
 $u({\bar x}) \le u(x)$. Indeed, let $o$ be the point at which the central
 geodesic crosses the past light cone of $x$, and ${\bar x}$ be any point 
 belonging to this  cone or its interior. The past light cone of $x$ lies
 entirely outside the light cone of $o$ as illustrated in Fig.2. We have
 $u(x) = u(o)$,
 and $u({\bar x}) \le u(o)$ since all points for which $u$ is bigger than
 $u(o)$ are inside the future light cone of $o$.

 When calculating only in the  asymptotically flat domain
 of infinite luminosity distance, the compact domain may be cut out. The
 choice of the central geodesic is then immaterial, the retarded time
 is measured by an observer at infinity, and $\phi$ labels the points
 of the celestial 2-sphere ${\cal S}$ to which the radial light rays come
 as $r\to\infty$.

 To every radial ray there corresponds a 2-parameter bundle of parallel
 rays that come simultaneously to the same point of the celestial
  sphere and are, therefore, indistinguishable at infinity.
(They generate a null hyperplane discussed in Sec.4 .) Since
  the radial rays make a 3-parameter family, there is all together a
  5-parameter family of light rays in spacetime but only three parameters
  are distinguishable in the principal approximation at infinity:
  $u, \phi$.
  Changing the central geodesic signifies going over to the parallel rays.

  We shall denote ${\cal I}^+[u,\phi,r\to\infty]$, or ${\cal I}^+[u,\phi]$,
   or ${\cal I}^+$, the limit of infinite luminosity distance $r$ along
   the null geodesic that, when traced to the future, comes at the instant 
   $u$ of retarded time to the point $\phi$ of the celestial 2-sphere
   ${\cal S}$. By using the arbitrariness $u\to f(u)$, the retarded time
   will be 
   normalized to coincide with the proper time of a  distant observer at
   rest that registers the outgoing rays. This normalization condition
   reads
   \begin{equation}
   (\nabla u,\nabla r)\biggl|_{{\cal I}^+} = -1\quad ,
   \end{equation}
   and the asymptotic form of the metric in the Bondi-Sachs frame is 
   \begin{equation}
   ds^2\biggl|_{{\cal I}^+} = -du^2 - 2dudr + r^2 (d\theta^2 + 
   \sin^2 \theta d\varphi^2)
   \end{equation}
   where 
   $(\theta,\varphi) = \phi \in {\cal S}$. 
   The integral over the 2-sphere ${\cal S}$ (normalized to have the area
   $4\pi$) will be denoted
   \begin{equation}
   \int d^2 {\cal S} (\phi) = \int\limits_0^{2\pi} d\varphi 
   \int\limits_0^{\pi} d\theta \sin \theta \quad .
   \end{equation}

   To build a finite vector basis at ${\cal I}^+$ define the scalar
   \begin{equation}
   v\biggl|_{{\cal I}^+} = 2r + u + O\Bigl(\frac{1}{r}\Bigr)
   \end{equation}
   and the complex null vector tangent to the sphere ${\cal S}$
   \begin{equation}
   m_{\alpha}\biggl|_{{\cal I}^+} = r ( \nabla_{\alpha}\theta + 
   \mbox{i}\sin \theta \nabla_{\alpha}\varphi) + O(r^0)\quad .
   \end{equation}
   The resultant null tetrad at ${\cal I}^+$
   \begin{equation}
   e_{\alpha}(\mu) = \nabla_{\alpha}u,\;\nabla_{\alpha}v,\;m_{\alpha},
   \;m^*_{\alpha} \qquad (\mu = 1,2,3,4)
   \end{equation}
   with $\mu$ labelling the vectors of the tetrad, and $m^*$
   complex conjugate to $m$ satisfies the orthonormality relations
   $$
   (\nabla u)^2 = (\nabla v)^2 = m^2 = (\nabla u,\; m) = 
   (\nabla v,\; m) = 0\quad,\nonumber
$$
\begin{equation}
 (\nabla u,\;\nabla v) = -2\quad ,\quad
   (m,\;m^*) = 2
   \end{equation}
   and can be used to expand the metric as follows:
  \begin{equation}
  g_{\mu\nu}\biggl|_{{\cal I}^+} = - \frac{1}{2}
  (\nabla_{\mu}u\nabla_{\nu}v + \nabla_{\mu}v\nabla_{\nu}u) + 
  \frac{1}{2} (m_{\mu}m^*_{\nu} + m^*_{\mu}m_{\nu})\quad .
  \end{equation}

  Consider now a timelike geodesic, and let $s$ be the proper 
  time along this geodesic. As $s\to\infty$, the particle moving along the geodesic 
  will go out of the domain of nonstationarity of external fields
  with the energy $E$ (per unit rest mass). Only the geodesics with 
  $E>1$ that reach the asymptotically flat infinity are relevant to
  the present discussion. We shall replace $E$ with the {\it boost}
  parameter
  \begin{equation}
  \gamma = \frac{\sqrt{E^2 - 1}}{E}\quad , \quad 0 < \gamma < 1\;\;
  \end{equation}
  and denote $i^+[\gamma,\phi,s\to \infty],$ or $ i^+[\gamma,\phi],$
  or $i^+,$ the limit $s\to\infty$ along the geodesic that comes to 
  infinity with a given value of $\gamma$ to a given point $\phi$ of the 
  celestial 2-sphere.

  A distinction of this case from the case of null geodesics (apart from
  $\gamma = 1$ in the latter case) is that the timelike geodesics differing
  by translations including the time translations are indistinguishable
  at infinity. There are again three parameters that register at infinity but
  the parameters are now $\gamma$ and $\phi$, and there is a 3-parameter 
  congruence of the geodesics that come to infinity with one and the 
  same values of $\gamma$ and $\phi$. (This congruence will be considered in
  Sec.7 .) All together, there is  a 6-parameter family of timelike geodesics
  that come to infinity.

  The Bondi-Sachs frame can be used also at $i^+$ with the 
  asymptotic form of the metric in (1.15). For the 
  asymptotically flat metric at $i^+(\gamma>0)$ one may use the vector
  basis (1.21) or the related basis
  \begin{equation}
  g_{\mu\nu}\biggl|_{i^+} = - \nabla_{\mu}t \nabla_{\nu}t + 
  \nabla_{\mu}r \nabla_{\nu}r + \frac{1}{2}(m_{\mu}m^*_{\nu} + 
  m^*_{\mu}m_{\nu})
  \end{equation}
  with
  \begin{equation}
  t\biggl|_{i^+} = r + u + \mbox{const.}\quad ,
  \quad \nabla u = \nabla t - \nabla r\quad , 
  \quad \nabla v = \nabla t + \nabla r\quad .
  \end{equation}

  The parameters $\phi$ of a point at $i^+$ or ${\cal I}^+$ can be 
  replaced by a unit direction vector at infinity whose spatial
  components in the Minkowski frame at $i^+$
  \begin{equation}
  ds^2\biggl|_{i^+} = -\; dt^2 + \delta_{ik} d\x^i d\x^k
  \end{equation}
  will be denoted $n_i = n_i(\phi),\;\;i=1,2,3.$ 
  With the Euler parametrization $\phi = (\theta, \varphi)$ in (1.15), 
  \begin{equation}
  n_1(\phi) = \sin\theta\sin\varphi\quad , \quad n_2(\phi) = 
  \sin\theta\cos\varphi\quad , \quad n_3(\phi) = \cos\theta\quad,
  \end{equation}
  and
  \begin{equation}
  \x^i = r\;n^i(\phi)\quad , \quad n^i = \delta^{ik}n_k\quad , 
  \quad n^in_k = 1\quad .
  \end{equation}
  The basis vectors in (1.23) have in the Minkowski frame (1.25) the 
  components
  \begin{equation}
  \nabla_{\mu}r = \delta_{\mu}^{\;\;i}n_i\quad ,\quad
  \frac{1}{2}(m_{\mu}m_{\nu}^* + m_{\mu}^*m_{\nu}) = 
  \delta_{\mu}^{\;\;i}\delta_{\nu}^{\;\;k}(\delta_{ik} - n_in_k)\quad .
  \end{equation}

  The limits $i^+$ and ${\cal I}^+$ are related. As $\gamma\to 1$ the
  geodesic at $i^+$ approaches the null geodesic that comes to
  ${\cal I}^+$ at late time. Therefore, for an analytic function
  $X$, the sequence of limits $i^+$ and $\gamma \to 1$ should
  coincide with the future of ${\cal I}^+$:
  \begin{equation}
  \Bigl(X\biggl|_{i^+[\gamma,\phi,s\to\infty]}\Bigr)_{\gamma\to 1} = 
  \Bigl(X\biggl|_{{\cal I}^+[u,\phi,r\to\infty]}\Bigr)_{u\to\infty} 
  \end{equation}
  under the substitutions
  \begin{equation}
  s = \frac{\sqrt{1 - \gamma^2}}{\gamma}\;r\quad , 
  \quad \gamma = 1 - \frac{u}{r} + O\Bigl(\frac{u^2}{r^2}\Bigr)\quad .
  \end{equation}

  \subsection*{\bf Vacuum radiation in the expectation-value spacetime.}

 $$ $$

  In the framework of the problem of particle creation by an external
  field, it is impossible to answer the question where the energy
  of the created real particles comes from. Clearly, the vacuum
  particle production is only a mechanism of the energy transfer.
  The energy comes ultimately from the external field but this 
  answer assumes that the external field should stop being external.

  In the self-consistent setting of the problem, the external fields become
  expectation values evolving from an initial quantum state. Specifically 
  in the presently considered model, $\varphi^A$ is the full set of quantum fields.
  Some of these fields have nonvanishing  expectation values represented by the
  set of curvatures in (1.4). The expectation-value equations are obtained
  by the rules of Ref. [14] from the action
  \begin{equation}
  \frac{1}{16\pi} \int dx g^{1/2} R + S_{\mbox{\scriptsize source}} + 
  S_{\mbox{\scriptsize vac}}
  \end{equation}
  where $S_{\mbox{\scriptsize source}}$ is the (renormalized) classical action
  for ${\hat {\cal R}}_{\mu\nu}$ and ${\hat P}$, and 
  $S_{\mbox{\scriptsize vac}}$ is the sum of all vacuum loops. The equations
  for the expectation value of the metric can be written in the form of the
  Einstein equations
  \begin{equation}
  R^{\mu\nu} - \frac{1}{2} g^{\mu\nu} R = 8\pi 
  \Bigl( T^{\mu\nu}_{\mbox{\scriptsize source}} + 
  T^{\mu\nu}_{\mbox{\scriptsize vac}}\Bigr) \quad ,
  \end{equation}
  \begin{equation}
  T^{\mu\nu}_{\mbox{\scriptsize source}} = \frac{2}{g^{1/2}}\;
  \frac{\delta S_{\mbox{\scriptsize source}}}{\delta g_{\mu\nu}}
  \end{equation}
  in which there appears some extra source 
  $T^{\mu\nu}_{\mbox{\scriptsize vac}}$ ,
  the energy-momentum tensor of the  in-vacuum. This source is a subject
  of the calculation below. Eq. (1.32) should be supplemented with similar
   equations for ${\hat {\cal R}}_{\mu\nu}$ and ${\hat P}$
   ( and, of course, there should be equations for the higher-order 
   correlation functions, also derivable from (1.31), which will not be 
   discussed here).

   The energy conservation law in the expectation-value spacetime is a 
   consequence of Eq. (1.32) with the asymptotically flat boundary conditions.
   It is expressed in the existence of a conserved ADM mass 
   $M_{\mbox{\scriptsize ADM}}$ which equals the total energy of all
   sources and waves in the initial state, and a non-conserved Bondi mass
   $M(u)$ defined by the metric at ${\cal I}^+$ and accounting for radiation
   [25]. The difference $M(-\infty) - M(u)$ is the energy radiated from an
   isolated system by the instant $u$ of retarded time, and
   \begin{equation}
   M(-\infty) = M_{\mbox{\scriptsize ADM}}\quad .
   \end{equation}
   
The equation of the energy balance, or the Bondi-Sachs equation [26-27,23],
is of the form
\begin{eqnarray}
- \frac{d\mbox{M}(u)}{du} &=&\frac{1}{4\pi} \int d^2{\cal S}(\phi)\,
\biggl|\frac{\partial}{\partial u}\;\mbox{\bf C}_{
\mbox{\scriptsize grav}}(u,\phi)\biggr|^2     
 \\
& + &\int d^2{\cal S}(\phi)\,\Bigl(\frac{1}{4}\;r^2\;
T^{\mu\nu}_{\mbox{\scriptsize source}}\nabla_{\mu}v\nabla_{\nu}v
+\frac{1}{4}\;r^2 \;
T^{\mu\nu}_{\mbox{\scriptsize vac}}\nabla_{\mu}v\nabla_{\nu}v
\Bigr)\Biggl|_{{\cal I}^+[u,\phi,r\to\infty]}\nonumber
\end{eqnarray}
and is to be solved with the initial condition (1.34). The first
term in (1.35) is the energy flux of the outgoing gravitational
waves ($\partial \mbox{\bf C}_{\mbox{\scriptsize grav}}/\partial u$
is the  complex gravitational news function [26,27] defined by
the metric at ${\cal I}^+$), and the remaining terms are the energy fluxes
through ${\cal I}^+$ of all sources of the {\it mean} gravitational
field including the vacuum. The news function also contains a
contribution induced by the vacuum [23].

We are presently interested only in the contribution of 
$T^{\mu\nu}_{\mbox{\scriptsize vac}}$ to the radiation flux. This
contribution will be nonvanishing only when the fields solving
the expectation-value equations create real massless particles from the 
vacuum. If the other contributions are absent
\footnote{
As is the case, for example, if the radiation of waves, both 
gravitational and matter,  is banned by the symmetry of the initial state.} :
\begin{equation}
- \frac{d\mbox{M}(u)}{du} =
\int d^2{\cal S}(\phi)\;\Bigl(\frac{1}{4}r^2 
T^{\mu\nu}_{\mbox{\scriptsize vac}}\nabla_{\mu}v\nabla_{\nu}v
\Bigr)\Biggl|_{{\cal I}^+[u,\phi,r\to\infty]}
\end{equation}
then the total radiated energy equals the total energy of created particles:
 \begin{equation}
 \mbox{M}(-\infty) - \mbox{M}(\infty) = \int\limits^{\infty}_{-\infty}
 \biggl(-\frac{d\mbox{M}}{du}\biggr) du = \sum_{p}\epsilon_p < \mbox{in vac}\; |
 a^{+p}_{\mbox{\scriptsize out}} a^p_{\mbox{\scriptsize out}} 
 |\; \mbox{in vac} >
 \end{equation}
 (see, e.g., [22]). Here 
 $a^{+p}_{\mbox{\scriptsize out}},a^p_{\mbox{\scriptsize out}}$ 
 are the creation and annihilation operators for the quanta of the field
 $\varphi$ in the out-vacuum, 
 and $\epsilon_p$ is the energy in the out-mode $p$.

The inference from Eq. (1.37) is that the massless particles
created from the vacuum are radiated through the future null infinity
of the expectation-value spacetime. As seen from the initial condition
(1.34), this vacuum radiation takes its energy from the ADM mass of
 the  expectation value of the metric. Since the ADM mass is conserved, it 
 can be calculated on a spacelike hypersurface taken in the remote
 past. The $T^{\mu\nu}_{\mbox{\scriptsize vac}}$ is a retarded functional
 of the mean fields, and it vanishes in the remote past ( see Sec.5 below).
 Therefore, the ADM mass remains unaffected by quantum corrections and
 equal to the energy in $T^{\mu\nu}_{\mbox{\scriptsize source}}$ 
 calculated on an initial hypersurface ( plus the energy of an incoming
 gravitational wave if any)
 \footnote{
 This has an important consequence that the ADM mass of the mean metric
 remains positive if $T^{\mu\nu}_{\mbox{\scriptsize source}}$
  is energy-dominant [16].}.
It thus equals the energy of the "external" fields in the past from
their domain of nonstationarity. In this way the energy conservation
is restored.

Eq. (1.36) with
$T^{\mu\nu}_{\mbox{\scriptsize vac}}$
obtained from the effective action will be used below to calculate the
quantity (1.37). Special attention will be payed to the positivity of this
quantity.
}

\newpage

{\renewcommand{\theequation}{2.\arabic{equation}}

\begin{center}
\section{\bf    The vacuum energy-momentum tensor}
\end{center}
\subsection*{\bf The effective action.}

$$ $$

The one-loop vacuum action for the field $\varphi$ is 
\begin{equation}
S_{\mbox{\scriptsize vac}} = \frac{\mbox{i}}{2} \mbox{Tr} \log {\hat H}
\end{equation}
where Tr denotes the functional trace, and ${\hat H}$ is the operator
(1.2). Upon the calculation of the loop [14-18], this action takes the
form
\begin{eqnarray}
S_{\mbox{\scriptsize vac}} &=& S(2) + S(3) + O[\Re^4]\quad,\\
S(2) &=& \frac{1}{2(4\pi)^2} \int dx g^{1/2} \,\tr \,\sum_{i=1}^{5}
\gamma_i(-\Box_2)\Re_1\Re_2(i)\quad , \\
S(3) &=& \frac{1}{2(4\pi)^2} \int dx g^{1/2} \,\tr \,\sum_{i=1}^{29}
\Gamma_i(-\Box_1,-\Box_2,-\Box_3)\Re_1\Re_2\Re_3(i)\quad , \\
\Box &\equiv & g^{\mu\nu}\nabla_{\mu}\nabla_{\nu}
\end{eqnarray}
which is the general form of a functional of the field strengths (1.4)
expanded over a basis of nonlocal invariants [20]. The term $S(2)$ is
a linear combination of five basis invariants $\Re_1\Re_2(i)$ of
 second order in $\Re$ with the operator coefficients (form factors) 
 $\gamma_i(-\Box)$. The term $S(3)$ is a linear combination of twenty
 nine basis invariants $\Re_1\Re_2\Re_3(i)$ of third order in $\Re$
 with the form factors $\Gamma_i(-\Box_1,-\Box_2,-\Box_3).$ A
 calculation of the effective action in any specific model or
 approximation boils down to obtaining the form factors in (2.2).
 The expansion (2.2) can also be assumed as a basis of a phenomenological
 theory of the vacuum [19-22].

 The basis invariants of second order in $\Re$ and their respective
 one-loop form factors are of  the form [15]
 \begin{eqnarray}
 \Re_1\Re_2(1)  & = &    R_{1\mu\nu}R_2^{\;\mu\nu} {\hat 1}\quad ,\\
 \Re_1\Re_2(2)  & = &    R_1R_2 {\hat 1}\quad ,                   \\
 \Re_1\Re_2(3)  & = &    {\hat P}_1 R_2 \quad ,                   \\
 \Re_1\Re_2(4)  & = &    {\hat P}_1 {\hat P}_2 \quad ,            \\
 \Re_1\Re_2(5)  & = &    \R_{1\mu\nu} \R_2^{\;\mu\nu}\quad ,       
\end{eqnarray}
\begin{eqnarray}
 \gamma_1(-\Box) &=&  \frac{1}{60}\biggl( - \log
 \Bigl(-\frac{\Box}{c^2}\Bigr) + \frac{16}{15}\biggr) \quad , \\
 \gamma_2(-\Box) &=&  \frac{1}{180}\biggl(  \log
 \Bigl(-\frac{\Box}{c^2}\Bigr) - \frac{37}{30}\biggr) \quad , \\
 \gamma_3(-\Box) &=& - \frac{1}{18} \quad ,\\
 \gamma_4(-\Box) &=&  - \frac{1}{2} \log
 \Bigl(-\frac{\Box}{c^2}\Bigr) \quad , \\
 \gamma_5(-\Box) &=&  \frac{1}{12}\biggl( - \log
 \Bigl(-\frac{\Box}{c^2}\Bigr) + \frac{2}{3}\biggr) 
 \end{eqnarray}
where the parameter $c^2>0$ accumulates the ultraviolet arbitrariness.
This parameter doesn't affect the vacuum current at null infinity
(see [21] and the calculation below).

The third-order action contains no arbitrary parameters. Among the basis
invariants of third order in $\Re$, eleven contain no derivatives, for
example
\begin{equation}
\Re_1\Re_2\Re_3(1) = {\hat P}_1{\hat P}_2{\hat P}_3\quad ,
\end{equation}
fourteen contain two derivatives, for example
\begin{equation}
\Re_1\Re_2\Re_3(12) = \R_1^{\;\alpha\beta}\nabla^{\mu}\R_{2\mu\alpha}
\nabla^{\nu}\R_{3\nu\beta}\quad ,
\end{equation}
three contain four derivatives, for example
\begin{equation}
\Re_1\Re_2\Re_3(26) = \nabla_{\alpha}\nabla_{\beta}
R_1^{\;\mu\nu}\nabla_{\mu}\nabla_{\nu}R_{2}^{\;\alpha\beta}{\hat P}_3\quad ,
\end{equation}
and one contains six derivatives :
\begin{equation}
\Re_1\Re_2\Re_3(29) = \nabla_{\lambda}\nabla_{\sigma}
R_1^{\;\alpha\beta}\nabla_{\alpha}\nabla_{\beta}R_{2}^{\;\mu\nu}
\nabla_{\mu}\nabla_{\nu}R_3^{\;\lambda\sigma}{\hat 1} \quad .
\end{equation}
The full table of third-order invariants and their one-loop form
factors is given in [17].

Since the operator arguments of the form factors in (2.4) commute, the form
factors themselves are ordinary functions \footnote{
Beyond third order in $\Re$ this is no more the case [19,20].} .
As analytic functions, they are defined by their spectral forms
(in each argument) with the resolvents $1/(\Box - m^2)$.

The action in the form (2.2)-(2.4) (i.e. with the loop done) determines 
both the matrix elements between the in- and out- vacua and the expectation
values in the in-vacuum [14].
The difference is in the boundary conditions for the resolvents of the nonlocal
operators. When the action (2.2) is varied, the resolvents are regarded as
obeying the variational rule
\begin{equation}
\delta\; \frac{1}{\Box - m^2} = - \frac{1}{\Box - m^2}\;\delta\Box\;
\frac{1}{\Box - m^2}\quad ,
\end{equation}
and, after the variation has been completed, they are identified with
the Feynman Green functions in the case of matrix elements and with the
retarded Green functions in the case of expectation values [14].
Introducing a notation for the result of this procedure,
we may write for $T^{\mu\nu}_{\mbox{\scriptsize vac}}$ in (1.32) the 
expression
\begin{equation}
T^{\mu\nu}_{\mbox{\scriptsize vac}} = \frac{2}{g^{1/2}} 
\frac{\delta S_{\mbox{\scriptsize vac}}}{\delta g_{\mu\nu}}\biggl|_{\Box 
\to \Box_{\mbox{\scriptsize ret}}} \quad .
\end{equation}

For obtaining $T^{\mu\nu}_{\mbox{\scriptsize vac}}$ from the action
(2.2) one needs to know the variational  derivatives of the commutator 
curvature and potential with respect to the metric. The dependence
of $\R_{\mu\nu}$ and ${\hat P}$ on the metric is different in different
models [24]. The calculation below is carried out for the case where 
$\R_{\mu\nu}$ and ${\hat P}$ are metric independent but, as explained
in conclusion, the final result is valid for arbitrary local 
$\R_{\mu\nu}$ and ${\hat P}$ .

\subsection*{\bf Obtaining $T^{\mu\nu}_{\mbox{\scriptsize vac}}$ at
 ${\cal I}^+$ .}

 $$  $$

The $T^{\mu\nu}_{\mbox{\scriptsize vac}}(x)$ is here to be calculated
only for $x$ at ${\cal I}^+$, and only the terms that contribute to
the energy flux through ${\cal I}^+$ are to be retained. These are the
terms of order $r^{-2}(x),\;\; x\to{\cal I}^+$ with  $r$ the luminosity
distance. Below, the notation
 $T^{\mu\nu}_{\mbox{\scriptsize vac}}\Bigl|_{{\cal I}^+}$
 is used for $T^{\mu\nu}_{\mbox{\scriptsize vac}}$
calculated up to terms $O(1/r^3)$, and the terms $O(1/r^3)$ are referred to
as vanishing at ${\cal I}^+$.

When computing $T^{\mu\nu}_{\mbox{\scriptsize vac}}\Bigl|_{{\cal I}^+}$
from the action (2.2) it is useful to have at hand the following 
behaviours that we quote from Refs.[21-23] and  Sec.4 below:
\begin{equation}
\frac{1}{\Box} X\biggl|_{{\cal I}^+} = O\Bigl(\frac{1}{r}\Bigr)\quad ,
\quad \log(-\Box) X\biggl|_{{\cal I}^+} = O\Bigl(\frac{1}{r^2}\Bigr)\quad ,
\end{equation}
\begin{equation}
\frac{\log(\Box_1/\Box_2)}{\Box_1 - \Box_2} X_1\;X_2\biggl|_{{\cal I}^+} = 
O\Bigl(\frac{1}{r^3}\Bigr)\quad .
\end{equation}
The behaviours of these and other form factors including the higher-order
ones can be obtained from the behaviour of the resolvent operator [21,22]:
\begin{equation}
\frac{1}{\Box - m^2}\;X\biggl|_{{\cal I}^+} \propto \frac{1}{r}\;
\exp(-|\mbox{const.}|m\sqrt{r})\quad .
\end{equation}
The behaviours of the field strengths (1.4) and their sources (1.8)
are in the most general case (see, e.g., Sec.8 below)
\begin{equation}
\Re\biggl|_{{\cal I}^+} = O\Bigl(\frac{1}{r}\Bigr)\quad , \quad
J\biggl|_{{\cal I}^+} = O\Bigl(\frac{1}{r^2}\Bigr)
\end{equation}
and, with the present assumptions, these behaviours soften. Hence it is
seen that the terms of $T^{\mu\nu}_{\mbox{\scriptsize vac}}(x)$ in
which the curvature $\Re$  appears at the observation point $x$ can
often be discarded at ${\cal I}^+.$ Also the terms in which there
forms a positive integer power of the operator $\Box$ acting at the
observation point can often be discarded since
\begin{equation}
\Box X\biggl|_{{\cal I}^+} = O\Biggl(\frac{1}{r}\;X\Bigr)\quad .
\end{equation}
Another useful fact is
\begin{equation}
g^{\mu\nu}\nabla_{\mu}X_1\nabla_{\nu}X_2\biggl|_{{\cal I}^+} =
O\Bigl(\frac{1}{r}X_1X_2\Bigr)\quad .
\end{equation}
Both (2.26) and (2.27) follow from the form of the metric at ${\cal I}^+$,
Eq. (1.15).

Using Eq. (2.23) it has been shown in Ref. [23] that the variations
of second-order form factors $\gamma$ make a vanishing contribution
at ${\cal I}^+$. This fact makes an  essential difference with the case
of two dimensions (see below). Owing to this fact, the only contribution
of the action $S(2)$ to $T^{\mu\nu}_{\mbox{\scriptsize vac}}$ that survives
at ${\cal I}^+$ comes from varying the curvatures $\Re$ in the  products
$\Re_1\Re_2(i).$

Since varying destroys the curvature, the effective action of third order in
$\Re$ enables one to obtain $T^{\mu\nu}_{\mbox{\scriptsize vac}}$ only
to second order in $\Re.$
Therefore, in the action $S(3)$ only the curvatures $\Re$ in the products
$\Re_1\Re_2\Re_3(i)$ need to be varied. The contributions of variations
 of the third-order form factors $\Gamma$ are already $O[\Re^3].$
 It is also useful to keep in mind that any commutator of the derivatives
 $\nabla$ with each other or with the form factors is $O[\Re].$

 Using the rules above to discard the terms vanishing at ${\cal I}^+$ we
 obtain
 \begin{eqnarray}
 T^{\mu\nu}_{\mbox{\scriptsize vac}}\biggl|_{{\cal I}^+} &=&  
 \frac{1}{(4\pi)^2}\;\tr \,\nabla^{\mu}\nabla^{\nu}\Bigl(
 \gamma_1(-\Box) + 2\gamma_2(-\Box)\Bigr)R{\hat 1}   \nonumber \\
&+& \frac{1}{(4\pi)^2}\;\tr\,\Bigl\{\nabla^{\mu}
[\nabla_{\alpha},\gamma_1(-\Box)]
 R^{\alpha\nu} + 
 \nabla^{\nu}[\nabla_{\alpha},\gamma_1(-\Box)]R^{\alpha\mu}  \nonumber\\
 & & - \frac{1}{2}\nabla^{\mu}[\nabla^{\nu}, \gamma_1(-\Box)]R - 
 \frac{1}{2}\nabla^{\nu}[\nabla^{\mu},\gamma_1(-\Box)]R\Bigr\}{\hat 1}
\nonumber \\ & + &
 \frac{2}{g^{1/2}}\;\frac{\delta S(3)}{\delta g_{\mu\nu}} + O[\Re^3]
 \end{eqnarray}
 where the variation of the action $S(3)$ is not yet calculated but there appear
 the commutators which are of the same order $\Re^2$ as the variation
 of $S(3).$ The commutators emerge when the terms linear in $\Re$ are
 brought to their form in (2.28) with the aid of the Bianchi identities.
 Taking these commutators into account is necessary for 
 $T^{\mu\nu}_{\mbox{\scriptsize vac}}$ to have the correct form at 
 ${\cal I}^+$ (Eq. (2.36) below). Both commuting and varying of the
 operator functions is accomplished with the aid of their spectral
 forms [17,23]. By (2.21), all nonlocal operators in (2.28) have the
 retarded boundary conditions.

 Consider now the variation of the action $S(3).$ A generic term in 
 $\delta S(3)$ is of the form
 \begin{equation}
 \int dx g^{1/2} \Gamma(-\Box_1,-\Box_2,-\Box_3) \delta \Re_1\;\Re_2\;\Re_3
 + O[\Re^3]
 \end{equation}
 where $\delta \Re$ has the structure
 \begin{equation}
 \delta \Re(x) = \nabla\nabla\delta g(x) + O[\Re]\quad .
 \end{equation}
 Since, when going over to the variational derivative, the operator acting
  on $\delta g(x)$ transposes, the respective term in the variational
  derivative at a point $x$ has the following structure:
  \begin{equation}
  \nabla\nabla\Gamma(-\Box,-\Box_2,-\Box_3)\Re(x_2)\Re(x_3)\biggl|_{
  x_2=x_3=x}
  \end{equation}
  where the operators $\nabla\nabla$ and the first argument $-\Box$ of the
  form factor act at the observation point $x$. In (2.31), first $\Box_2$ 
  acts on $\Re(x_2)$ and $\Box_3$ acts on $\Re(x_3)$ with subsequently
  making the points $x_2$ and $x_3$ coincident with the observation point
  $x$, and next the argument ${-\Box}$ of the form factor and the
  derivatives $\nabla\nabla$ act on the thus obtained function of the
  observation point.

  The important fact is that one of the operator arguments of $\Gamma$,
  the one that in (2.29) acts on $\delta \Re$, becomes an overall
  operator in (2.31). Therefore, the theorem in [21,22] applies by which
  the behaviour of the function (2.31) as $x\to {\cal I}^+$ is determined
  by the behaviour of the function
  $ \Gamma(-\Box,-\Box_2,-\Box_3) $ as $-\Box\to 0$ with $-\Box$ the 
  argument that becomes an overall operator in (2.31). For the function
  (2.31) to be $O(1/r^2)$, the form factor should be [21]
  \begin{equation}
  \Gamma(-\Box,-\Box_2,-\Box_3) = O\Bigl(\log(-\Box)\Bigr)\quad ,
  \quad -\Box \to 0\quad.
  \end{equation}
  {\it The} $\log(-\Box)$ {\it terms of the form factors at small}
  $\Box$ {\it determine the } ${1/r^2}$ {\it terms of}
  $T^{\mu\nu}_{\mbox{\scriptsize vac}}$ {\it at} ${\cal I}^+$
  [21]. Eq. (2.22) anticipates this fact.

  Thus, for the calculation of
  $T^{\mu\nu}_{\mbox{\scriptsize vac}}$ at ${\cal I}^+$, one
  doesn't need the exact form factors. One needs only their
  asymptotic behaviours in each argument at the limit where this
  argument is small, and the remaining arguments are fixed.
  These asymptotic behaviours are presented in [18]. All the form
  factors $\Gamma$ behave as follows $(k = 1,2,3)$:
  \begin{eqnarray}
  \Gamma_i(-\Box_1,-\Box_2,-\Box_3) &=& 
  \frac{1}{\Box_k}A_i^{\;k}(\Box_m,\Box_n)
  + \log(-\Box_k)B_i^{\;k}(\Box_m,\Box_n) 
 + O(\Box_k^{\;0})\;\;,\nonumber \\
  & &-\Box_k\to 0 \;\;,\;\; m\ne k,\;\;n\ne k,\;\;m<n
  \end{eqnarray}
where $A_i^{\;k}$ and $B_i^{\;k}$ are functions of the two arguments $\Box$ 
other than $\Box_k$. These behaviours differ from (2.32) since, in addition
to the expected $\log(-\Box)$ terms, they contain senior $1/\Box$ terms.
However, as shown in [23], the $1/\Box$ terms exactly cancel in the 
energy-momentum tensor. The variational derivative of $S(3)$ is a sum of
contributions of the form (2.31), and the form factors $\Gamma$ enter
this sum only in certain linear combinations. The $1/\Box$ terms cancel
in these combinations leaving the $\log(-\Box)$ terms as the leading
asymptotic terms at $\Box\to 0$ [23].

Thus we infer that the contributions of the form factors
$\Gamma$ to $T^{\mu\nu}_{\mbox{\scriptsize vac}}\Bigl|_{{\cal I}^+}$
are completely determined by the coefficient $B_i^{\;k}$ in
(2.33). The table of these coefficients is given in [18].
The simplest ones are the coefficients in the form factor $\Gamma_1$
of the basis invariant (2.16). They are of the form
\begin{equation}
B_1^{\;k}(\Box_m,\Box_n) = 
\frac{1}{3}\;\frac{\log(\Box_m/\Box_n)}{
\Box_m - \Box_n}
\end{equation}
for all the three values of $k$. The remaining $B_i^{\;k}$ are obtained 
by differentiating the generating expression
\begin{equation}
\frac{\log(j_m \Box_m/j_n \Box_n)}{
j_m\Box_m - j_n \Box_n}
\end{equation}
with respect to the auxiliary variables $\;j_m,j_n\;$ and subsequently
setting $\;j_m = j_n = 1\;$ [17,18]. All $B_i^{\;k}$ are the thus obtained
functions having also rational coefficients of the form $1/\Box_n$
or $\Box_m/\Box_n$ and rational additions.

The table of the functions $B_i^{\;k}$ in [18] can be considerably
simplified with the aid of  the technique outlined in Appendix A.
The use of this technique makes it possible to bring
$T^{\mu\nu}_{\mbox{\scriptsize vac}}\Bigl|_{{\cal I}^+}$ to the
final form below.

\subsection*{\bf The result for $T^{\mu\nu}_{\mbox{\scriptsize vac}}$
 at ${\cal I}^+$ .}

$$ $$

The result for $T^{\mu\nu}_{\mbox{\scriptsize vac}}\Bigl|_{{\cal I}^+}$ 
is of the form
\begin{equation}
T^{\mu\nu}_{\mbox{\scriptsize vac}}\biggl|_{{\cal I}^+} = 
\frac{1}{(4\pi)^2}\;\nabla^{\mu}\nabla^{\nu}\log(-\Box)\;I(x)
\end{equation}
and is determined by a single scalar $I(x)$:
\begin{equation}
I(x) = \tr {\hat I}_1(x) + \tr {\hat I}_2(x) + O[\Re^3]
\end{equation}
where ${\hat I}_1(x)$ and ${\hat I}_2(x)$ are the contributions of
first and second order in $\Re.$ The contribution of first order obtains
directly from (2.28) and from the expressions (2.11)-(2.12) for the  
second-order form factors:
\begin{equation}
{\hat I}_1 = - \frac{1}{180}\;R{\hat 1}\quad .
\end{equation}
 The contribution of second order obtains from the results in [18] with the 
 aid of the technique of Appendix A. It is of the following form:
 \begin{equation}
 {\hat I}_2 = \nabla_{\alpha}\nabla_{\beta}\nabla_{\mu}\nabla_{\nu}
 \;{\hat V}_{\mbox{\scriptsize grav}}^{\alpha\beta\mu\nu} + 
 \nabla_{\alpha}\nabla_{\beta}{\hat V}_{
 \mbox{\scriptsize vect}}^{\alpha\beta} +
 \nabla_{\alpha}\nabla_{\beta}{\hat V}_{
 \mbox{\scriptsize cross}}^{\alpha\beta} + 
 {\hat V}_{\mbox{\scriptsize scalar}} + \mbox{TREES}\quad ,
\end{equation}
\begin{eqnarray}
 {\hat V}^{\alpha\beta\mu\nu}_{\mbox{\scriptsize grav}} &=&
 \frac{1}{\Box_1\Box_2}\;\Bigl(\frac{1}{180}F(3,3) - \frac{1}{36}F(2,2)\Bigr) 
 J_1^{\;\alpha\beta}J_2^{\;\mu\nu}{\hat 1}\quad , \\
 {\hat V}^{\alpha\beta}_{\mbox{\scriptsize vect}} &=&
 - \frac{1}{\Box_1\Box_2}\;\Bigl(\frac{1}{6}F(2,2) - \frac{1}{3}F(1,1)\Bigr) 
 {\hat J}_1^{\;\alpha}{\hat J}_2^{\;\beta}\quad , \\
 {\hat V}^{\alpha\beta}_{\mbox{\scriptsize cross}} &=&
 - \frac{1}{\Box_2}\;\Bigl(\frac{1}{6}F(2,2) - \frac{1}{3}F(1,1)\Bigr) 
 J_1^{\;\alpha\beta}{\hat Q}_2\quad , \\
 {\hat V}_{\mbox{\scriptsize scalar}} &=&
 \Bigl(\frac{1}{2}F(1,1) - \frac{1}{6}F(0,0)\Bigr) 
 {\hat Q}_1{\hat Q}_2\quad 
 \end{eqnarray}
  where $J^{\alpha\beta},\;{\hat J}^{\alpha},\;{\hat Q}$ are the
  sources of external fields in (1.8) and (1.12), and $F(n,n)$ with
  $n=0,1,2,3$ are specific cases of the {\it vertex} operator
  \begin{equation}
  F(m,n) = \Bigl(\frac{\partial}{\partial j_1}\Bigr)^m
  \Bigl(\frac{\partial}{\partial j_2}\Bigr)^n
  \frac{\log(j_1\Box_1/j_2\Box_2)}{j_1\Box_1 - j_2\Box_2}\biggl|_{j_1=j_2=1}
  \quad .
  \end{equation}
  The $V$ 's in (2.39) will be referred to as the vertex terms. 
As distinct from the  $V$ 's, 
the trees \footnote{ The name "tree" is used here not quite in the
  usual sense. It should also be clear that both the vertices and trees are 
  terms of {\it the calculated loop}.}
   in (2.39) are the terms whose operator coefficients factorize into
   functions of one variable:
   \begin{equation}
   \mbox{TREES} = \frac{1}{90}\Bigl(\nabla_{\beta}\frac{1}{\Box}J^{
   \alpha\lambda}\Bigr)\Bigl(\nabla_{\alpha}\frac{1}{\Box}J^{\beta}_{\;\lambda}
   \Bigr){\hat 1} + \frac{1}{180}\;J^{\alpha\beta}\Bigl(\frac{1}{\Box}
   J_{\alpha\beta}\Bigr){\hat 1} - \frac{1}{360} R \Bigl(\frac{1}{\Box}R\Bigr)
   {\hat 1}\quad .
   \end{equation}

   The fact that only the sources of external fields appear in 
   $T^{\mu\nu}_{\mbox{\scriptsize vac}}$ is a result of the use of the
   Jacobi and Bianchi identities in the calculation of the nonlocal effective 
   action [15-20]. As will be seen, the conservation laws (1.9) for the
   vector and tensor sources play a crucial role in the consistency of the 
   result above.

   The present calculation reveals a number of nontrivial properties of the
   effective action such as Eq. (2.36). When working with the 
   explicit one-loop form factors these properties emerge as a result
   of mysterious cancellations. In fact they can be predicted on the basis
   of the axiomatic approach to the effective action [19-22].
   In particular, the behaviours (2.33) follow from the requirement of
   asymptotic flatness of the expectation-value spacetime and should hold
    for all form factors to all loop orders. The fact that the $1/\Box$
    terms of these behaviours cancel in the energy-momentum tensor is 
    predictable on the same grounds. As shown in [23], these terms
    stand for a vacuum generation of the gravitational waves.
    Eq. (2.36) is also a consequence of the asymptotic flatness. For the correct
    behaviour at ${\cal I}^+$ , 
$T^{\mu\nu}_{\mbox{\scriptsize vac}}$ {\it must}
    have the form (2.36). Finally, the  structure of the vertex terms in 
    (2.39)-(2.44) should, as the results below suggest, be a consequence
    of the condition of unitarity encoded in the one-loop
    triangular diagrams.

    \subsection*{\bf The problem of quantum noise.}

    $$ $$

    The vacuum energy-momentum tensor
    $ T^{\mu\nu}_{\mbox{\scriptsize vac}}$ does not obey the dominant 
    energy condition. Therefore, the energy flux $\Bigl(-dM/du\Bigr)$
    in (1.36) is not positive definite but the total radiation
    energy in (1.37) is positive. The point here is that the quantity
    in (1.36) is an expectation value over a quantum state rather
    than a classical observable. The indefinite oscillations in
    $\Bigl(-dM/du\Bigr)$ should be within the quantum uncertainty, i.e.
    they represent a quantum noise which is present in
    $ T^{\mu\nu}_{\mbox{\scriptsize vac}}$ even at the asymptotically flat
    infinity ${\cal I}^+$ (see [22] for a detailed discussion).
    The problem is in discovering the  mechanism by which the quantum
    noise gets averaged and the positive total energy emerges.

    For the effective action in two dimensions [28]
    \begin{equation}
     S_{\mbox{\scriptsize vac}} = \mbox{const.}\int d^2x g^{1/2} 
     R\;\frac{1}{\Box}\;R
     \end{equation}
     this problem has a simple solution [5]. The positive radiation energy
     comes from the variation of the  second-order form factor $1/\Box$
     in (2.46). The contribution of this variation to 
     $ T^{\mu\nu}_{\mbox{\scriptsize vac}}$ is quadratic in $R$ and
     energy-dominant whereas  the contribution of the variation $\delta R$
     is linear in $R$ and sign indefinite. However, this latter contribution 
     has the form of a total derivative and vanishes in the full integral
     over time [5]. This is the mechanism by which the quantum noise
     sums to zero for the whole history. A counterpart of Eq. (2.46) in
     four dimensions is [15]
     \begin{equation}
      S_{\mbox{\scriptsize vac}} = \mbox{const.} \int d^4x g^{1/2}
      \Re\log(-\Box)\Re\;\; + \; O[\Re^3]\quad ,
     \end{equation}
     and in this case the variation of the second-order form factor,
     $\log(-\Box)$, makes no contribution to 
     $ T^{\mu\nu}_{\mbox{\scriptsize vac}}$ at ${\cal I}^+$ [23].
     The apparent problem in four dimensions is that
     $ T^{\mu\nu}_{\mbox{\scriptsize vac}}\Bigl|_{{\cal I}^+}$
     is entirely a total derivative, Eq. (2.36).

     The insertion of (2.36) into (1.36) yields
     \begin{equation}
     - \frac{dM(u)}{du} = \frac{1}{(4\pi)^2}\;\frac{d^2}{du^2}
     \int d^2{\cal S}(\phi)\Bigl(r^2\log(-\Box)I\Bigr)\biggl|_{{\cal I}^+
     [u,\phi,r\to\infty]}\quad .
     \end{equation}
     Since this is a total derivative, the integrated energy flux is determined 
     only by the limits
of late time and early time $u = \pm \infty$. By the retardation property
of the form factors , the contribution of early time vanishes (see Sec.5 for the 
proof). There remains only the contribution of late time:
     \begin{equation}
     M(-\infty) - M(\infty) = \lim_{u\to\infty} 
     \frac{1}{(4\pi)^2}\;\frac{d}{du}
     \int d^2{\cal S}(\phi)\Bigl(r^2\log(-\Box)I\Bigr)\biggl|_{{\cal I}^+
     [u,\phi,r\to\infty]}\quad ,
     \end{equation}
and, for it to be finite and nonvanishing, the integrand in (2.49)
should have a linear growth at late time
     \begin{equation}
\Bigl(r^2\log(-\Box)I\Bigr)\biggl|_{{\cal I}^+[u,\phi,r\to\infty]}
\propto u\quad ,\quad  u \to \infty\quad .
    \end{equation}
Since the setting of the problem with external fields assumes that these fields
become asymptotically stationary as $u\to \pm \infty$ [1], the growth in time
required in (2.50) puzzles.

As we shall show, there can be no such growth if $I$ in (2.50) is local in the 
field strength. The $I_1$ in Eq. (2.38) is local, and, therefore,
the vacuum radiation of first order in $\Re$ is pure quantum noise
[21,22]. The $I_2$ in Eq. (2.39) is quadratic in $\Re$ and nonlocal but
the contribution of the trees is also pure quantum noise. The solution of 
the problem  in four dimensions is that the growth in time is provided 
by the vertex operator (2.44). The kernel of this operator grows like
$u^{m+n-3},\;u\to\infty,$ and the highest exponents $m$ and $n$ in all
terms of (2.39) are precisely such that the result is (2.50). If
the original quantum field contains no ghosts, the proportionality
coefficient in (2.50) is positive.
}

\newpage

{\renewcommand{\theequation}{3.\arabic{equation}}

\begin{center}
\section{\bf    Retarded kernels of the nonlocal operators}
\end{center}

\subsection*{\bf The retarded resolvent.}

$$ $$

With the form factors in the spectral forms, the only nonlocal
operator in the current (2.21) is the resolvent 
$1/(\Box - m^2)$ which, in the case of the expectation-value equations,
 is the retarded  Green function [14]. It admits an expansion of the same
  nature as the action (2.2), i.e. the expansion up to terms $O[\Re^n]$
  by  covariant perturbation theory. To lowest order in $\Re$,  the retarded
operator $1/(\Box - m^2)$ acting on an arbitrary tensor   
$X^{\mu_1\ldots\mu_n}$ is of the form
\begin{eqnarray}
-\frac{1}{(\Box - m^2)}\; X^{\mu_1\ldots\mu_n}(x)  \hspace{11cm}\\
=\frac{1}{4\pi}\int\limits_{\mbox{\scriptsize past of}\; x} d{\bar x}
{\bar g}^{1/2}\Bigl(\delta(\sigma) - \theta(-\sigma)\;\frac{
mJ_1(m\sqrt{-2\sigma})}{\sqrt{-2\sigma}}\Bigr)
g^{\mu_1}_{\;\;{\bar \mu}_1}\ldots g^{\mu_n}_{\;\;{\bar \mu}_n}
X^{{\bar \mu}_1\ldots {\bar \mu}_n}({\bar x}) + 
O[\Re\times X] \nonumber
\end{eqnarray}
where $\sigma = \sigma(x,{\bar x})$ is the world function [29,30]
(one half of the square of the geodetic distance between
$x$ and ${\bar x}$ ), 
$g^{\mu}_{\;\;{\bar \mu}} = g^{\mu}_{\;\;{\bar \mu}}(x, {\bar x})$ 
is the propagator of the  geodetic parallel transport for a vector [30],
$J_1$ is the order-1 Bessel function, and the integration point
${\bar x}$ is in the past of the observation point $x$. Here and below,
the bar over a symbol means that this symbol refers to the point 
${\bar x}$.

The geometrical two-point functions entering the Green function (3.1)
satisfy the equations [30]
\begin{equation}
g^{\mu\nu}\nabla_{\mu}\sigma\nabla_{\nu}\sigma = 2\sigma
\quad , \quad 
{\bar g}^{{\bar \mu}{\bar \nu}}
{\bar \nabla}_{{\bar \mu}}\sigma{\bar \nabla}_{{\bar \nu}}\sigma = 2\sigma
\quad,
\end{equation}
\begin{equation}
\sigma^{\mu}\nabla_{\mu}g^{\alpha}_{\;\;{\bar \alpha}} = 0\quad, \quad
\sigma^{{\bar \mu}}{\bar \nabla}_{\bar \mu}
g^{\alpha}_{\;\;{\bar \alpha}} = 0\quad,\quad 
g^{\alpha}_{\;\;{\bar \alpha}}\biggl|_{x={\bar x}} = \delta^{\alpha}_{
\;\;{\bar \alpha}}\quad ,
\end{equation}
\begin{equation}
\sigma^{\mu} = - g^{\mu}_{\;\;{\bar \mu}}\;\sigma^{\bar \mu}
\end{equation}
with
\begin{equation}
\sigma^{\mu} = g^{\mu\nu}\sigma_{\nu}\quad , \quad 
\sigma^{\bar \mu} = {\bar g}^{{\bar \mu}{\bar \nu}}\sigma_{\bar \nu}\quad , 
\quad \sigma_{\nu} =  \nabla_{\nu} \sigma \quad , 
\quad \sigma_{\bar \nu} = {\bar \nabla}_{\bar \nu}\sigma\quad .   
\end{equation}
To lowest order in $\Re$ one can use the relations
\begin{eqnarray}
g^{\mu}_{\;\;{\bar \mu}} &=&  - \nabla^{\mu}{\bar \nabla}_{\bar \mu}\sigma +
O[\Re]\quad , \\
\nabla_{\alpha}g^{\mu}_{\;\;{\bar \mu}} &=&  O[\Re]\quad , \quad
{\bar \nabla}_{\bar \alpha}g^{\mu}_{\;\;{\bar \mu}} = O[\Re]\quad .
\end{eqnarray}

As seen from (3.1), we are always dealing with some scalar source
\begin{equation}
{\bar X} = 
g^{\mu_1}_{\;\;{\bar \mu}_1}(x,{\bar x})\ldots
g^{\mu_n}_{\;\;{\bar \mu}_n}(x,{\bar x})
X^{{\bar \mu}_1\ldots{\bar \mu}_n}({\bar x})
\end{equation}
which may depend parametrically on the observation point. Since
the resolvent will be used only in the approximation (3.1) \footnote{
The curvature correction to the lowest-order form factor is calculated
in Sec.4.} , we shall
omit the symbol $O[\Re\times X]$ and use the short notation
\begin{equation}
- \frac{1}{(\Box - m^2)} X(x) = 
\frac{1}{4\pi}\int\limits_{\mbox{\scriptsize past of}\;x} 
d{\bar x}{\bar g}^{1/2}\Bigl(\delta(\sigma) - \theta(-\sigma)
\frac{mJ_1(m\sqrt{-2\sigma})}{\sqrt{-2\sigma}}\Bigr){\bar X}
\end{equation}
which will always assume that the source on the right-hand
side is transported in a parallel fashion to the observation point.

For the integration over masses in the spectral integrals, the Green
function should first be transformed by using the relation for the 
Bessel functions
\begin{equation}
{\bar \xi}^{\mu}{\bar \nabla}_{\mu}J_0(m\sqrt{-2\sigma}) = 
\frac{mJ_1(m\sqrt{-2\sigma})}{\sqrt{-2\sigma}} 
({\bar \xi}\cdot{\bar \nabla}\sigma)
\end{equation}
with an arbitrary timelike vector field $\xi^{\mu}$, and the Gauss theorem 
in the form valid for null boundaries:
\begin{equation}
\int\limits_{\Omega} dx g^{1/2} \nabla_{\mu}A^{\mu} = 
\int dx g^{1/2} A^{\mu}\Bigl(\nabla_{\mu}\Sigma\Bigl) \delta(\Sigma)\quad .
\end{equation}
Here $\Omega$ is an integration domain, $\Sigma = 0$ is the  equation of 
its boundary, and $\Sigma<0$ holds inside $\Omega$. Integrating by
parts in (3.9) one obtains the expression 
\begin{equation}
- \frac{1}{(\Box - m^2)} X(x) = 
\frac{1}{4\pi}\int\limits_{\mbox{\scriptsize past of}\;x} 
d{\bar x}{\bar g}^{1/2} \theta(-\sigma)
J_0(m\sqrt{-2\sigma}){\bar \nabla}_{\alpha}
\Bigl(\frac{{\bar \xi}^{\alpha}}{({\bar \xi}\cdot{\bar \nabla}
\sigma)}{\bar X}\Bigr)
\end{equation}
which holds with any choice of $\xi^{\alpha}$, and from which
the massless contribution proportional to $\delta(\sigma)$ is absent. 

\subsection*{\bf The kernels of $1/\Box$ , $\log(-\Box)$ , and of the vertex
operators.}

$$ $$

The retarded kernel of the massless operator $1/\Box$ is given
by Eq. (3.9) with $m=0.$ Up to $O[\Re\times X]$ ,
\begin{equation}
- \frac{1}{\Box}X(x) = \frac{1}{4\pi} \int\limits_{
\mbox{\scriptsize past of}\;x} d{\bar x}{\bar g}^{1/2}
\delta\Bigl (\sigma(x,{\bar x})\Bigr ){\bar X}
\end{equation}
where the integration is over the past light cone of the observation 
point $x$ .

The  retarded kernel of the operator $\log(-\Box)$ has been calculated
in [21,22]. To lowest order in $\Re$ it is of the form
\begin{equation}
 \log(-\Box)X(x) = \frac{1}{2\pi} \int\limits_{
\mbox{\scriptsize past of}\;x} d{\bar x}{\bar g}^{1/2}
\delta'\Bigl (\sigma(x,{\bar x})\Bigr ){\bar X} + \mbox{ a local term }
\end{equation}
where the local term, i.e. a term proportional to $X$ at the observation
 point $x$, is irrelevant to the present consideration \footnote{
 Expression (3.14) without a further specification is valid only for
  $x$ located outside the support of $X$. This includes the case where
  $x\to{\cal I}^+$ provided that $X\Bigl|_{{\cal I}^+} = O(1/r^3).$
  The latter case is the one that we presently consider. 
In the general case, the
  integral on the right-hand side of (3.14) is improper. For its precise 
  definition and the form of the local term see [21,22].}.

  The kernels of the vertex operators (2.44) are obtained by using
   the following spectral form of their generating function:
   \begin{equation}
   -\frac{\log(j_1\Box_1/j_2\Box_2)}{j_1\Box_1 - j_2\Box_2} = 
   \frac{1}{j_1j_2} \int\limits_0^{\infty}\;\frac{dm^2}{
   (m^2/j_1 - \Box_1)(m^2/j_2 - \Box_2)}\quad .
   \end{equation}
   Combining (3.15) and (3.12) one finds
   \begin{eqnarray}
   -\frac{\log(j_1\Box_1/j_2\Box_2)}{j_1\Box_1 - j_2\Box_2}X_1X_2(x) 
   &= &\frac{1}{4\pi}\int\limits_{\mbox{\scriptsize past of}\;x} 
d{\bar x}_1{\bar g}^{1/2}_1 \theta(-\sigma_1)
{\bar \nabla}_{\alpha}
\Bigl(\frac{{\bar \xi}^{\alpha}}{({\bar \xi}\cdot{\bar \nabla}
\sigma_1)}{\bar X}_1\Bigr) \\
&\times &\frac{1}{4\pi}\int\limits_{\mbox{\scriptsize past of}\;x} 
d{\bar x}_2{\bar g}^{1/2}_2 \theta(-\sigma_2)
{\bar \nabla}_{\alpha}
\Bigl(\frac{{\bar \xi}^{\alpha}}{({\bar \xi}\cdot{\bar \nabla}
\sigma_2)}{\bar X}_2\Bigr) \nonumber\\
&\times &\frac{1}{j_1j_2}
\int\limits_0^{\infty} dm^2 
J_0(m\sqrt{-2\sigma_1/j_1})J_0(m\sqrt{-2\sigma_2/j_2})  \nonumber
\end{eqnarray}
where
\begin{equation}
\sigma_1 = \sigma(x,\;{\bar x}_1)\quad ,\quad 
\sigma_2 = \sigma(x,\;{\bar x}_2)\quad .
\end{equation}

The spectral-mass integral in (3.16) can be calculated with the aid
of the Fourier-Bessel relation
\begin{equation}
\int\limits_0^{\infty} dm^2 J_0(m\sqrt{-2\sigma_1/j_1})J_0(m\sqrt{-2\sigma_2/j_2})  
= 2 \delta\Bigl(\frac{\sigma_1}{j_1} - \frac{\sigma_2}{j_2}\Bigr)\quad .
\end{equation}
It is convenient to factorize the vertex function by introducing
an auxiliary integration over a parameter. We write
\begin{equation}
\theta(-\sigma_1)\theta(-\sigma_2) 
\delta\Bigl(\frac{\sigma_1}{j_1} - \frac{\sigma_2}{j_2}\Bigr) = 
\int\limits^0_{-\infty} dq\;\delta\Bigl(q - \frac{\sigma_1}{j_1}\Bigr)
\delta\Bigl(q - \frac{\sigma_2}{j_2}\Bigr)\quad ,
\end{equation}
and introduce the operator ${\cal H}_q$ depending on the
parameter $q$ and defined as follows:
\begin{equation}
{\cal H}_qX(x) = \frac{1}{4\pi} \int\limits_{\mbox{\scriptsize past of}\;x}
d{\bar x}{\bar g}^{1/2} \delta\Bigl(\sigma(x,{\bar x}) - q\Bigr){\bar X}
\quad , \quad q\le 0\quad .
\end{equation}
Here the integration is over the past sheet of the hyperboloid
$\sigma(x,{\bar x}) = q$ associated with the point $x$ (the past hyperboloid
of $x$ , Fig.3). The operator 
${\cal H}_q$ is a generalization of the retarded operator $-1/\Box$ ,
Eq. (3.13). A calculation of the derivative
\begin{eqnarray}
\frac{d}{dq}{\cal H}_qX(x) & = &
\frac{1}{4\pi} \int\limits_{\mbox{\scriptsize past of}\;x}
d{\bar x}{\bar g}^{1/2} \delta'\Bigl(q - \sigma(x,{\bar x})\Bigr){\bar X}
 \\
& = & \frac{1}{4\pi} \int\limits_{\mbox{\scriptsize past of}\;x}
d{\bar x}{\bar g}^{1/2} \delta\Bigl(\sigma(x,{\bar x}) - q\Bigr)
{\bar \nabla}_{\alpha}\Bigl(\frac{{\bar \xi}^{\alpha}}{({\bar \xi}\cdot
{\bar \nabla}\sigma)}{\bar X}\Bigr)\nonumber
\end{eqnarray}
with the result containing an arbitrary timelike vector field $\xi^{\alpha}$
shows that this derivative with $q$ replaced by $jq$ will appear in (3.16)
upon the use of (3.18) and (3.19).

In this way we obtain
\begin{equation}
-\frac{\log(j_1\Box_1/j_2\Box_2)}{j_1\Box_1 - j_2\Box_2}\;X_1X_2(x) =
2 \int\limits^0_{-\infty} dq 
\Bigl[ \frac{d}{d(qj_1)}{\cal H}_{qj_1}X_1(x)\Bigr]
\Bigl[ \frac{d}{d(qj_2)}{\cal H}_{qj_2}X_2(x)\Bigr]\quad .
\end{equation}
Finally, the result for the kernel of the vertex operator (2.44) is
\begin{equation}
F(m,n)X_1X_2(x) = 
- 2 \int\limits^0_{-\infty} dq q^{m+n}
\Bigl[ \Bigl(\frac{d}{dq}\Bigr)^{m+1}{\cal H}_{q}X_1(x)\Bigr]
\Bigl[ \Bigl(\frac{d}{dq}\Bigr)^{n+1}{\cal H}_{q}X_2(x)\Bigr]\quad .
\end{equation}

\subsection*{\bf The kernels of the vertex operators superposed with
$1/\Box$ .}

$$ $$

There are two more types of the vertex operators in (2.39):
\begin{equation}
(a) \;\;\;\frac{1}{\Box_2}\;F(m,n)\quad,\quad 
(b) \;\;\;\frac{1}{\Box_1\Box_2}\;F(m,n)\quad .
\end{equation}
It is important that in $(a)\;\; n \ge 1$, and in $(b)\;\; 
m \ge 1,\;n \ge 1,$
i.e.  the appearance of a factor $1/\Box_1$ or $1/\Box_2$ in the vertex
is necessarily accompanied by the appearance of a derivative
$\partial/\partial j$ bearing the same number \footnote{
This is a manifestation of the general property of the form
factors established in [17] with the aid of the integral 
$\alpha$-representation and called there "rule of the like $\alpha$".
}.
Therefore, it suffices to consider the generating expressions
\begin{equation}
 -\frac{1}{\Box_2}
 \frac{\partial}{\partial j_2}
 \frac{\log(j_1\Box_1/j_2\Box_2)}{j_1\Box_1 - j_2\Box_2}X_1X_2  = 
\frac{1}{j_1j_2{}^2} \int\limits_0^{\infty}\;\frac{dm^2}{
(m^2/j_1 - \Box_1)(m^2/j_2 - \Box_2)^2}X_1X_2 \; ,
\end{equation}
\begin{equation}
-\frac{1}{\Box_1\Box_2}
\frac{\partial}{\partial j_1}\frac{\partial}{\partial j_2}
\frac{\log(j_1\Box_1/j_2\Box_2)}{j_1\Box_1 - j_2\Box_2}X_1X_2  = 
\frac{1}{j_1{}^2j_2{}^2} \int\limits_0^{\infty}\;\frac{dm^2}{
(m^2/j_1 - \Box_1)^2(m^2/j_2 - \Box_2)^2}X_1X_2 \; .
\end{equation}

The square (and, generally, any power) of the massive Green
function is a well defined operator whose kernel can be obtained
by differentiating the Green function with respect to the mass:
\begin{equation}
\frac{1}{(m^2 - \Box)^2}X = - \frac{\partial}{\partial m^2}
\frac{1}{(m^2 - \Box)}X\quad .
\end{equation}
The behaviour of the function (3.27) as $m^2 \to 0$ and
hence  the convergence of the integrals (3.25),(3.26)
at the lower limits depends on the class of the sources $X$.
This question will be considered in Sec.5.

The differentiation of the Green function with respect to the mass
 is readily  accomplished in the original expression (3.9) by using
 the relation for the Bessel functions
 \begin{equation}
 \frac{\partial}{\partial m^2}\;\frac{mJ_1(m\sqrt{-2\sigma})}{
 \sqrt{-2\sigma}} = \frac{1}{2}\;J_0(m\sqrt{-2\sigma})\quad .
 \end{equation}
 The result is
 \begin{equation}
 \frac{1}{(m^2 - \Box)^2}X(x) = \frac{1}{8\pi} 
 \int\limits_{\mbox{\scriptsize past of }\;x} d {\bar x}
 {\bar g}^{1/2}\theta(-\sigma)J_0(m\sqrt{-2\sigma}){\bar X}\quad .
 \end{equation}

 The insertion of (3.29) and (3.12) into (3.25), (3.26) leads
 to the same spectral integral and the same factorization procedure as
 in the previous case. For the generating expressions we obtain
\begin{equation}
-\frac{1}{\Box_2}\frac{\partial}{\partial j_2}
\frac{\log(j_1\Box_1/j_2\Box_2)}{j_1\Box_1 - j_2\Box_2}\;X_1X_2(x) =
 \int\limits^0_{-\infty} dq 
\Bigl[ \frac{d}{d(qj_1)}{\cal H}_{qj_1}X_1(x)\Bigr]
\Bigl[ \frac{1}{j_2}{\cal H}_{qj_2}X_2(x)\Bigr] \; ,
\end{equation}
\begin{equation}  
-\frac{1}{\Box_1\Box_2}\frac{\partial}{\partial j_1}\frac{\partial}{\partial j_2}
\frac{\log(j_1\Box_1/j_2\Box_2)}{j_1\Box_1 - j_2\Box_2}\;X_1X_2(x) =
\frac{1}{2} \int\limits^0_{-\infty} dq 
\Bigl[ \frac{1}{j_1}{\cal H}_{qj_1}X_1(x)\Bigr]
\Bigl[ \frac{1}{j_2}{\cal H}_{qj_2}X_2(x)\Bigr] \; ,  
\end{equation}
and, finally, the results for the kernels of the vertex operators are 
\begin{eqnarray}
\frac{1}{\Box_2}F(m,n)X_1X_2(x)  \hspace{110mm} \\ 
=-  \int\limits^0_{-\infty} dq q^{m+n}
\Bigl[ \Bigl(\frac{d}{dq}\Bigr)^{m+1}{\cal H}_{q}X_1(x)\Bigr]
\Bigl[ \Bigl(\frac{d}{dq}\Bigr)^{n-1}\frac{1}{q}
{\cal H}_{q}X_2(x)\Bigr]\; , \; \; n \ge 1, \nonumber \\
\frac{1}{\Box_1\Box_2}F(m,n)X_1X_2(x)  \hspace{110mm} \\ 
=- \frac{1}{2} \int\limits^0_{-\infty} dq q^{m+n}
\Bigl[ \Bigl(\frac{d}{dq}\Bigr)^{m-1}\frac{1}{q}{\cal H}_{q}X_1(x)\Bigr]
\Bigl[ \Bigl(\frac{d}{dq}\Bigr)^{n-1}\frac{1}{q}
{\cal H}_{q}X_2(x)\Bigr]\; , \; \; m \ge 1, \; n \ge 1.    \nonumber
\end{eqnarray}
The question of convergence of the integrals (3.25),(3.26) at $m^2=0$
transfers now to the integrals (3.32),(3.33) with respect to $q$ whose 
convergence at $q = - \infty$ should be provided by the properties
of the sources $X$.
}

\newpage

{\renewcommand{\theequation}{4.\arabic{equation}}

\begin{center}
\section{\bf    The asymptotic behaviours at ${\cal I}^+$}
\end{center}

\subsection*{\bf Null hyperplanes.}

$$ $$

The power of growth of the world function as one of its points
tends to ${\cal I}^+$ and the other one stays in a compact
 domain should be the same as in flat spacetime. Therefore, we 
  may write
  \begin{equation}
  \sigma(x,\;{\bar x})\biggl|_{x\in{\cal I}^+[u,\phi,r\to\infty]}
  = rZ(u,\phi\, ;\;{\bar x}) + O(r^0)
  \end{equation}
  with some coefficient function $Z$.

  Inserting the behaviour (4.1) into the equation (3.2) for 
  $\sigma$ with respect to the point $x$ and using the asymptotic
  form of the metric at ${\cal I}^+$ we obtain
  \begin{equation}
  -2rZ\frac{\partial}{\partial u}Z + O(r^0) = 2rZ + O(r^0)
  \end{equation}
  whence
  \begin{equation}
  \frac{\partial}{\partial u} Z(u,\phi\, ; \;{\bar x}) = -1\quad .
  \end{equation}
  Therefore, we introduce a new notation to write down the solution
  for $Z$
  \begin{equation}
  Z(u,\phi\, ; \;{\bar x}) = -u + U_{\phi}({\bar x})\quad ,
  \end{equation}
  and rewrite Eq. (4.1) in its final form
  \begin{equation}
  \sigma(x,\;{\bar x})\biggl|_{x\in{\cal I}^+[u,\phi,r\to\infty]} = 
  r\Bigl( - u + U_{\phi}({\bar x})\Bigr) + O(r^0)\quad .
  \end{equation}

  Inserting the behaviour (4.5) into the equation (3.2) for 
  $\sigma$ with respect to the point ${\bar x}$ we obtain
  \begin{equation}
  \Bigl({\bar \nabla}U({\bar x})\Bigr)^2 = 0 \quad , \quad 
  U({\bar x}) \equiv U_{\phi}({\bar x})\quad .
  \end{equation}  
  Hence we infer that the equation
  \begin{equation}
  U_{\phi}({\bar x}) = u
  \end{equation}
  with fixed $\phi$ defines a family of null hypersurfaces labelled
   by the retarded time $u$ but different from the future light cones of
   the Bondi-Sachs frame. The hypersurfaces (4.7) will be called hyperplanes
\footnote{ More specifically, they should have been called {\it future}
hyperplanes as distinct from the {\it past} hyperplanes defined
similarly by the conditions at ${\cal I}^-$ . However, the past
hyperplanes do not figure in the present consideration.} .
  To different directions $\phi$ at infinity there correspond different 
  families of null hyperplanes.

  An important property of the function $U_{\phi}({\bar x})$ follows from 
  Eq. (3.4). Inserting the behaviour (4.5) in (3.4) one obtains
\begin{equation}
{\bar \nabla}_{\bar \mu}U_{\phi}({\bar x}) = 
g_{\bar \mu}^{\;\;\mu}({\bar x},x)\nabla_{\mu}u - 
g_{\bar \mu}^{\;\;\mu}({\bar x},x)\nabla_{\mu}\phi^{a}
\frac{\partial}{\partial\phi^a}U_{\phi}({\bar x})  
\end{equation}
where $a=1,2,$ and $x\to{\cal I}^+.$ By (1.18),
\begin{equation}
\nabla_{\mu}\phi \;\;
\propto \;\;\frac{1}{r}m_{\mu}\biggl|_{{\cal I}^+}
\end{equation}
while the contraction $g_{\bar \mu}^{\;\;\mu}({\bar x},x)m_{\mu}$
remains finite as $x\to{\cal I}^+$ [23]. Therefore,
\begin{equation}
g_{\bar \mu}^{\;\;\mu}({\bar x},x)\nabla_{\mu}\phi^a\biggl|_{x\to{\cal I}^+}
= O\Bigl(\frac{1}{r}\Bigr)\quad .
\end{equation}
As a result, one obtains the following law of parallel transport:
\begin{equation}
{\bar \nabla}_{\bar \mu}U_{\phi}({\bar x}) = 
g_{\bar \mu}^{\;\;\mu}({\bar x},x)
\nabla_{\mu}u(x)\biggl|_{x\to{\cal I}^+[u,\phi]}\quad .
\end{equation}

The geometric meaning of the equations above will clarify if one answers 
the following question: what becomes of the past light cone of a point
$x$ at the limit $x\to{\cal I}^+$? At this limit, one of  the null
generators of the cone merges with ${\cal I}^+$ i.e. disappears from any
compact domain. Therefore, the resultant limiting surface is no more
a cone although it remains a null hypersurface. As follows from Eq. (4.5),
 this limiting surface is none other than the hyperplane (4.7) whose 
 parameters $u,\phi$ label the point of ${\cal I}^+$ to which
 the vertex of the prelimiting cone comes (Fig.4). The null generators
 of the prelimiting cone all but one become the generators of the limiting
 hyperplane. It follows that the generators of each hyperplane in (4.7)
  are the null geodesics that, when traced towards the future, come  to
  one and the same point $\phi$ of the  celestial sphere at one and the same
  instant $u$ of retarded time. This can be taken for a definition of  
  light rays parallel in the future. 
Since the null generators of a hyperplane merge at infinity,
  they have a common tangent vector at ${\cal I}^+$ which is $\nabla u$.
  On the other hand, at a point ${\bar x}$ of a compact domain the
  vector tangent to the generator is ${\bar \nabla}_{\bar \mu}U_{\phi}
  ({\bar x})$ . Eq. (4.11) is, therefore, the law of parallel transport
  of the tangent vector along the null geodesic emanating from a given
  point and belonging to a given hyperplane.  
  
  Further properties of
  the function $U_{\phi}({\bar x})$ derive under the assumption that
  for both the point ${\bar x}$ and the point at ${\cal I}^+$ one can
  use one and the same global Bondi-Sachs frame:
  \begin{equation}
  U_{\phi}({\bar x}) = U_{\phi}({\bar u},{\bar \phi},{\bar r})\quad .
  \end{equation}
  Then we have
  \begin{equation}
U_{\phi}({\bar u},{\bar \phi},{\bar r})\ge {\bar u}\quad ,
  \end{equation}
and
  \begin{equation}
U_{\phi}({\bar u},{\bar \phi},{\bar r})\biggl|_{{\bar \phi} = \phi}
= {\bar u}\quad ,\quad \frac{\partial}{\partial {\bar \phi}}
U_{\phi}({\bar u},{\bar \phi},{\bar r})\biggl|_{{\bar \phi} = \phi}=0\quad .
\end{equation}
The inequality (4.13) is a consequence of the general fact mentioned
in Sec.1 that, for the points ${\bar x}$ belonging to the past light cone of 
$x$ , $u({\bar x})\le u(x)$ . With $x$ at ${\cal I}^+$ , the latter inequality
holds for the points ${\bar x}$ belonging to the limiting hyperplane
(4.7). Eqs. (4.14) follow from the fact that the radial geodesic 
${\bar u} = u$ , ${\bar \phi}=\phi$ along which the point $x$ in (4.5)
tends to ${\cal I}^+$ serves at the same time as a generator of the  past
light cone of $x$ and, therefore, belongs to the limiting hyperplane
(4.7). The second equation in (4.14) implies
${\bar \nabla}U_{\phi}\Bigl |_{{\bar \phi}=\phi}={\bar \nabla}{\bar u}$
and follows from the first equation and Eq. (4.6).

Finally, in the case where ${\bar x}$ is at the future  asymptotically
flat infinity ${\cal I}^+$ or $i^+$, one can use the flat-spacetime formula
for $U_{\phi}({\bar x}).$
Indeed, in this case the geodesic connecting the points $x$ and ${\bar x}$
in (4.5) passes entirely through the asymptotically flat domain.\footnote{
The geometric two-point functions $\sigma(x,\;{\bar x})$ and
$g^{\mu}_{\;\;{\bar \mu}}(x,\;{\bar x})$ are nonlocal objects involving
the metric on the geodesic connecting the two points. Therefore, even
if both points are at the asymptotically flat infinity but one is in the 
future and the other in the past, the flat-spacetime approximation is
invalid since the geodesic connecting the two points passes through
a domain of strong field.
}
One has
\begin{eqnarray}
U_{\phi}({\bar x})\biggl|_{{\bar x}\to{\cal I}^+,i^+} & = &
{\bar u} + {\bar r}\Bigl(1 - \cos\omega(\phi,\;{\bar \phi})\Bigr)\nonumber\\
&=&{\bar t} - n_i(\phi){\bar \x}^i
\end{eqnarray}
where the first form  refers to the  Bondi-Sachs parametrization (4.12) and
the second to the Minkowski coordinates (1.25) for ${\bar x}$. In (4.15),
$\omega(\phi,\;{\bar \phi})$ is the arc length between the points $\phi$ 
and ${\bar \phi}$ on the unit 2-sphere:
\begin{equation}
\cos\omega(\phi,\;{\bar \phi}) = n_i(\phi)n^i({\bar \phi})\quad .
\end{equation}
Here $n_i(\phi)$ is the direction vector introduced in (1.26).

\subsection*{\bf The $1/\Box$ and $\log(-\Box)$ at ${\cal I}^+$ .}

$$ $$

The behaviours of the kernels of $1/\Box$ and $\log(-\Box)$ at
${\cal I}^+$ follow immediately from (3.13),(3.14) and (4.5):
\begin{eqnarray}
\frac{1}{\Box}X\biggl|_{{\cal I}^+[u,\phi,r\to\infty]} &=& 
- \frac{1}{r}D_{\1}(u,\phi|X)\quad,                    \\
\log(-{\Box})X\biggl|_{{\cal I}^+[u,\phi,r\to\infty]} &=& 
- \frac{2}{r^2}\frac{\partial}{\partial u}D_{\1}(u,\phi|X)\quad,
\end{eqnarray}
and the coefficient of these behaviours is an integral over the 
null hyperplane (4.7):
\begin{equation}
D_{\1}(u,\phi|X) = \frac{1}{4\pi}\int d{\bar x}{\bar g}^{1/2}
\delta\Bigr(U_{\phi}({\bar x}) - u\Bigr){\bar X}\quad .
\end{equation}
As discussed in Sec.7 below, this integral is the ultrarelativistic
limiting case of a more general hypersurface integral $D$ called radiation
moment of the source $X$. The subscript $\1$ that $D_{\1}$ in (4.19)
bears is to distinguish the ultrarelativistic moment.

It follows from Eqs. (4.18) and (2.48) that the moment $D_{\1}$
determines the flux of vacuum energy through ${\cal I}^+$. We obtain
\begin{equation}
\frac{dM(u)}{du} = \frac{2}{(4\pi)^2}\;\frac{\partial^3}{\partial u^3}
\;\int d^2{\cal S}(\phi)\, D_{\1}(u,\phi|I)
\end{equation}
where $I$ is the scalar in (2.36). Since $D_{\1}$ serves at the same
 time as a coefficient in (4.17), it governs also the classical
 radiation (see, e.g., Sec.8 below). Of special importance is, therefore,
 the question of  convergence of the integral (4.19). It is only with
 this convergence that the behaviours (4.17) and (4.18) hold.

 Note that the convergence of the original integral (3.13) defining
 the retarded Green function imposes a restriction only on the behaviour
 of the source $X$ at the past null infinity whereas the convergence
 of the moment (4.19) requires certain behaviours of $X$ also at the future 
 null infinity and spatial infinity.
 The  behaviour at the future null infinity is critical.

 To study the convergence of the integral (4.19) at ${\cal I}^+$
 consider a portion of the integration domain defined by the following 
 inequalities in the Bondi-Sachs frame:
 \begin{equation}
 \Omega\;\;\;:\;\;\; u_2 > {\bar u} > u_1\quad , \quad \phi + \Delta > 
 {\bar \phi} > \phi - \Delta\quad , \quad {\bar r} > r_0
 \end{equation}
 with large $r_0$, small $\Delta$, and any fixed $u_1,u_2.$
In the integral restricted to $\Omega$, the integration point ${\bar x}$
is near ${\cal I}^+$. Therefore, one can use the flat-spacetime
expressions for the metric and the function $U_{\phi}({\bar x}),$ 
Eqs. (1.15)
and (4.15). One finds
\begin{eqnarray}
\int\limits_{\Omega} d{\bar x} {\bar g}^{1/2}\delta\Bigl(
U_{\phi}({\bar x}) - u\Bigr){\bar X} &=&
\int\limits_{u_1}^{u_2}d{\bar u}
\int\limits_{\phi - \Delta}^{\phi + \Delta}
d^2{\cal S}({\bar \phi})\int\limits_{r_0}^{\infty}d{\bar r}\,{\bar r}^2
\delta\Bigl({\bar u} - u + {\bar r}(1 - \cos\omega)\Bigr){\bar X}  
\nonumber \\
&= &\int\limits_{u_1}^{u_2}d{\bar u}
\int\limits_{\phi - \Delta}^{\phi + \Delta}
\frac{d^2{\cal S}({\bar \phi})}{1 - \cos\omega}
\Bigl({\bar r}^2{\bar X}\Bigr)_{{\cal I}^+}[
{\bar u},{\bar \phi},{\bar r} = \frac{u - {\bar u}}{1 - \cos\omega}]
\end{eqnarray}
with $\omega = \omega(\phi,{\bar \phi})$ in (4.16), and ${\bar r}$ large i.e.
$(1 - \cos\omega)$ small. In agreement with the fact that 
the null hyperplane reaches 
the future infinity at only one value of ${\bar \phi}$ , ${\bar \phi}=\phi$ ,
the question of convergence concerns the angle integral. The source
${\bar X}$ may have any power of decrease or even increase at ${\cal I}^+$
\begin{equation}
{\bar X}\biggl|_{{\cal I}^+[{\bar u},{\bar \phi},{\bar r}\to\infty]}
= \kappa({\bar u},{\bar \phi}){\bar r}^n
\end{equation}
provided that
\begin{equation}
\kappa({\bar u},{\bar \phi})\biggl|_{{\bar \phi}\to\phi} = 
O\Bigl(({\bar \phi} - \phi)^{5+2n}\Bigr)\quad .
\end{equation}
However, to have $5 + 2n \ne 0$ in (4.24), the source  ${\bar X}$
should depend on the external angles $\phi$. This is possible in
the case of a tensor $X$ since in this case ${\bar X}$ depends 
parametrically on the observation point through the propagators
of parallel transport, Eq. (3.8). For a scalar source that has no
special relation to the integration hyperplane in (4.19), the convergence
 condition is $5 + 2n = 0$ , or, with the analyticity of $X$ at ${\cal I}^+$ ,
\begin{equation}
X\biggl|_{{\cal I}^+} = O\Bigl(\frac{1}{r^3}\Bigr)\quad .
\end{equation}
                                                        
The $I$ in Eqs. (2.36)-(2.39) and (4.20) is a scalar. We must, therefore,
make sure that it satisfies the criterion (4.25):
\begin{equation}
I\biggl|_{{\cal I}^+} = O\Bigl(\frac{1}{r^3}\Bigr) \quad .
\end{equation}
Let us check if this criterion is fulfilled for the vertex contributions.

\subsection*{\bf The vertex operators at ${\cal I}^+$ .}

$$ $$

The  asymptotic behaviours of the vertex functions (3.23) as $x\to{\cal I}^+$
are obtained by making the following replacement of the integration 
variable in (3.23):
\begin{equation}
q = r(\tau - u)
\end{equation}
where $\tau$ is the new integration variable, and $u,r$ are the parameters
of
$$
x\in{\cal I}^+[u,\phi,r\to\infty] \quad .
$$ 
With $q$ replaced as in (4.27), and
$\tau$ fixed,
\begin{equation}
{\cal H}_qX(x)\biggl|_{x\in{\cal I}^+[u,\phi,r\to\infty]} = \frac{1}{r}
D_{\1}(\tau,\phi|X)\quad .
\end{equation}
As a result we obtain
\begin{eqnarray}
F(m,n)X_1X_2\biggl|_{{\cal I}^+[u,\phi,r\to\infty]} \hspace{75mm}\\ 
=-\frac{2}{r^3}\int\limits_{-\infty}^{u}d\tau(\tau - u)^{m+n}
\Bigl[\Bigl(\frac{d}{d\tau}\Bigr)^{m+1}D_{\1}(\tau,\phi|X_1)\Bigr]
\Bigl[\Bigl(\frac{d}{d\tau}\Bigr)^{n+1}D_{\1}(\tau,\phi|X_2)\Bigr]
\; . \nonumber
\end{eqnarray}
In the same way, for the vertex functions (3.32) and (3.33) we obtain
\begin{eqnarray}
\frac{1}{\Box_2}F(m,n)X_1X_2\biggl|_{{\cal I}^+[u,\phi,r\to\infty]}  
\hspace{9cm}\\
=-\frac{1}{r^2}\int\limits_{-\infty}^{u}d\tau(\tau - u)^{m+n}
\Bigl[\Bigl(\frac{d}{d\tau}\Bigr)^{m+1}D_{\1}(\tau,\phi|X_1)\Bigr]
\Bigl[\Bigl(\frac{d}{d\tau}\Bigr)^{n-1}
\frac{1}{(\tau - u)}D_{\1}(\tau,\phi|X_2)\Bigr]
\; ,\nonumber\\
\frac{1}{\Box_1\Box_2}F(m,n)X_1X_2\biggl|_{{\cal I}^+[u,\phi,r\to\infty]}  
\hspace{9cm}\\
=-\frac{1}{2r}\int\limits_{-\infty}^{u}d\tau(\tau - u)^{m+n}
\Bigl[\Bigl(\frac{d}{d\tau}\Bigr)^{m-1}\frac{1}{(\tau - u)}
D_{\1}(\tau,\phi|X_1)\Bigr]
\Bigl[\Bigl(\frac{d}{d\tau}\Bigr)^{n-1}
\frac{1}{(\tau - u)}D_{\1}(\tau,\phi|X_2)\Bigr]
\; .\nonumber
\end{eqnarray}

Thus we infer that, with the general sources, the operator in (4.29)
does have the needed power of decrease at ${\cal I}^+$ but the operators
in (4.30) and (4.31) do not. In Eqs. (2.39)-(2.42) these latter operators
act only on the vector and tensor sources and are accompanied by extra
derivatives but the derivatives alone do not help since they have the
projections tangential to ${\cal I}^+.$ What helps is the conservation
of the vector and tensor sources. Owing to this conservation, the
tangential derivatives vanish, and the behaviours in (4.30) and (4.31) soften
down by at least one power of $1/r.$ As will be seen below, in the case
of the vector vertex this is still insufficient and, to ensure condition
(4.26), we shall have to impose a  special limitation on the vector
source.

\subsection*{\bf The conserved charges.}

$$ $$

Below, for the moments $D_{\1}$ of the sources in (1.8) we use the short
notations
\begin{eqnarray}
D_{\1}(\tau,\phi|{\hat P}) &=& {\hat D}_{\1}(\tau,\phi) = {\hat D}_{\1}\quad ,
\\
D_{\1}(\tau,\phi|{\hat J}^{\mu}) &=& 
{\hat D}_{\1}^{\;\;\mu}(\tau,\phi) = {\hat D}^{\;\;\mu}_{\1}\quad ,
\\
D_{\1}(\tau,\phi|{J}^{\mu\nu}) &=& 
{D}_{\1}^{\;\;\mu\nu}(\tau,\phi) = {D}^{\;\;\mu\nu}_{\1}\quad .
\end{eqnarray}
To distinguish the moments of the Ricci scalar $R$ and the matrix ${\hat Q}$
in (1.12), they will be denoted
\begin{eqnarray}
D_{\1}(\tau,\phi|{R}) &=& 
{D}_{\1}^{\;\;R}(\tau,\phi) = {D}^{\;\;R}_{\1}\quad ,
\\
D_{\1}(\tau,\phi|{\hat Q}) &=& 
{\hat D}_{\1}^{\;\;Q}(\tau,\phi) = {\hat D}^{\;\;Q}_{\1}\quad .
\end{eqnarray}
We have
\begin{equation}
g_{\mu\nu}D_{\1}^{\;\;\mu\nu} = - D_{\1}^{\;\;R}\quad .
\end{equation}

The non-scalar moments contain the propagators of parallel transport.
Specifically \footnote{The matrix moments contain implicitly the propagators 
of parallel transport for the matrix indices.},
\begin{eqnarray}
{\hat D}_{\1}^{\;\;\mu} &=& \frac{1}{4\pi}\int d{\bar x}{\bar g}^{1/2}
\delta\Bigl(U_{\phi}({\bar x}) - \tau\Bigr)g^{\mu}_{\;\;{\bar \mu}}(x,{\bar x})
{\hat J}^{\bar \mu}({\bar x})\biggl|_{x\to{\cal I}^+[\tau,\phi]}\quad ,\\
{D}_{\1}^{\;\;\mu\nu} 
&=& \frac{1}{4\pi}\int d{\bar x}{\bar g}^{1/2}
\delta\Bigl(U_{\phi}({\bar x}) - \tau\Bigr)
g^{\mu}_{\;\;{\bar \mu}}(x,{\bar x})g^{\nu}_{\;\;{\bar \nu}}(x,{\bar x})
{J}^{{\bar \mu}{\bar \nu}}({\bar x})\biggl|_{x\to{\cal I}^+[\tau,\phi]}\quad ,
\end{eqnarray}
and it is important that the propagators 
$g^{\alpha}_{\;\;{\bar \alpha}}(x,{\bar x})$ are taken at the point $x$ of
${\cal I}^+$ with the same parameters $(\tau,\phi)$ as in the 
$\delta-$function. The moments ${\hat D}_{\1},{\hat D}_{\1}^{\;\;\alpha},
{D}_{\1}^{\;\;\alpha\beta},$ etc. have the transformation properties
suggested by their  indices  but it should be remembered that they are 
defined only for a point at ${\cal I}^+.$ Thus, the metric $g_{\mu\nu}$ in
Eq. (4.37) can only be at the  point $(\tau, \phi)$ of ${\cal I}^+.$
Eq. (4.37) itself is a consequence of the relation [30]
\begin{equation}
g_{\mu\nu}(x)g^{\mu}_{\;\;{\bar \mu}}(x,{\bar x})
g^{\nu}_{\;\;{\bar \nu}}(x,{\bar x}) = g_{{\bar \mu}{\bar \nu}}({\bar x})
\quad .
\end{equation}

Using the law of parallel transport in (4.11) we may calculate
\begin{eqnarray}
\nabla_{\alpha}u \,\frac{d}{d\tau} {\hat D}_{\1}^{\;\;\alpha} & = &
- \frac{1}{4\pi}\int d{\bar x}{\bar g}^{1/2}
\delta'\Bigl(U_{\phi}({\bar x}) - \tau\Bigr){\bar \nabla}_{\bar \mu}
U_{\phi}({\bar x})
{\hat J}^{\bar \mu}({\bar x})\nonumber\\ 
&=&- \frac{1}{4\pi}\int d{\bar x}{\bar g}^{1/2}{\bar \nabla}_{\bar \mu}
\delta\Bigl(U_{\phi}({\bar x}) 
- \tau\Bigr){\hat J}^{\bar \mu}({\bar x})\quad .
\end{eqnarray}
Since, by assumption, the support of ${\hat J}^{\mu}$ is confined to
a spacetime tube, the integration by parts in (4.41) gives rise to
no boundary term. Hence, using the conservation law (1.9) we obtain
\begin{equation}
\nabla_{\alpha}u \,\frac{d}{d\tau}\;{\hat D}_{\1}^{\;\;\alpha} = 0 \quad .
\end{equation}

The conserved quantity in (4.42)
\begin{equation}
{\hat e} \equiv \nabla_{\alpha}u \,{\hat D}_{\1}^{\;\;\alpha}  
\end{equation}
is the full charge of the vector source since the charge can be written
as the integral
\begin{equation}
{\hat e} = \frac{1}{4\pi}\int d{\bar x} {\bar g}^{1/2}
\delta\Bigl(\Sigma({\bar x})\Bigr){\bar \nabla}_{\bar \mu}\Sigma\;
{\hat J}^{\bar \mu}({\bar x})
\end{equation}
over any hypersurface $\Sigma = 0$ crossing the support tube of 
${\hat J}^{\mu}$ ($\nabla\Sigma$ is past directed). In expression (4.41)
this hypersurface is a null hyperplane. If ${\hat J}^{\mu}$ is  a pure
vector like in the case of the electromagnetic current, Eqs. (4.42)-(4.44)
are exact. In the general case where the integral (4.44) contains the matrix
propagators of parallel transport, they are valid with accuracy 
$O[\Re^2]$ since the proof of conservation makes use of Eq. (3.7).

The significance of the assumption about the support of ${\hat J}^{\mu}$
is in the fact that it excludes a radiation of the charge. In the general
case, the full conserved charge is given by Eq. (4.44) in which the 
hypersurface $\Sigma = 0$ is spacelike. With this hypersurface null,
 the charge is generally not  conserved. It particular, if this hypersurface
  is the radial  light cone $u =$ const., Eq. (4.44) defines the Bondi charge
\begin{equation}
{\hat e}(u) = \frac{1}{4\pi}\int d{\bar x} {\bar g}^{1/2}
\delta\Bigl({\bar u} - u\Bigr){\bar \nabla}_{\bar \mu}{\bar u}\;
{\hat J}^{\bar \mu}({\bar x})\quad .
\end{equation}
Then we have by using Eq. (3.11) :
\begin{eqnarray}
\frac{d}{du}{\hat e}(u) 
&=& - \frac{1}{4\pi}\int d{\bar x} {\bar g}^{1/2}{\bar \nabla}_{\bar \mu}
\delta\Bigl({\bar u} - u\Bigr){\hat J}^{\bar \mu}({\bar x})\nonumber\\
&=&- \lim_{v_0\to\infty}
 \frac{1}{4\pi}\int d{\bar x} {\bar g}^{1/2}
\delta\Bigl({\bar u} - u\Bigr)\delta\Bigl({\bar v} - v_0\Bigr)
{\bar \nabla}_{\bar \mu}{\bar v}
{\hat J}^{\bar \mu}({\bar x}) 
\end{eqnarray}
with $v$ in (1.17). The boundary $v=v_0$ with $v_0\to\infty$ is ${\cal I}^+$.
Since at this boundary one can use the asymptotic form of the Bondi-Sachs
metric (1.15), the result is
\begin{equation}
- \frac{d{\hat e}(u)}{du} = \frac{1}{4\pi} \int d^2{\cal S}(\phi)\,\Bigl(
\frac{1}{2}r^2\nabla_{\mu}v{\hat J}^{\mu}\Bigr)\biggl|_{{\cal I}^+[u,
\phi,r\to\infty]}
\end{equation}
which is an equation analogous to (1.35). If, on the other hand, the support
of ${\hat J}^{\mu}$ is confined to a tube, we have 
${\hat J}^{\mu}\biggl|_{{\cal I}^+}\equiv 0$ , and
$d{\hat e}/du= 0$ .

In the case of the tensor source, a calculation similar to
(4.41)-(4.42) and valid with accuracy $O[\Re^2]$ since Eq. (3.7)
has to be used yields
\begin{equation}
\nabla_{\mu}u\,\frac{d}{d\tau}D_{\1}^{\;\;\mu\nu} = 0\quad .
\end{equation}
The respective conserved charge
\begin{equation}
p^{\nu} \equiv \frac{1}{2}\nabla_{\mu}u\, D_{\1}^{\;\;\mu\nu}
\end{equation}
is the full momentum
\begin{equation}
p^{\nu} = 
\frac{1}{8\pi}\int d{\bar x}{\bar g}^{1/2}\delta\Bigl(\Sigma({\bar x})\Bigr)
g^{\nu}_{\;\;{\bar \nu}}(x,{\bar x}){\bar \nabla}_{\bar \mu}
\Sigma\;J^{{\bar \mu}{\bar \nu}}({\bar x})\biggl|_{x\to{\cal I}^+}\quad ,
\end{equation}
and a radiation of the charge is again excluded by the assumption
about the support of $J^{\mu\nu}.$

\subsection*{\bf The vector and cross vertices at ${\cal I}^+$ .}

$$ $$

In Eq. (4.29) the moments $D_{\1}$ always appear differentiated with respect
to time. This is no more the case in Eqs. (4.30) and (4.31) but in the
specific combination of $F(m,n)$ entering the  vertex functions (2.41)
and (2.42) the terms with the undifferentiated $D_{\1}$ cancel.
Using integration by parts we obtain
\begin{eqnarray}
{\hat V}^{\alpha\beta}_{\mbox{\scriptsize cross}}
\biggl|_{{\cal I}^+[u,\phi,r\to\infty]} & = &
- \frac{1}{6r^2}\int\limits^u_{-\infty}d\tau\Bigl\{
(\tau - u)^3
\Bigl[\frac{d^2}{d\tau^2}\;{ D}_{\1}^{\;\;\alpha\beta}(\tau,\phi)\Bigr]
\Bigl[\frac{d^2}{d\tau^2}\;{\hat D}_{\1}^{\;\;Q}(\tau,\phi)\Bigr] \\
& &\hspace{2cm}+2(\tau - u)^2
\Bigl[\frac{d^2}{d\tau^2}\;D_{\1}^{\;\;\alpha\beta}(\tau,\phi)\Bigr]
\Bigl[\frac{d}{d\tau}\;{\hat D}_{\1}^{\;\;Q}(\tau,\phi)\Bigr]\Bigr\}\quad , 
\nonumber\\
{\hat V}^{\alpha\beta}_{\mbox{\scriptsize vect}}
\biggl|_{{\cal I}^+[u,\phi,r\to\infty]} & = &
 \frac{1}{12r}\int\limits^u_{-\infty}d\tau
(\tau - u)^2
\Bigl[\frac{d}{d\tau}\;{\hat D}_{\1}^{\;\;\alpha}(\tau,\phi)\Bigr]
\Bigl[\frac{d}{d\tau}\;{\hat D}_{\1}^{\;\;\beta}(\tau,\phi)\Bigr] 
\quad .
\end{eqnarray}
Owing to the presence of the derivatives acting on the $D_{\1}$'s in
all terms of these expressions, we have
\begin{equation}
\nabla_{\alpha}u\,{\hat V}^{\alpha\beta}_{\mbox{\scriptsize cross}}\biggl|_{ 
{\cal I}^+} = 0\quad , \quad
\nabla_{\alpha}u\,{\hat V}^{\alpha\beta}_{\mbox{\scriptsize vect}}\biggl|_{ 
{\cal I}^+} = 0
\end{equation}
by (4.42) and (4.48).

For a tensor $V^{\alpha\beta}$ with the property (4.53) we are to calculate
the behaviour at ${\cal I}^+[u,\phi,r\to\infty]$ of the quantity 
$\nabla_{\alpha}\nabla_{\beta}V^{\alpha\beta}$ appearing in (2.39).
For that, we expand $V^{\alpha\beta}$ over the null-tetrad basis in (1.19):
\begin{equation}
V^{\alpha\beta} = e^{\alpha}(\mu)e^{\beta}(\nu)V(\mu\nu)
\end{equation}
and use that
\begin{equation}
\nabla_{\alpha}u\; e^{\alpha}(\mu)V(\mu\nu)\biggl|_{{\cal I}^+} = 0
\end{equation}
by (4.53), and
\begin{equation}
\nabla e^{\alpha}(\mu)\biggl|_{{\cal I}^+} = O\Bigl(\frac{1}{r}\Bigr)\quad ,
\quad
\nabla\nabla e^{\alpha}(\mu)\biggl|_{{\cal I}^+} = 
O\Bigl(\frac{1}{r^2}\Bigr)\quad ,\;\;\;\mbox{etc.}
\end{equation}
by the result quoted in [23].

Since $V(\mu\nu)$ is a collection of scalars, one has
\begin{equation}
\nabla_{\alpha}V(\mu\nu) = \Bigl(
\nabla_{\alpha}u\;\frac{\partial}{\partial u} + 
\nabla_{\alpha}r\;\frac{\partial}{\partial r} +
\nabla_{\alpha}\phi^a\frac{\partial}{\partial \phi^a}\Bigr)
V(\mu\nu)
\end{equation}
whence, in view of (4.9),
\begin{equation}
\nabla_{\alpha}V(\mu\nu)\biggl|_{{\cal I}^+} =
\nabla_{\alpha}u\;\frac{\partial}{\partial u}V(\mu\nu) +  
O\Bigl(\frac{1}{r}V\Bigr)\quad .
\end{equation}
Similarly,
\begin{eqnarray}
\nabla_{\alpha}\nabla_{\beta}V(\mu\nu)\biggl|_{{\cal I}^+} & = &
\nabla_{\alpha}u\nabla_{\beta}u\frac{\partial^2}{\partial u^2}V(\mu\nu) +
(\nabla_{\alpha}u\nabla_{\beta}r + 
\nabla_{\beta}u\nabla_{\alpha}r)
\frac{\partial^2}{\partial u \partial r}V(\mu\nu) \hspace{2cm}\nonumber\\
&+& (\nabla_{\alpha}u\nabla_{\beta}\phi^a + 
\nabla_{\beta}u\nabla_{\alpha}\phi^a)
\frac{\partial^2}{\partial u \partial \phi^a}V(\mu\nu) +
(\nabla_{\alpha}\nabla_{\beta}u)\frac{\partial}{\partial u}V(\mu\nu)
\nonumber \\ & + & O\Bigl(\frac{1}{r^2}V\Bigr) \quad .
\end{eqnarray}
Upon the insertion of these expressions in 
$\nabla_{\alpha}\nabla_{\beta}V^{\alpha\beta}$ and the use of (4.54)-(4.56)
along with the condition
$\nabla(e^{\alpha}(\mu)\nabla_{\alpha}u) = 0$ , one is left with
\begin{equation}
\nabla_{\alpha}\nabla_{\beta}V^{\alpha\beta}\biggl|_{{\cal I}^+} = 
-(\nabla_{\alpha}\nabla_{\beta}u)\frac{\partial}{\partial u}V^{\alpha\beta} +
O\Bigl(\frac{1}{r^2}V\Bigr)\quad .
\end{equation}

An explicit calculation in the metric (1.15) yields
\begin{equation}
\nabla_{\alpha}\nabla_{\beta}u\biggl|_{{\cal I}^+} = 
-\frac{1}{2r}(m_{\alpha}m_{\beta}^* + m^*_{\alpha}m_{\beta})
\end{equation}
whence finally
\begin{equation}
\nabla_{\alpha}\nabla_{\beta}V^{\alpha\beta}\biggl|_{{\cal I}^+} = 
\frac{1}{r}\frac{\partial}{\partial u}\Bigl(m_{\alpha}m^*_{\beta}
V^{\alpha\beta}\Bigr) +
 O\Bigl(\frac{1}{r^2}V\Bigr)\quad .
\end{equation}
By using (1.21) and (4.53), this result can also be written in the form
\begin{equation}
\nabla_{\alpha}\nabla_{\beta}V^{\alpha\beta}\biggl|_{{\cal I}^+} = 
\frac{1}{r}\frac{\partial}{\partial u}\Bigl(g_{\alpha\beta}
V^{\alpha\beta}\Bigr) +
 O\Bigl(\frac{1}{r^2}V\Bigr)\quad .
\end{equation}
One power of $1/r$ is thus gained.

For ${\hat V}^{\alpha\beta}_{\mbox{\scriptsize  cross}}$ in (4.51) this gain
is sufficient to satisfy the criterion (4.26):
\begin{eqnarray}
\nabla_{\alpha}\nabla_{\beta}
{\hat V}^{\alpha\beta}_{\mbox{\scriptsize cross}}
\biggl|_{{\cal I}^+[u,\phi,r\to\infty]} & = &
- \frac{1}{6r^3}\int\limits^u_{-\infty}d\tau\Bigl\{
3(u - \tau)^2
\Bigl[\frac{d^2}{d\tau^2}\;{ D}_{\1}^{\;\;R}(\tau,\phi)\Bigr]
\Bigl[\frac{d^2}{d\tau^2}\;{\hat D}_{\1}^{\;\;Q}(\tau,\phi)\Bigr]  
\nonumber\\
& &-4(u - \tau)
\Bigl[\frac{d^2}{d\tau^2}\;D_{\1}^{\;\;R}(\tau,\phi)\Bigr]
\Bigl[\frac{d}{d\tau}\;{\hat D}_{\1}^{\;\;Q}(\tau,\phi)\Bigr]\Bigr\} 
\end{eqnarray}
but in the case of ${\hat V}^{\alpha\beta}_{\mbox{\scriptsize  vect}}$
in (4.52) we wind up with the behaviour $1/r^2$:
\begin{equation}
\nabla_{\alpha}\nabla_{\beta}{\hat V}^{\alpha\beta}_{\mbox{\scriptsize vect}
}\biggl|_{{\cal I}^+[u,\phi,r\to\infty]}  = 
 \frac{1}{6r^2}\int\limits^u_{-\infty}d\tau
(u - \tau)g_{\alpha\beta}
\Bigl[\frac{d}{d\tau}\;{\hat D}_{\1}^{\;\;\alpha}(\tau,\phi)\Bigr]
\Bigl[\frac{d}{d\tau}\;{\hat D}_{\1}^{\;\;\beta}(\tau,\phi)\Bigr] 
+ O\Bigl(\frac{1}{r^3}\Bigr) \; .
\end{equation}
The integrand in the latter expression 
\begin{equation}
g_{\alpha\beta}\Bigl(\frac{d}{d\tau}{\hat D}_{\1}^{\;\;\alpha}\Bigr)
\Bigl(\frac{d}{d\tau}{\hat D}_{\1}^{\;\;\beta}\Bigr) = 
\Bigl|\frac{d}{d\tau}D_{\1}^{\;\;\alpha}m_{\alpha}\Bigr|^2
\end{equation}
is none other than the energy flux of the outgoing waves of the
vector connection field (see Sec.8). To ensure the fulfilment
of the criterion (4.26) we are compelled to impose a limitation
on the vector source, namely that this source does not radiate
classically; then the quantity (4.66) vanishes. The energy
of the vacuum radiation is obtained in the present paper under 
this limitation, and  the limitation itself is discussed in conclusion.

\subsection*{\bf The gravitational vertex at ${\cal I}^+$ .}

$$ $$

Using expression (4.31) for the $F(m,n)$ entering the vertex function
(2.40), and integrating by parts, we obtain
\begin{eqnarray}
{\hat V}^{\alpha\beta\mu\nu}_{\mbox{\scriptsize grav}}\biggl|_{{\cal I}^+[u,
\phi,r\to\infty]}
& = &
- \frac{\hat 1}{360r}\int\limits^u_{-\infty}d\tau
\biggl\{(\tau - u)^4
\Bigl[\frac{d^2}{d\tau^2}D_{\1}^{\;\;\alpha\beta}(\tau,\phi)\Bigr]
\Bigl[\frac{d^2}{d\tau^2}D_{\1}^{\;\;\mu\nu}(\tau,\phi)\Bigr]  \\
& &+(\tau - u)^2
\Bigl[\frac{d}{d\tau}D_{\1}^{\;\;\alpha\beta}(\tau,\phi)\Bigr]
\Bigl[\frac{d}{d\tau}D_{\1}^{\;\;\mu\nu}(\tau,\phi)\Bigr] + 
2D_{\1}^{\;\;\alpha\beta}(\tau,\phi)D_{\1}^{\;\;\mu\nu}(\tau,\phi)
\biggl\}\; . \nonumber
\end{eqnarray}
Here, as distinct from the previous case, the  terms with the 
undifferentiated $D_{\1}$ do not cancel. For the completely symmetrized
${\hat V}^{\alpha\beta\mu\nu}_{\mbox{\scriptsize grav}}$ we have
\begin{equation}
\nabla_{\alpha}u\,
{\hat V}^{(\alpha\beta\mu\nu)}_{\mbox{\scriptsize grav}}\biggl|_{{\cal I}^+[u,
\phi,r\to\infty]} = 
- \frac{\hat 1}{90r}p^{(\beta}\int\limits^u_{-\infty}d\tau
D_{\1}^{\;\;\mu\nu)}(\tau,\phi)
\end{equation}
by (4.48) and (4.49). Two more time derivatives and two more contractions with 
$\nabla u$ are needed for this quantity to vanish:
\begin{equation}
\nabla_{\alpha}u\,\frac{\partial^2}{\partial u^2}
{\hat V}^{(\alpha\beta\mu\nu)}_{\mbox{\scriptsize grav}}\biggl|_{{\cal I}^+} 
= 
- \frac{\hat 1}{270r}\Bigl\{
p^{\beta}\frac{\partial}{\partial u}D_{\1}^{\;\;\mu\nu}(u,\phi) + 
p^{\mu}\frac{\partial}{\partial u}D_{\1}^{\;\;\nu\beta}(u,\phi) + 
p^{\nu}\frac{\partial}{\partial u}D_{\1}^{\;\;\beta\mu}(u,\phi) \Bigr\}\; ,
\end{equation}
\begin{equation}
\nabla_{\mu}u\nabla_{\beta}u\nabla_{\alpha}u\,\frac{\partial^2}{\partial u^2}
{\hat V}^{(\alpha\beta\mu\nu)}_{\mbox{\scriptsize grav}}\biggl|_{{\cal I}^+} 
= 0 \quad .
\end{equation}

Above, the tensor ${\hat V}^{\alpha\beta\mu\nu}_{\mbox{\scriptsize grav}}$
 is considered symmetrized because, with the appropriate accuracy,
 it is symmetrized in the quantity appearing in (2.39):
 \begin{equation}
 \nabla_{\alpha}\nabla_{\beta}\nabla_{\mu}\nabla_{\nu}\,
 {\hat V}^{\alpha\beta\mu\nu}_{\mbox{\scriptsize grav}} = 
 \nabla_{\alpha}\nabla_{\beta}\nabla_{\mu}\nabla_{\nu}\,
 {\hat V}^{(\alpha\beta\mu\nu)}_{\mbox{\scriptsize grav}} + O[\Re^3]\quad .
 \end{equation}
 For the calculation of this quantity at ${\cal I}^+$, we introduce the 
 projection of 
 ${\hat V}^{(\alpha\beta\mu\nu)}_{\mbox{\scriptsize grav}}$
on the null tetrad
\begin{equation}
{\hat V}^{(\alpha\beta\mu\nu)}_{\mbox{\scriptsize grav}} = 
e^{\alpha}(\gamma)e^{\beta}(\sigma)e^{\mu}(\rho)e^{\nu}(\delta)\,
V(\gamma\sigma\rho\delta)
\end{equation}
and use that by (4.70)
\begin{equation}
\nabla_{\mu}u\nabla_{\beta}u\nabla_{\alpha}u
\;e^{\alpha}(\gamma)e^{\beta}(\sigma)e^{\mu}(\rho)\frac{\partial^2}{\partial u^2}
V(\gamma\sigma\rho\delta)\biggl|_{{\cal I}^+} = 0 \quad .
\end{equation}

One needs only the third and fourth derivatives of 
$V(\gamma\sigma\rho\delta)$ since by (4.56)
\begin{eqnarray}
\nabla_{\alpha}\nabla_{\beta}\nabla_{\mu}\nabla_{\nu}
{\hat V}^{(\alpha\beta\mu\nu)}_{\mbox{\scriptsize grav}}\biggl|_{{\cal I}^+}
&=&
4\nabla_{\alpha}\Bigl(
e^{\alpha}(\gamma)
e^{\beta}(\sigma)e^{\mu}(\rho)e^{\nu}(\delta)\Bigr)
\nabla_{\beta}\nabla_{\mu}\nabla_{\nu}\, V(\gamma\sigma\rho\delta) \\
&+ &
e^{\alpha}(\gamma)e^{\beta}(\sigma)e^{\mu}(\rho)e^{\nu}(\delta)
\nabla_{\alpha}\nabla_{\beta}\nabla_{\mu}\nabla_{\nu}\,
V(\gamma\sigma\rho\delta) +
O\Bigl(\frac{1}{r^2}V\Bigr)\nonumber\;\; .
\end{eqnarray}
The third derivative is needed only to lowest order,
\begin{equation}
\nabla_{\beta}\nabla_{\mu}\nabla_{\nu}\,
V(\gamma\sigma\rho\delta)\biggl|_{{\cal I}^+} = 
\nabla_{\beta}u\nabla_{\mu}u\nabla_{\nu}u
\frac{\partial^3}{\partial u^3}
V(\gamma\sigma\rho\delta) +
O\Bigl(\frac{1}{r}V\Bigr)
\end{equation}
whereas of the fourth derivative one needs two orders:
\begin{eqnarray}
\nabla_{\alpha}\nabla_{\beta}\nabla_{\mu}\nabla_{\nu}\,
V(\gamma\sigma\rho\delta)\biggl|_{{\cal I}^+} &=& 
\nabla_{\alpha}u\nabla_{\beta}u\nabla_{\mu}u\nabla_{\nu}u
\frac{\partial^4}{\partial u^4}
V(\gamma\sigma\rho\delta)  \hspace{5cm} \nonumber\\
&+ &
\Bigl(
\nabla_{\alpha}u\nabla_{\beta}u\nabla_{\mu}u\nabla_{\nu}r +
\nabla_{\alpha}u\nabla_{\beta}u\nabla_{\nu}u\nabla_{\mu}r  \nonumber\\
& &
+\nabla_{\alpha}u\nabla_{\mu}u\nabla_{\nu}u\nabla_{\beta}r +
\nabla_{\beta}u\nabla_{\mu}u\nabla_{\nu}u\nabla_{\alpha}r    \Bigr)
\frac{\partial^4}{\partial u^3\partial r}
V(\gamma\sigma\rho\delta)  \nonumber\\
&+ &
\Bigl(
\nabla_{\alpha}u\nabla_{\beta}u\nabla_{\mu}u\nabla_{\nu}\phi^a +
\nabla_{\alpha}u\nabla_{\beta}u\nabla_{\nu}u\nabla_{\mu}\phi^a  
\nonumber \\ 
 & &
+\nabla_{\alpha}u\nabla_{\mu}u\nabla_{\nu}u\nabla_{\beta}\phi^a +
\nabla_{\beta}u\nabla_{\mu}u\nabla_{\nu}u\nabla_{\alpha}\phi^a    \Bigr)
\frac{\partial^4}{\partial u^3\partial \phi^a}
V(\gamma\sigma\rho\delta)    \nonumber\\
&+ &
\Bigl(
\nabla_{\mu}u\nabla_{\nu}u\cdot\nabla_{\alpha}\nabla_{\beta}u +
\nabla_{\mu}u\nabla_{\alpha}u\cdot\nabla_{\nu}\nabla_{\beta}u \nonumber\\
& & + 
\nabla_{\mu}u\nabla_{\beta}u\cdot\nabla_{\nu}\nabla_{\alpha}u  
+\nabla_{\nu}u\nabla_{\alpha}u\cdot\nabla_{\mu}\nabla_{\beta}u  \nonumber\\
& & +  
\nabla_{\nu}u\nabla_{\beta}u\cdot\nabla_{\mu}\nabla_{\alpha}u + 
\nabla_{\alpha}u\nabla_{\beta}u\cdot\nabla_{\mu}\nabla_{\nu}u 
\Bigr)\frac{\partial^3}{\partial u^3}
V(\gamma\sigma\rho\delta) \nonumber\\
&+&O\Bigl(\frac{1}{r^2}V\Bigr)\quad .            
\end{eqnarray}
All terms in the two latter expressions have a sufficient number of 
$\partial/\partial u$ for Eq. (4.73) to work but not all have a 
sufficient number of $\nabla u$. As a result, one is left with
\begin{equation}
\nabla_{\alpha}\nabla_{\beta}\nabla_{\mu}\nabla_{\nu}
{\hat V}^{\alpha\beta\mu\nu}_{\mbox{\scriptsize grav}}\biggl|_{{\cal I}^+} =
-6\nabla_{\alpha}u\nabla_{\beta}u(\nabla_{\mu}\nabla_{\nu}u)
\frac{\partial^3}{\partial u^3}{\hat V}^{(\alpha\beta\mu\nu)}_{
\mbox{\scriptsize grav}} +
O\Bigl(\frac{1}{r^2}V\Bigr)\quad .
\end{equation}
It is already seen that the present case is dissimilar to the previous
one (cf. Eq. (4.60)). Since $\nabla\nabla u = O(1/r)$ , there can be and is
only one factor of $\nabla\nabla u$ in (4.77). The remaining indices
have to be contracted with $\nabla u$. Therefore, the completely transverse
projection of ${\hat V}^{\alpha\beta\mu\nu}_{\mbox{\scriptsize grav}}$
drops out of (4.77).

Using (4.61) we have
\begin{equation}
\nabla_{\alpha}\nabla_{\beta}\nabla_{\mu}\nabla_{\nu}
{\hat V}^{\alpha\beta\mu\nu}_{\mbox{\scriptsize grav}}\biggl|_{{\cal I}^+} =
\frac{6}{r}\frac{\partial^3}{\partial u^3}
\biggl(\nabla_{\alpha}u\nabla_{\beta}u\;m_{\mu}m^*_{\nu}
{\hat V}^{(\alpha\beta\mu\nu)}_{\mbox{\scriptsize grav}}\biggr) +
O\Bigl(\frac{1}{r^2}V\Bigr)\quad ,
\end{equation}
or, by (1.21) and (4.70),
\begin{equation}
\nabla_{\alpha}\nabla_{\beta}\nabla_{\mu}\nabla_{\nu}
{\hat V}^{\alpha\beta\mu\nu}_{\mbox{\scriptsize grav}}\biggl|_{{\cal I}^+} =
\frac{6}{r}\frac{\partial^3}{\partial u^3}
\Bigl(\nabla_{\alpha}u\nabla_{\beta}u\;g_{\mu\nu}
{\hat V}^{(\alpha\beta\mu\nu)}_{\mbox{\scriptsize grav}}\Bigr) +
O\Bigl(\frac{1}{r^2}V\Bigr)\quad .
\end{equation}
Finally, using (4.69),(4.37), and denoting
\begin{equation}
\mu = \nabla_{\nu}u\, p^{\nu}\quad ,
\end{equation}
we obtain
\begin{equation}
\nabla_{\alpha}\nabla_{\beta}\nabla_{\mu}\nabla_{\nu}
{\hat V}^{\alpha\beta\mu\nu}_{\mbox{\scriptsize grav}}\biggl|_{
{\cal I}^+[u,\phi,r\to\infty]} =
\frac{\hat 1}{r^2}\;\frac{\mu}{45}\;\frac{d^2}{du^2}\;D_{\1}^{\;\;R}(u,\phi)
+ O\Bigl(\frac{1}{r^3}\Bigr)\quad .
\end{equation}
The result is that the $1/r^2$ term survives also here but there is nothing
like a radiation flux in its coefficient. The most important fact is that 
this coefficient vanishes at late time as distinct from the coefficient in 
(4.65) which grows at late time (Eqs. (6.53) and (6.54) below). Since only
the late-time behaviour is relevant to the calculation of the total
vacuum energy, the $1/r^2$ term in (4.81) will have no effect on this 
calculation and no special limitation on the tensor source will be
required.

The convergence condition (4.26) should, of course, be fulfilled everywhere
at ${\cal I}^+$. The full solution of this problem including a removal
of the limitation on the vector source remains beyond the scope of the 
present work (see the conclusion).

\subsection*{\bf The trees at ${\cal I}^+$ .}

$$ $$ 

There remain to be considered the terms in (2.38)-(2.39) whose operator
coefficients factorize into a product of massless Green functions.
The simplest such terms are ${\hat I}_1$ in (2.38) and the trees in 
${\hat I}_2$ that contain a curvature at the observation point :
\begin{equation}
R\quad,\quad J^{\alpha\beta}\Bigl(\frac{1}{\Box}J_{\alpha\beta}\Bigr)\quad,
\quad R\Bigl(\frac{1}{\Box}R\Bigr)\quad .
\end{equation}
Generally, in an asymptotically flat spacetime one has (see, e.g., [23])
\begin{equation}
J^{\alpha\beta}\biggl|_{{\cal I}^+} = 
O\Bigl(\frac{1}{r^2}\Bigr)\quad , 
\quad R\biggl|_{{\cal I}^+} = O\Bigl(\frac{1}{r^3}\Bigr)\quad ,
\end{equation}
and the power of decrease of $(1/\Box)X$ at ${\cal I}^+$ is $1/r$.
Therefore, in the general case all the trees in (4.82) are
$O(1/r^3),$ and with the present assumption about the support of the
sources they vanish identically outside a spacetime tube. 

The remaining tree in (2.45) can be written in the form
\begin{equation}
\Bigl(\nabla_{\beta}\frac{1}{\Box}J^{\alpha\lambda}\Bigr)
\Bigl(\nabla_{\alpha}\frac{1}{\Box}J^{\beta}_{\;\;\lambda}\Bigr) =
\nabla_{\alpha}\nabla_{\beta}A^{\alpha\beta} + O[\Re^3]
\end{equation}
where
\begin{equation}
A^{\alpha\beta} = \Bigl(\frac{1}{\Box}J^{\alpha\lambda}\Bigr)
\Bigl(\frac{1}{\Box}J^{\beta}_{\;\;\lambda}\Bigr)
\end{equation}
and use is made of the conservation law (1.9).
By (4.17)
\begin{equation}
A^{\alpha\beta}\biggl|_{{\cal I}^+[u,\phi,r\to\infty]} = 
\frac{1}{r^2}g_{\mu\nu}D_{\1}^{\;\;\alpha\mu}(u,\phi)
D_{\1}^{\;\;\beta\nu}(u,\phi)\quad ,
\end{equation}
and, to lowest order in $1/r$,
\begin{equation}
\nabla_{\alpha}\nabla_{\beta}A^{\alpha\beta}\biggl|_{{\cal I}^+} = 
\nabla_{\alpha}u\nabla_{\beta}u\frac{\partial^2}{\partial u^2}A^{\alpha\beta}
+ O\Bigl(\frac{1}{r}A\Bigr)\quad .
\end{equation}
Hence
\begin{equation}
\nabla_{\alpha}\nabla_{\beta}A^{\alpha\beta}\biggl|_{{\cal I}^+} = 
\frac{4}{r^2}\;\frac{d^2}{du^2}g_{\mu\nu}p^{\mu}p^{\nu} + 
O\Bigl(\frac{1}{r^3}\Bigr) = 
O\Bigl(\frac{1}{r^3}\Bigr)
\end{equation}
where $p^{\mu}$ is the conserved momentum (4.49).
Thus, also this tree is $O(1/r^3)$ at ${\cal I}^+.$

There is, however, one more tree which thus far has not been taken into
account. The point is that we are using everywhere the lowest-order
approximation (3.1) for the resolvent. This is correct in the terms of 
second order in $\Re$ but in the terms of  first order in $\Re$ the
resolvent itself should be taken with a higher accuracy.
Therefore, we must come back to expression (2.36) with the first-order $I$ ,
$I=I_1$ ,
and calculate the curvature correction to the form factor \footnote{
The curvature corrections to the derivatives $\nabla^{\mu}\nabla^{\nu}$
in (2.36) vanish at ${\cal I}^+.$}
in this expression:
\begin{equation}
\delta\log(-\Box)I_1\biggl|_{{\cal I}^+}\quad .
\end{equation}

By using the spectral form
\begin{equation}
\log(-\Box/c^2) = - \int\limits_0^{\infty}dm^2\Bigl(\frac{1}{m^2 - \Box} - 
\frac{1}{m^2 + c^2}\Bigr)
\end{equation}
and the variational law for the retarded Green function we obtain
\begin{equation}
\delta\log(-\Box)X(x) = 
- \int\limits_0^{\infty}dm^2\frac{1}{m^2 - \Box_3}\delta\Box_2 
\frac{1}{m^2 - \Box_1}X(x) = 
\frac{\log(\Box_3/\Box_1)}{\Box_3-\Box_1}\delta\Box_2X(x)
\end{equation}
where the numbers on the $\Box$'s indicate the order in which the 
operators act on $X(x).$ We may now use the theorem in [21,22]
by which the limit of (4.91) as $x\to{\cal I}^+$ is determined by the
 limit $\Box\to 0$ in the operator $\Box$ acting the last i.e. in $\Box_3.$
 We obtain
\begin{equation}
\delta\log(-\Box)X\biggl|_{{\cal I}^+} = 
- \frac{\log(-\Box_3)}{\Box_1}\delta\Box_2 X = 
-\log(-\Box)\delta\Box\frac{1}{\Box}X\quad .
\end{equation}
As could be expected, the variation of $\log(-\Box)$ in the background fields
behaves at ${\cal I}^+$ in the same way as $\log(-\Box)$ itself, i.e. like
$1/r^2,$ and the result (4.92) is valid up to $O(1/r^3).$

Using the latter result in expression (4.89), we may write
\begin{equation}
\Bigl(\log(-\Box) + \delta\log(-\Box)\Bigr)I_1\biggl|_{{\cal I}^+} = 
\log(-\Box)\Bigl(I_1 - \delta\Box\frac{1}{\Box}I_1\Bigr)\quad .
\end{equation}
This means that introducing a correction to the form factor boils down to
the replacement of the scalar $I$ by $I + \delta I $ with
\begin{equation}
\delta I = - \delta \Box \frac{1}{\Box} I_1\quad .
\end{equation}
Upon this replacement one may use everywhere the lowest-order approximation
for the resolvent. In (4.94) one is to insert the expression for $I_1$
from Eq. (2.38) and the expression
\begin{equation}
\delta \Box = 2\Bigl(\frac{1}{\Box}R^{\alpha\beta}\Bigr)
\nabla_{\alpha}\nabla_{\beta}
\end{equation}
from Ref. [15]. As a result, the correction under the sign of trace
in (2.37) takes the form
\begin{equation}
\delta {\hat I} = \frac{\hat 1}{90}\Bigl(\frac{1}{\Box}R^{\alpha\beta}\Bigr)
\nabla_{\alpha}\nabla_{\beta}\frac{1}{\Box}R\quad ,
\end{equation}
or, in terms of the conserved current $J^{\alpha\beta},$
\begin{equation}
\delta {\hat I} = \frac{\hat 1}{90}\nabla_{\alpha}\nabla_{\beta}B^{\alpha\beta}
+ \frac{\hat 1}{180} R\Bigl(\frac{1}{\Box}R\Bigr)
\end{equation}
with
\begin{equation}
B^{\alpha\beta} = \Bigl(\frac{1}{\Box} J^{\alpha\beta}\Bigr)
\Bigl(\frac{1}{\Box}R\Bigl)\quad .
\end{equation}
Expression (4.97) is one more tree.

The second term in (4.97) is of the type (4.82), and its contribution at
${\cal I}^+$ is $O(1/r^3).$ The calculation of the first term at ${\cal I}^+$
is similar to the one in Eqs. (4.86)-(4.87):
\begin{eqnarray}
B^{\alpha\beta}\biggl|_{{\cal I}^+[u,\phi,r\to\infty]} &=& 
\frac{1}{r^2}D_{\1}^{\;\;\alpha\beta}(u,\phi)D_{\1}^{\;\;R}(u,\phi)\quad ,\\
\nabla_{\alpha}\nabla_{\beta}
B^{\alpha\beta}\biggl|_{{\cal I}^+} &=& 
\nabla_{\alpha}u\nabla_{\beta}u\frac{\partial^2}{\partial u^2}
B^{\alpha\beta} + O\Bigl(\frac{1}{r}B\Bigr)
\end{eqnarray}
but the result is different. We obtain
\begin{equation}
\delta {\hat I}\biggl|_{{\cal I}^+[u,\phi,r\to\infty]} = 
\frac{\hat 1}{r^2}\;\frac{\mu}{45}\;\frac{d^2}{du^2}D_{\1}^{\;\;R}(u,\phi)
+ O\Bigl(\frac{1}{r^3}\Bigr)
\end{equation}
with $\mu$ in (4.80). Here the $1/r^2$ term survives but it
has the same form as (4.81), and the same inference applies.
}

\newpage

{\renewcommand{\theequation}{5.\arabic{equation}}  

\begin{center}
\section{\bf    The early-time behaviours}
\end{center}

$$ $$

Let ${\cal T}$ be the support tube of the physical sources in (1.8).
By the original assumption about the asymptotic stationarity of 
external fields, in the past and future of tube ${\cal T}$ (${\cal T}^-$ 
and ${\cal T}^+$ respectively) there exist asymptotic timelike Killing 
vectors such that all sources $J$ in (1.8) are conserved along
their integral curves. 
Let $\xi^{\alpha}(x)$ be a timelike vector field that interpolates between 
the Killing vectors at ${\cal T}^-$ and ${\cal T}^+$, and up to $O[\Re]$
is the Killing vector for the whole of ${\cal T}$. Denote 
${\cal L}_{\xi}$ the Lie derivative in the direction of $\xi^{\alpha}$,
and
\begin{equation}
\J = \Bigl(\LXI J^{\mu\nu},\;\LXI{\hat J}^{\mu},\;\LXI{\hat P}\Bigr)\;. 
\end{equation}

For simplicity, the support of $\J$ will be assumed compact. Then, 
on the central geodesic of any Bondi-Sachs frame, there will be two points
$o^-$ and $o^+$ with $u(o^-) = u^-,\;u(o^+)= u^+,$ and $u^- < u^+,$
such that the support of $\J$ is entirely inside the future light cone of
$o^-$ and the past light cone of $o^+$. In the approximation (3.13)
the fields of nonstationary sources propagate only between the cones
$u=u^-$ and $u=u^+$ rather than in the whole causal future of 
$\mbox{supp}\J$. We have
\begin{equation}
\nabla^{\mu}\xi^{\nu} + \nabla^{\nu}\xi^{\mu} = 0\quad,\quad\J = 0\;,
\quad\;\; u<u^-,\;\;u>u^+\quad.
\end{equation}

For what follows we need some properties that the functions involved in
the calculation possess in presence of the Killing vector field.

\subsection*{\bf Geometrical two-point functions in presence of the Killing\\
vector.}

$$ $$

If $\xi^{\alpha}$ is a Killing vector field, then the Lie derivative
$\LXI$ commutes with the covariant derivative $\nabla^{\beta}$
when acting on any object:
\begin{equation}
\LXI \nabla^{\beta}X^{\alpha\ldots} = \nabla^{\beta}\LXI X^{\alpha\ldots}
\quad .
\end{equation}
Indeed, by definition,
\begin{equation}
\LXI X^{\alpha\ldots} = \xi^{\mu}\nabla_{\mu}X^{\alpha\ldots} - 
\Sigma X^{\mu\ldots}\nabla_{\mu}\xi^{\alpha}
\end{equation}
whence making use of the Killing equation one obtains
\begin{equation}
\LXI \nabla^{\beta}X^{\alpha\ldots} - \nabla^{\beta}\LXI X^{\alpha\ldots} =
\Sigma X^{\mu\ldots}\Bigl(\nabla^{\beta}\nabla_{\mu}\xi^{\alpha} - 
R^{\alpha\;\beta}_{\cdot\mu\cdot\nu}\xi^{\nu}\Bigr)\quad ,
\end{equation}
and this commutator vanishes by the known property of the Killing
vector field [25]
\begin{equation}
\nabla_{\alpha}\nabla_{\beta}\;\xi_{\gamma} = 
R_{\gamma\beta\alpha\cdot}^{\;\;\;\;\;\;\;\sigma}\xi_{\sigma}\quad .
\end{equation}

Consider now the world function $\sigma=\sigma(x,{\bar x})$ for a 
timelike or spacelike separation of the points $x$ and ${\bar x}$.
The vector $n=\nabla\sigma/\sqrt{2|\sigma|}$ is a unit tangent at
the point $x$ to the geodesic connecting $x$ and ${\bar x}$, and the 
vector ${\bar n}={\bar \nabla}\sigma/\sqrt{2|\sigma|}$ is the oppositely
directed unit tangent to the same geodesic at the point ${\bar x}.$
Since the quantity $\xi^{\alpha}n_{\alpha}$ is conserved along
the geodesic [25], we have
\begin{equation}
\xi^{\alpha}n_{\alpha} = -{\bar \xi}^{\alpha}{\bar n}_{\alpha}\quad ,
\end{equation}
or 
\begin{equation}
\xi^{\alpha}\nabla_{\alpha}\sigma(x,{\bar x}) + 
{\bar \xi}^{\alpha}{\bar \nabla}_{\alpha}\sigma(x,{\bar x}) = 0\quad .
\end{equation}
This is the conservation law for the world function. By continuity it 
holds also for a null separation of the points $x$ and ${\bar x}.$

Eq. (5.8) can be written in the form
\begin{equation}
\Bigl(\LXI + {\cal L}_{\bar \xi}\Bigr)\sigma = 0
\end{equation}
whence using (5.3) one obtains
\begin{equation}
\Bigl(\LXI + {\cal L}_{\bar \xi}\Bigr)
\nabla_{\alpha_1}\ldots\nabla_{\alpha_n}
{\bar \nabla}_{\beta_1}\ldots{\bar \nabla}_{\beta_m}\sigma = 0
\end{equation}
--- a conservation law for the derivatives of the world function.

A similar law for the propagator of the geodetic parallel transport
can be obtained as follows. From the equations (3.3) for 
$g^{\alpha}_{\;\;{\bar \alpha}}$ we have
\begin{eqnarray}
\Bigl(\LXI\sigma^{\mu}\Bigr)\nabla_{\mu}g^{\alpha{\bar \alpha}} +
\sigma^{\mu}\nabla_{\mu}
\Bigl(\LXI g^{\alpha{\bar \alpha}}\Bigr) &=& 0\quad , \\
\Bigl({\cal L}_{\bar \xi}\sigma^{\mu}\Bigr)\nabla_{\mu}g^{\alpha{\bar \alpha}} +
\sigma^{\mu}\nabla_{\mu}
\Bigl({\cal L}_{\bar \xi}g^{\alpha{\bar \alpha}}\Bigr) &=& 0
\end{eqnarray}
where use is made of the commutation law (5.3). Since
\begin{equation}
\Bigl(\LXI + {\cal L}_{\bar \xi}\Bigr)\sigma^{\mu} = 0
\end{equation}
by (5.10), combining Eqs. (5.11) and (5.12) yields
\begin{equation}
\sigma^{\mu}\nabla_{\mu}
\Bigl(\LXI + {\cal L}_{\bar \xi}\Bigr)g^{\alpha{\bar \alpha}} = 0\quad.
\end{equation}
We have also
\begin{equation}
\Bigl(\LXI + {\cal L}_{\bar \xi}\Bigr)g^{\alpha{\bar \alpha}}
\biggl|_{{\bar x}=x} = - \nabla^{\bar \alpha}\xi^{\alpha} - 
\nabla^{\alpha}\xi^{\bar \alpha} = 0
\end{equation}
which follows from the initial condition in (3.3) and the condition
\begin{equation}
\nabla_{\mu}g^{\alpha{\bar \alpha}}\biggl|_{{\bar x}=x} = 0\quad .
\end{equation}
The  latter condition is obtained by differentiating the equation 
for $g^{\alpha}_{\;\;{\bar \alpha}}$ in (3.3)
and setting ${\bar x} = x.$ With
the initial condition (5.15), the solution of Eq. (5.14) is
\begin{equation}
\Bigl(\LXI + {\cal L}_{\bar \xi}\Bigr) g^{\alpha{\bar \alpha}} = 0.
\end{equation}
This is the desired result.

Finally, consider Eq. (5.8) with a timelike $\xi^{\alpha}$ and go over
to the limit $x\to{\cal I}^+$ in this equation. The insertion of the 
asymptotic behaviour of the world function (4.5) yields
\begin{equation}
- \xi^{\mu}\nabla_{\mu}u(x) + \xi^{\mu}\nabla_{\mu}\phi^a(x)
\frac{\partial}{\partial \phi^a}U_{\phi}({\bar x}) = 
-{\bar \xi}^{\mu}{\bar \nabla}_{\mu}U_{\phi}({\bar x})\quad , 
\quad x\to{\cal I}^+\;\;.
\end{equation}
Here the term with  $\partial/\partial\phi$ drops out by (4.9), and, with
the usual normalization of the timelike Killing vector at infinity, we have
\begin{equation}
\xi^{\mu}\nabla_{\mu}u\biggl|_{{\cal I}^+} = 1\quad .
\end{equation}
Thus we obtain
\begin{equation}
{\bar \xi}^{\mu}{\bar \nabla}_{\mu}U_{\phi}({\bar x}) = 1
\end{equation}
which is the conservation law for the null hyperplanes.
Using (5.3) we obtain also
\begin{equation}
{\cal L}_{\bar \xi}{\bar \nabla}_{\alpha_1}\ldots{\bar \nabla}_{\alpha_n}
U_{\phi}({\bar x}) = 0\quad , \quad n\ge 1\quad .
\end{equation}
In the Bondi-Sachs frame with the retarded time normalized as in (1.14),
the solution of Eq. (5.20) is
\begin{equation}
U_{\phi}({\bar u},{\bar \phi},{\bar r}) = {\bar u} + 
L_{\phi}({\bar \phi},{\bar r})\quad ,
\end{equation}
where the function $L_{\phi}({\bar \phi},{\bar r})$ possesses
the properties
\begin{equation}
L_{\phi}({\bar \phi},{\bar r})\ge 0\quad , \quad 
L_{\phi}({\bar \phi},{\bar r})\biggl|_{{\bar \phi}=\phi} = 0\quad , \quad
\frac{\partial}{\partial {\bar \phi}}L_{\phi}({\bar\phi},{\bar r})
\biggl|_{{\bar\phi}=\phi}=0
\end{equation}
by (4.13) and (4.14).

It should be emphasized that the relations above for the two-point functions
hold only in the case where the geodesic connecting the two points lies 
entirely in the Killing domain.

\subsection*{\bf Retarded kernels in presence of the Killing vector.}

$$ $$

That an arbitrary timelike vector in Eq. (3.12) and similar equations 
has been denoted $\xi^{\alpha}$ is no mere coincidence. As will be seen
from the calculations below, it is advantageous to choose for this arbitrary
vector the vector $\xi^{\alpha}$ defined in the beginning of the present
section.

Consider an integral of some scalar source $X$ over the past hyperboloid 
of a point $x$, and calculate
\begin{eqnarray}
\xi^{\alpha}\nabla_{\alpha}
\int\limits_{\mbox{\scriptsize past of}\;x} d{\bar x}{\bar g}^{1/2} 
\delta\Bigl(\sigma(x,{\bar x}) - q\Bigr){\bar X} \hspace{8cm}\\  =
\int\limits_{\mbox{\scriptsize past of}\;x} d{\bar x}{\bar g}^{1/2} 
\frac{(\xi\cdot\nabla\sigma)}{({\bar \xi}\cdot{\bar \nabla}\sigma)}
{\bar \xi}^{\alpha}{\bar \nabla}_{\alpha}
\delta\Bigl(\sigma(x,{\bar x}) - q\Bigr){\bar X}\hspace{2cm} \nonumber \\ =
-\int\limits_{\mbox{\scriptsize past of}\;x} d{\bar x}{\bar g}^{1/2} 
\delta\Bigl(\sigma(x,{\bar x}) - q\Bigr)
{\bar \nabla}_{\alpha}\biggl({\bar \xi}^{\alpha}
\frac{(\xi\cdot\nabla\sigma)}{({\bar \xi}\cdot{\bar \nabla}\sigma)}
{\bar X}\biggr)\quad .\hspace{1cm}\nonumber
\end{eqnarray}
Let now the observation point $x$ belong to the past Killing domain
$u(x)<u^-$. Since the hyperboloid in (5.24) lies entirely in the
causal past of $x$, the integration point ${\bar x}$ also belongs to
the past Killing domain. Furthermore, the timelike geodesic connecting
$x$ and ${\bar x}$ lies inside the past light cone  of $x$ and, therefore, 
passes  entirely through the Killing domain. Then, using Eq. (5.8) and
the corollary 
${\bar \nabla}_{\alpha}{\bar \xi}^{\alpha}=0$ of the Killing equation,
we obtain
\begin{eqnarray}
\xi^{\alpha}\nabla_{\alpha}
\int\limits_{\mbox{\scriptsize past of}\;x} d{\bar x}{\bar g}^{1/2} 
\delta\Bigl(\sigma(x,{\bar x}) - q\Bigr){\bar X} 
=
\int\limits_{\mbox{\scriptsize past of}\;x} d{\bar x}{\bar g}^{1/2} 
\delta\Bigl(\sigma(x,{\bar x}) - q\Bigr)
{\bar \xi}^{\alpha}{\bar \nabla}_{\alpha}{\bar X} \quad ,\nonumber\\
\quad u(x) < u^-\quad .
\end{eqnarray}

For obtaining a similar result in the case of a tensor source $X$,
consider ${\bar X}$ in Eq.(3.8). Write
\begin{equation}
{\bar \xi}^{\alpha}{\bar \nabla}_{\alpha}{\bar X} = 
{\cal L}_{\bar \xi}
\Bigl(g^{\mu_1}_{\;\:{\bar \mu}_1}\ldots g^{\mu_n}_{\;\:{\bar \mu}_n}\Bigr)
X^{{\bar \mu}_1\ldots{\bar \mu}_n}({\bar x}) + 
g^{\mu_1}_{\;\:{\bar \mu}_1}\ldots g^{\mu_n}_{\;\:{\bar \mu}_n}
{\cal L}_{\bar \xi}X^{{\bar \mu}_1\ldots{\bar \mu}_n}({\bar x})
\end{equation}
and use the law (5.17). It is then seen that Eq. (5.25) 
generalizes as follows:
\begin{eqnarray}
\LXI
\int\limits_{\mbox{\scriptsize past of}\;x} d{\bar x}{\bar g}^{1/2} 
\delta\Bigl(\sigma(x,{\bar x}) - q\Bigr) 
g^{\mu_1}_{\;\:{\bar \mu}_1}\ldots g^{\mu_n}_{\;\:{\bar \mu}_n}
{\bar X}^{{\bar \mu}_1\ldots{\bar \mu}_n}
=              \hspace{6cm} \\
= \int\limits_{\mbox{\scriptsize past of}\;x} d{\bar x}{\bar g}^{1/2} 
\delta\Bigl(\sigma(x,{\bar x}) - q\Bigr)
g^{\mu_1}_{\;\:{\bar \mu}_1}\ldots g^{\mu_n}_{\;\:{\bar \mu}_n}
{\cal L}_{\bar \xi}{\bar X}^{{\bar \mu}_1\ldots{\bar \mu}_n}
\quad ,
\quad u(x) < u^-\quad .\nonumber
\end{eqnarray}

Eq. (5.27) implies that, in the past Killing domain, the Lie derivative
$\LXI$ commutes with the operator ${\cal H}_q$ in (3.20):
\begin{equation}
\LXI {\cal H}_q X(x) = {\cal H}_q \LXI X(x)\quad , \quad u(x)<u^-\quad .
\end{equation}
It then commutes with all the vertex operators:
\begin{equation}
\LXI F(m,n)X_1X_2(x) = 
F(m,n)\Bigl({\LXI}X_1\Bigr)X_2 + 
F(m,n)X_1\Bigl({\LXI}X_2\Bigr)\quad ,
\end{equation}
\begin{equation}
\LXI \frac{1}{\Box_2}F(m,n)X_1X_2(x) = 
\frac{1}{\Box_2}F(m,n)\Bigl({\cal L}_{\xi}X_1\Bigr)X_2 + 
\frac{1}{\Box_2}F(m,n)X_1\Bigl(\LXI X_2\Bigr)\quad ,
\end{equation}
\begin{equation}
\LXI \frac{1}{\Box_1\Box_2}F(m,n)X_1X_2(x) = 
\frac{1}{\Box_1\Box_2}F(m,n)\Bigl(\LXI X_1\Bigr)X_2 + 
\frac{1}{\Box_1\Box_2}F(m,n)X_1\Bigl(\LXI X_2\Bigr)\quad ,
\end{equation}
$$ 
\quad\quad\quad \qquad \qquad \qquad \qquad u(x)<u^- 
$$
and, since (5.28) is valid for $q=0$ as well, it commutes with
the retarded operator $1/\Box$:
\begin{equation}
\LXI\frac{1}{\Box}X(x) = \frac{1}{\Box}\LXI X(x)\quad , \quad
u(x)<u^-\quad .
\end{equation}

\subsection*{\bf Causality of the vacuum radiation.}

$$ $$

For any of the physical sources $J$ the commutation relations
above yield
\begin{eqnarray}
\LXI\frac{1}{\Box}J(x) &=&
\frac{1}{\Box}\J(x) = 0\quad ,\\
\LXI{\cal H}_qJ(x) &=& {\cal H}_q\J(x) = 0\quad ,\\
\LXI F(m,n)J_1J_2(x) &=& 
F(m,n)\J_1J_2(x) + F(m,n)J_1\J_2(x) = 0\;\;,\;\;\\ &&
\quad\qquad\qquad\qquad\qquad u(x)<u^- \nonumber
\end{eqnarray}
and similarly with the other vertex operators. Here use is made
of Eq. (5.2) and of the retardation of all kernels.
The relations above and the commutativity of $\LXI$ with the covariant
derivative suffice to infer that the scalar $I(x)$ in Eq. (2.37)
possesses the property
\begin{equation}
\xi^{\alpha}\nabla_{\alpha}I(x) = 0\quad , \quad u(x)<u^-\quad .
\end{equation}

Consider now the moment $D_{\1}$ of the scalar $I$, and calculate
\begin{eqnarray}
\frac{\partial}{\partial u}D_{\1}(u,\phi|I) &=&
-\frac{1}{4\pi}\int d{\bar x} {\bar g}^{1/2}\delta'
\Bigl(U_{\phi}({\bar x}) - u\Bigr){\bar I}  \\
&=& \frac{1}{4\pi}\int d{\bar x}{\bar g}^{1/2}
\delta\Bigl(U_{\phi}({\bar x}) - u\Bigr){\bar \nabla}_{\alpha}
\Bigl(\frac{{\bar \xi}^{\alpha}}{({\bar \xi}\cdot{\bar \nabla}U_{\phi})}
{\bar I}\Bigr)\quad .\nonumber
\end{eqnarray}
On the integration hyperplane in (5.37) we have 
$u=U_{\phi}({\bar x})\ge u({\bar x})$ by (4.13).
Therefore, with $u<u^-$ we have $u({\bar x})<u^-$ for the whole of 
the hyperplane. Furthermore, the null geodesic connecting the point
${\bar x}$ of the hyperplane with the point $u,\phi$ at ${\cal I}^+$
belongs to this hyperplane itself. Therefore, we may use Eq. (5.20)
and the Killing equation to obtain
\begin{equation}
\frac{\partial}{\partial u}D_{\1}(u,\phi|I)\biggl|_{u<u^-} =
\frac{1}{4\pi}\int d{\bar x} {\bar g}^{1/2}\delta
\Bigl(U_{\phi}({\bar x}) - u\Bigr)
{\bar \xi}^{\alpha}{\bar \nabla}_{\alpha}{\bar I} \quad .
\end{equation}
Hence, by (5.36),
\begin{equation}
\frac{\partial}{\partial u}D_{\1}(u,\phi|I)\biggl|_{u<u^-} = 0\quad .
\end{equation}

Recalling that the flux of the vacuum energy is expressed through
the moment in (5.39) by Eq. (4.20), we arrive at the following final result:
\begin{equation}
\frac{dM(u)}{du}\biggl|_{u<u^-} = 0\quad .
\end{equation}
The classical radiation is also governed by the time derivatives of the
moments $D_{\1}$ but these are the moments of directly the physical
sources $J$. Also for these moments we have
\begin{equation}
\frac{\partial}{\partial u}D_{\1}(u,\phi|J)\biggl|_{u<u^-} = 0
\end{equation}
by (5.2).

We proved the following assertions. Stationary sources radiate neither 
classically nor quantum-mechanically. Moreover, they don't produce even
quantum noise. Radiation, including the uncertain oscillations of the energy
flux, starts not earlier than the first light signal from a nonstationary 
source reaches the observer at infinity.

The proof of these assertions given above is adjusted to the approximations made 
in the present paper but the assertions themselves are valid beyond these
approximations. By using the commutation law (5.3) and the boundary condition
of retardation, it is not difficult to show that, in the past Killing domain,
the operator $\LXI$ commutes with the exact retarded resolvent. Then it
commutes with all retarded form factors of the form
\begin{equation}
\Gamma(\Box_1,\ldots \Box_n) = 
\int \frac{dm_1^{\;2}\ldots dm^{\;2}_n \rho (m_1^{\;2},\ldots m_n^{\;2})}{
(\Box_1 - m_1^{\;2})\ldots(\Box_n - m_n^{\;2})}\quad .
\end{equation}
In other words, if the source and background fields in the equation 
$(\Box - m^2)\varphi = J$ are static in the past, then so is the retarded
solution $\varphi$ and so are all functions
\begin{equation}
\Gamma(\Box_1,\ldots \Box_n)J_1\ldots J_n \quad . 
\end{equation}
Furthermore, as remarked in Sec.2, expression (2.48) is exact, and $I$ in
this expression is generally a sum of terms (5.43).
Owing to the presence of the overall time derivatives in (2.48), Eq. (5.40)
is an exact fact.

\subsection*{\bf Convergence of the vertex operators.}

$$ $$

We may now come back to the question of convergence of the vertex operators
in (3.23) and (3.32)-(3.33). The vertex operators act directly on the physical
sources $J$ and are expressed through the operator ${\cal H}_q$ in (3.20).
By rewriting the integration measure in (3.20) in the form
\begin{equation}
d{\bar x}{\bar g}^{1/2} = \frac{d\sigma d{\bar \Sigma}}{
\sqrt{-({\bar \nabla}\sigma)^2}}
\end{equation}
where $d{\bar \Sigma}$ is the induced volume element on the hyperboloid
$\sigma(x,{\bar x}) = q$ , and using that on this hyperboloid
$({\bar \nabla}\sigma)^2 = 2\sigma = 2q$ , we obtain
\begin{equation}
{\cal H}_qJ(x) = \frac{1}{4\pi\sqrt{-2q}} 
\int\limits_{\mbox{\scriptsize past sheet of}\;\sigma(x,{\bar x})=q}
d{\bar \Sigma}{\bar J}\quad \quad , \quad q<0\quad.
\end{equation}
It is important that , with $x$ fixed and $q\to -\infty,$ the entire past
hyperboloid $\sigma(x,{\bar x})=q$ shifts to the past and finds itself
in the Killing domain where the sources $J$ are stationary. Since,
in addition, the support of $J$ is confined to a spacetime tube whose 
intersection with the hyperboloid $\sigma(x,{\bar x})=q$ remains compact
as $q\to - \infty$ , the integral in (5.45) tends to a finite limit.
Therefore, with  $x$ fixed and $q\to -\infty$ we have
\begin{equation}
{\cal H}_qJ(x)\biggl|_{q\to-\infty} \propto \frac{1}{\sqrt{-q}}\quad .
\end{equation}

It follows from (5.46) that, with the sources asymptotically static in the past,
the vertex functions (3.23) converge for all $m$ and $n$, and so do the
functions (3.32) but the functions (3.33) diverge logarithmically
for all $m$ and $n$. This is a consequence of the singularity at the zero-mass
 limit of the operator $1/(\Box - m^2)^2$ applied to static sources:
 the integral (3.26) with such sources diverges at $m^2=0.$

 The remedy is in the fact that the chain of retarded kernels connecting
 $T^{\mu\nu}_{\mbox{\scriptsize vac}}\Bigl|_{{\cal I}^+}$
with the sources $J$ contains time derivatives. This is, in
particular, the derivative $\partial/\partial u$ in Eq. (4.18).
Up to higher-order terms in $\Re$ it can be commuted with all kernels
and considered as acting directly on one of the $J$ 's in the guise of the
Lie derivative $\LXI$. This is seen from Eqs. (5.38),(5.31), etc. which,
up to higher-order terms in $\Re$, hold everywhere by the definition
of the vector field $\xi^{\alpha}$. When dealing with the vertex function
(3.33), this commutation should always be assumed done.

If at least one of the two sources in the vertex function has the support
properties of $\J$, all vertex functions converge including the function
(3.33).
Indeed, the support of $\J$ is compact and, therefore, no matter
where the point $x$ is located \footnote{
If the point $x$ is located at $u(x)<u^-$, all past hyperboloids of $x$
are outside the support of $\J$, Eq. (5.34).}
, at the limit $q\to -\infty$ the past hyperboloid of $x$ will go out
of this support (see Fig.5).
Let $\q(x)$ be the parameter $q$ of the earliest hyperboloid of $x$
that still crosses the support of $\J$.
Then Eq. (5.46) gets replaced with
\begin{equation}
 {\cal H}_q \J(x)\biggl|_{q<{\mbox{\scriptsize q}}(x)} = 0\;\;,
\end{equation}
 and all integrals (3.23),(3.32),(3.33) with one of the sources
 $X=\J$ acquire a cut off at the lower limit irrespectively
 of the nature of the other source.

That the vertex function (3.33) with one of the sources $X=\J$ converges is seen
 also from the fact that any power of the massless retarded operator $1/\Box$
 applied to $\J,(1/\Box^n)\J,$ is well defined since, by (5.33), all
 functions $(1/\Box^{n-1})\J$ vanish identically at ${\cal I}^-$. The
 integrand in (3.26) is then no more singular at $m^2=0$ than in the
  convergent integral (3.25).

  The presence of $\J$ among the sources ensures the convergence of an
   arbitrary superposition of retarded kernels since any convergent integral
   of the form
   \begin{equation}
   Y(x) = \int dx_1g_1^{1/2}\ldots dx_ng_n^{1/2}G(x|x_1,\ldots x_n)
   \J_1X_2\ldots X_n
   \end{equation}
   where $G$ is a retarded kernel possesses the property
   \begin{equation}
   Y(x) = 0 \;\;,\;\; u(x)<u^-\;\;.
   \end{equation}
   Owing to (5.49), the function $Y(x)$ vanishes identically not
   only at the past timelike and null infinities ($i^-$ and ${\cal I}^-$)
    but also at spatial infinity ($i^0$):
    \begin{equation}
    Y(x)\biggl|_{x\to i^-\;\mbox{\scriptsize or}\; {\cal I}^-\; \mbox{\scriptsize or}\; i^0}
    \equiv 0 \;\;.
    \end{equation}
    This is illustrated in Fig.5 whence it is seen that, at any of these
    limits, the past light cone of $x$ will go out of the support of $\J.$
    Owing to the property (5.50), any integral of $Y(x)$ with a
    retarded kernel converges, and any convergent integral of the form
    \begin{equation}
    Z(x) =
    \int dx_1g_1^{1/2}\ldots dx_ng_n^{1/2}G(x|x_1,\ldots x_n)
   Y_1X_2\ldots X_n
   \end{equation}
   possesses again the property (5.49):
   \begin{equation}
   Z(x) = 0 \;\;,\;\; u(x)<u^-\;\;.
   \end{equation}

   It is worth emphasizing that the support of a function
   like the $Y(x)$ or $Z(x)$  above is no more compact and,
   moreover, is not even a spacetime tube since it has
   a null boundary. This support (shown with broken lines in Fig.5)
   is the causal future of the support of $\J.$ Therefore, the
   presence of time derivatives in the kernels has generally
   no effect on their behaviours at the future infinities
    beginning with the  future null infinity for $u>u^-.$
    The integrals (4.51),(4.52), etc. are, however, cut from
    below owing to (5.41).
    }

\newpage

{\renewcommand{\theequation}{6.\arabic{equation}}  

\begin{center}
\section{\bf    The late-time behaviours}
\end{center}

$$ $$

Eq. (5.40) proves that the total energy of vacuum radiation
is determined indeed by the limit of late time in (2.49).
From (4.20) and (5.39) we have
\begin{equation}
M(-\infty) - M(\infty) = - \lim_{u\to\infty}
\frac{2}{(4\pi)^2}\,\frac{\partial^2}{\partial u^2}\int
d^2{\cal S}(\phi)\, D_{\1}(u,\phi|I)
\end{equation}
whence it follows that we are to study the behaviour of the moment
$D_{\1}$ as $u\to\infty,$ i.e. the behaviour of the retarded Green
function (4.17) in the future of ${\cal I}^+$. For (6.1) to be
finite and nonvanishing, this behaviour should be
\begin{equation}
D_{\1}(u,\phi|I)\biggl|_{u\to\infty} \propto u^2\quad .
\end{equation}

\subsection*{\bf The retarded Green function in the future of 
${\cal I}^+$ .}

We shall show that the behaviour of 
\begin{equation}
D_{\1}(u,\phi|X) \equiv D_{\1}(u,\phi)
\end{equation}
as $u\to\infty$ is determined by competing behaviours of $X(x)$
at the following four limits.

i) The limit of $X(x)$ as $u(x)\to\infty$ along the timelike
lines filling a spacetime tube $({\cal T}).$ For such lines one can take
the lines $r=\mbox{const.}, \phi = \mbox{const.}$ of an arbitrarily
chosen Bondi-Sachs frame. This limit will be denoted
\begin{equation}
X_{{\cal T}^+}[r,\phi,u\to\infty]\quad .
\end{equation}

ii) The  limit of $X(x)$ as $x$ moves to the future along the timelike
geodesics that reach the asymptotically flat infinity. In the Bondi-Sachs 
coordinates these geodesics are asymptotically of the form
\begin{equation}
r=\frac{\gamma}{\sqrt{1-\gamma^2}}s\;\; , 
\;\; u=\frac{\sqrt{1-\gamma}}{\sqrt{1+\gamma}}s\;\; , 
\;\; \phi=\mbox{const.}\;\; ,\;\;s\to\infty\;\;,\;\;0<\gamma<1
\end{equation}
where $s$ is the proper time, and $\gamma$ is the boost parameter (1.22).
This limit has been denoted in Introduction
\begin{equation}
X_{i^+}[\gamma,\phi,s\to\infty]\quad.
\end{equation}

iii) The limit of $X(x)$ as $x$ moves to the future along the null geodesics
\begin{equation}
u=\mbox{const.}\;\;,\;\;\phi=\mbox{const.}\;\;,\;\;r\to\infty\;\;.
\end{equation}
This limit has been denoted
\begin{equation}
X_{{\cal I}^+}[u,\phi,r\to\infty]\;.
\end{equation}

iiii)
The limit of $X(x)$ as $x$ moves to the asymptotically flat infinity
along the spacelike geodesics.
In the Bondi-Sachs coordinates these geodesics are asymptotically
of the form
\begin{equation}
u = - (1-\beta)r\;\;, \;\; \phi = \mbox{const.}\;\;, \;\; r \to \infty \;\;,
\;\; -1<\beta<1 \;\;.
\end{equation}
This limit will be denoted 
\begin{equation}
X_{i^0}[\beta,\phi,r\to\infty]\;\;.
\end{equation}

Consider the integration hyperplane in (4.19). One of its null
generators is radial i.e. crosses the central geodesic of the Bondi-Sachs
frame at some point $o$. As follows from the consideration in Sec.4
(and as illustrated by Fig.4), the integration hyperplane lies outside
both sheets of the light cone of $o$. Therefore, the support of
${\bar X}$ in (4.19) is confined to the exterior of this cone.
In Fig.6 the light cone of $o$ is depicted with bold lines, and its
exterior is divided into four subdomains. The support of ${\bar X}$ in
(4.19) is their union
\begin{equation}
\mbox{supp}\;{\bar X} = \I\cup \2 \cup \3
\cup \4\;\;,
\end{equation}
and the integral (4.19) itself is a sum of the respective
four contributions
\begin{equation}
D_{\1} = D_{\1}^{\;\I} + D_{\1}^{\;\2} + D_{\1}^{\;\3} + D_{\1}^{\;\4}\;\;.
\end{equation}

Subdomain $\I$ belongs to a tube $r<r_0$ with $r_0$ sufficiently
large for subdomains $\2, \3, \4$ to be already in the 
asymptotically flat zone. Subdomain $\3$ is bounded by two future
light cones ${\bar u}=u_1$ and ${\bar u}=u_2$ with large positive
$u_2=|u_2|$ and large negative $u_1=-|u_1|.$ With the Bondi-Sachs 
parametrization of the integrand in (4.19) we have
\begin{eqnarray}
D_{\1}^{\;\I}(u,\phi) &=&
\frac{1}{4\pi}\int d^2{\cal S}({\bar \phi})
\int\limits_0^{r_0}d{\bar r}{\bar r}^2
\int\limits_{-\infty}^{\infty}d{\bar u}\Bigl|({\bar \nabla}{\bar u},
{\bar \nabla}{\bar r})\Bigr|^{-1}
\delta\Bigl(U_{\phi}({\bar u},{\bar \phi},{\bar r}) - u\Bigr){\bar X}\;\;,
\\
D_{\1}^{\;\2}(u,\phi) &=&
\frac{1}{4\pi}\int d^2{\cal S}({\bar \phi})
\int\limits^{\infty}_{r_0}d{\bar r}{\bar r}^2
\int\limits_{u_2}^{\infty}d{\bar u}\Bigl|({\bar \nabla}{\bar u},
{\bar \nabla}{\bar r})\Bigr|^{-1}
\delta\Bigl(U_{\phi}({\bar u},{\bar \phi},{\bar r}) - u\Bigr){\bar X}\;\;,
\\
D_{\1}^{\;\3}(u,\phi) &=&
\frac{1}{4\pi}\int d^2{\cal S}({\bar \phi})
\int\limits^{\infty}_{r_0}d{\bar r}{\bar r}^2
\int\limits_{u_1}^{u_2}d{\bar u}\Bigl|({\bar \nabla}{\bar u},
{\bar \nabla}{\bar r})\Bigr|^{-1}
\delta\Bigl(U_{\phi}({\bar u},{\bar \phi},{\bar r}) - u\Bigr){\bar X}\;\;,
\\
D_{\1}^{\;\4}(u,\phi) &=&
\frac{1}{4\pi}\int d^2{\cal S}({\bar \phi})
\int\limits^{\infty}_{r_0}d{\bar r}{\bar r}^2
\int\limits^{u_1}_{-\infty}d{\bar u}\Bigl|({\bar \nabla}{\bar u},
{\bar \nabla}{\bar r})\Bigr|^{-1}
\delta\Bigl(U_{\phi}({\bar u},{\bar \phi},{\bar r}) - u\Bigr){\bar X}\;\;.
\end{eqnarray}

The argument $u$ of the function $D_{\1}(u,\phi)$ labels the future
light cone of $o$. As $u\to\infty,$ the point $o$ moves along the 
central geodesic to the future. In addition, the parameters $r_0,u_1,u_2$
should be made functions of $u$ such that , as $u\to\infty,$
\begin{eqnarray}
r_0(u)\to\infty\;\;&,&\;\;u_1(u)\to - \infty\;\; ,\;\;
u_2(u)\to +\infty\;\;, \\
\frac{r_0(u)}{u}\to 0\;\;&,&\;\;\;\;\;\frac{u_1(u)}{u}\to 0 \;\;,
\;\; \frac{u_2(u)}{u}\to 0 \;\;.\nonumber
\end{eqnarray}
It is then seen from Fig.6 that, as $u\to\infty,$ subdomain $\I$
shifts to ${\cal T}^+$, subdomain $\2$ 
shifts to $i^+$, subdomain $\3$ shifts to ${\cal I}^+$, and subdomain
$\4$ shifts to $i^0$. Let us show this by a direct calculation.

Consider first the contribution of the tube, Eq. (6.13). At a however
late $u$ the hyperplane $U_{\phi}({\bar u},{\bar \phi},{\bar r}) = u$
will cross the nonstationary region between the cones ${\bar u}=u^-$
and ${\bar u}=u^+$, and, therefore, will not belong entirely to the
future Killing domain. Nevertheless, at a  sufficiently late $u$
the intersection of the hyperplane with the tube {\it will} be
at ${\bar u}>u^+.$
The null generators of the hyperplane emanating from this intersection to the
future will then also be at ${\bar u}>u^+.$ Therefore, in the integral (6.13)
with a sufficiently late $u$ one may use Eq. (5.22) owing to which
the integrand becomes restricted to
\begin{equation}
{\bar u} = u - L_{\phi}({\bar \phi},{\bar r})\;\;,\;\;u\to\infty\;\;.
\end{equation}
Since, in addition, the range of the integration variables
${\bar \phi},{\bar r}$ in (6.13) is compact, ${\bar X}$ turns out
to be at ${\cal T}^+$. We obtain
\begin{equation}
D_{\1}^{\;\I}(u,\phi)\biggl|_{u\to\infty} =
\frac{1}{4\pi}\int d^2{\cal S}({\bar \phi})
\int\limits_0^{r_0}d{\bar r}{\bar r}^2
\Bigl|({\bar \nabla}{\bar u},
{\bar \nabla}{\bar r})\Bigr|^{-1}
{\bar X}_{{\cal T}^+}[{\bar r},{\bar \phi},{\bar u}=u\to\infty]\quad ,
\end{equation}
and the geometrical factor in the measure is bounded by
virtue of the original assumption that the metric has no horizons.

In the contributions of the domains $\2,\3,\4$ one can use the 
flat-spacetime expressions for  the metric and the function
$U_{\phi}({\bar u},{\bar \phi},{\bar r}),$ Eqs. (1.15) and
(4.15). In the integral (6.14) go over to the integration variable $\gamma$
 according to the formula 
 \begin{equation}
 {\bar r} = \frac{\gamma}{1-\gamma}{\bar u}
 \end{equation}
i.e. parametrize this integral with the family of geodesics (6.5).
It follows that, when $u\to\infty$ , $\gamma$ ranges in the interval
$0<\gamma<1$ , and, with $\gamma$ fixed, ${\bar X}$ turns out to be
at ${i^+}$. In the integral (6.15), ${\bar u}$ ranges in a bounded interval, 
and, with ${\bar u}$ fixed, ${\bar X}$ turns out to be at ${\cal I}^+$.
In the integral (6.16) go over to the integration variable $\beta$
according to the formula
\begin{equation}
{\bar r} = -\frac{1}{1-\beta}{\bar u}
\end{equation}
i.e. parametrize this integral with the family of geodesics (6.9).
Then $\beta$ ranges in a bounded interval, and ${\bar X}$
turns out to be at $i^0$. In this way we obtain
\begin{eqnarray}
D_{\1}^{\;\2}(u,\phi)\biggl|_{u\to\infty} = \hspace{11cm}\\
=\frac{u^3}{4\pi}\int d^2{\cal S}({\bar \phi})
\int\limits_{r_0/u}^{1 - (1-\cos\omega)|u_2|/u}
d\gamma\;\frac{\gamma^2}{(1-\gamma\cos\omega)^4}
{\bar X}_{i^+}[\gamma,{\bar \phi},{\bar s}=\frac{\sqrt{1-\gamma^2}}{
1-\gamma\cos\omega}u\to\infty]\;\;,  \nonumber        \\
D_{\1}^{\;\3}(u,\phi)\biggl|_{u\to\infty} = \hspace{11cm}\\
=\frac{1}{4\pi}\int \frac{d^2{\cal S}({\bar \phi})}{1-\cos\omega}
\int\limits_{u_1}^{u_2} d{\bar u}\Bigl({\bar r}^2{\bar X}\Bigr)_{
{\cal I}^+}[{\bar u},{\bar \phi},{\bar r} = \frac{u}{1-\cos\omega}\to\infty]
\;\;,\nonumber\\
D_{\1}^{\;\4}(u,\phi)\biggl|_{u\to\infty} = \hspace{11cm}\\ 
=\frac{u}{4\pi}\int d^2{\cal S}({\bar \phi})
\int\limits_{\cos\omega}^{1-(1-\cos\omega)|u_1|/u}
d\beta\frac{1}{(\beta-\cos\omega)^2}
\Bigl({\bar r}^2{\bar X}\Bigr)_{i^0}[\beta,{\bar \phi},{\bar r}=
\frac{u}{\beta - \cos\omega} \to \infty]\;\;.            \nonumber
\end{eqnarray}
Here $\omega= \omega(\phi,{\bar \phi})$ is the function of the angles defined
in (4.16).

By (6.17), the integration limits in (6.22) reach as $u\to\infty$
the end points of the interval $0<\gamma<1.$ As explained in Sec.1,
the end point $\gamma=1$ supports the contribution of the future
of ${\cal I}^+$ and so  does the end point $u_2(u)\to\infty$ in (6.23).
The behaviours of $X$ at these end points are related by Eq. (1.29).
Similarly, the end point $\gamma=0$ supports the contribution of the
boundary of the expanding tube ${\cal T}^+$ and so does the end point
$r_0(u)\to\infty$ in (6.19). Finally, the upper limit in (6.24)
reaches as $u\to\infty$ the end point $\beta = 1$ which supports
the contribution of the past of ${\cal I}^+$.
Another such contribution comes from the end point $u_1(u)\to-\infty$
in (6.23). The end-point contributions are of measure zero only
if the  respective integrals converge at these end points;
otherwise, they may be essential.
(The end point $\beta=-1$ which corresponds to the future of ${\cal I}^-$
makes no contribution.)

For simplicity, we may confine ourselves to scalar sources $X$ which
are, moreover, stationary in the past:
\begin{equation}
\frac{\partial}{\partial u}X(u,\phi,r) = 0\;\;, \;\; u<u^-
\end{equation}
since the source $I$ in (6.1) possesses these properties.
The behaviour of $X$ at ${\cal I}^+$ is then restricted by the convergence
condition (4.25). A restriction on the behaviour of $X$ at $i^0$ can be
 obtained by considering the retarded Green function with a static source
 (the Coulomb potential).
 The convergence condition is in this case
 \begin{equation}
 X\biggl|_{i^0} = O\Bigl(\frac{1}{r^4}\Bigr)\;\;.
 \end{equation}
Therefore, we may put
 \begin{equation}
 X_{i^0}[\beta,\phi,r\to\infty]  = 
\frac{A(\phi)}{r^n}\;\;,\;\;\; n\ge 4 
\end{equation}
 \begin{equation}
 X_{{\cal I}^+}[u,\phi,r\to\infty]  = \left\{
A(\phi)/r^n\;\;\;\;,\;\;\; u< u^- \atop
B(u,\phi)/r^3\;\;,\;\;\; u> u^-\right.
\end{equation}
with some coefficients $A$ and $B$. For $X$ at ${\cal T}^+$ and $i^+$
it suffices to consider the power behaviours
\begin{eqnarray}
X_{{\cal T}^+}[r,\phi,u\to\infty] &=& C(r,\phi)u^k\;\;,\\
X_{i^+}[\gamma,\phi,s\to\infty] &=& W(\gamma,\phi)s^p
\end{eqnarray}
with arbitrary $k$ and $p$. However, to agree with (6.28) and (6.29),
 the coefficient of the latter behaviour should have appropriate
 singularities at $\gamma = 1$ and $\gamma = 0$:
 \begin{equation}
 W(\gamma,\phi) = (1-\gamma)^{3 + p/2 }\gamma^{p-k}w(\gamma,\phi)
 \end{equation}
 where $w(\gamma,\phi)$ is a regular function. By the correspondence 
 (1.29) we have also
 \begin{equation}
 B(u,\phi)\biggl|_{u\to\infty} = w(1,\phi)\;2^{p/2}\;u^{p+3}\;\;,
 \end{equation}
 and similarly
 \begin{equation}
 C(r,\phi)\biggl|_{r\to\infty} = w(0,\phi) r^{p-k}\;\;.
 \end{equation}

 Upon the insertion of the behaviours above in (6.19)-(6.24) and the 
 use of (6.17) the following results are obtained.
 The contribution of $i^0$ is
 \begin{equation}
 D_{\1}^{\;\4}(u,\phi)\biggl|_{u\to\infty} = O(u^{-1})\;\;.
 \end{equation}
 The contribution of ${\cal I}^+$ proper, i.e. of a finite range of retarded
 time along ${\cal I}^+$, is also
 \begin{equation}
 D_{\1}^{\;\3}(u,\phi)\biggl|_{u\to\infty} = O(u^{-1})\;\;.
 \end{equation}
 The contribution of $i^+$ proper, i.e. of the open interval
 $0<\gamma<1$ , is
 \begin{equation}
 D_{\1}^{\;\2}(u,\phi)\biggl|_{u\to\infty} = O(u^{p+3})\;\;.
 \end{equation}
Finally, the contribution of the tube proper, i.e. of a tube
with compact spatial sections, is
 \begin{equation}
 D_{\1}^{\;\I}(u,\phi)\biggl|_{u\to\infty} = O(u^{k})\;\;.
 \end{equation}
As far as the end-point contributions are concerned, they become essential
only in the case where the  behaviour (6.36) is comparable with (6.35)
or (6.37). If the power in (6.36) is different from the one in (6.35)
and from the one in (6.37), then the total result is a sum of these competing
powers:
 \begin{equation}
 D_{\1}(u,\phi)\biggl|_{u\to\infty} = O(u^{-1}) + 
 O(u^{p+3}) + O(u^{k})\;\; , \;\; p+3 \ne -1\;\;,\;\; p+3 \ne k\;,
 \end{equation}
 and one is to pick out the dominant one. If, however, $p+3 = -1$ or
 $p + 3 = k$ , then the joint contribution of the coincident powers
 gets amplified by a factor of $\log u$:
 \begin{equation}
D_{\1}(u,\phi)\biggl|_{u\to\infty} = 
\left\{O(u^{-1}\log u) +  O(u^k)\;,\;p+3 = -1\atop
O(u^{k}\log u) +  O(u^{-1})\;,\;p+3 = k\quad\!\right. \;.
\end{equation}

A comparison of the behaviours obtained with the one in (6.2) shows that
relevant contributions may come only from $X$ at $i^+$ or ${\cal T}^+$.
However, at $i^+$, there are three extra powers added to the exponent
of the behaviour of $X$ at late time, Eq. (6.36).
If $X$ doesn't grow at either of the  limits (as is normally the case)
or has one and the same power of growth at $i^+$ and ${\cal T}^+$, then
only $X$ at $i^+$ is capable of making a contribution to the real
vacuum energy. Therefore, the most important case is generally the one
where the contribution of $X$ at $i^+$ is dominant:
\begin{equation}
p>-4\;\;,\;\; p>k-3\;\;.
\end{equation}
In this case the result is
\begin{equation}
D_{\1}(u,\phi|X)\biggl|_{u\to\infty} = 
\frac{u^3}{4\pi}\int d^2{\cal S}({\bar \phi})
\int\limits_0^1 d\gamma
\frac{\gamma^2}{(1-\gamma\cos\omega)^4}
{\bar X}_{i^+}[\gamma,{\bar \phi},{\bar s}=\frac{u\sqrt{1-\gamma^2}}{
1-\gamma\cos\omega}\to\infty]
\end{equation}
with $\cos\omega$ in (4.16). With the specification in (6.31) this integral
converges.

\subsection*{\bf Formula for the energy of vacuum radiation.}

$$ $$

It will be shown below that for $X=I$ the conditions (6.40) are fulfilled
with $p=k$, and the exponent $p$ in (6.30) has the needed value $p=-1$:
\begin{equation}
I_{i^+}[\gamma,\phi,s\to\infty] = \frac{1}{s}\:W(\gamma,\phi)\;\;.
\end{equation}
Then Eq. (6.41) yields
\begin{equation}
D_{\1}(u,\phi|I)\biggl|_{u\to\infty} = 
\frac{u^2}{4\pi}\int d^2{\cal S}({\bar \phi})
\int\limits_0^1 d\gamma
\frac{\gamma^2}{\sqrt{1-\gamma^2}}
\Bigl(1 - \gamma\cos\omega(\phi,{\bar \phi})\Bigr)^{-3}
W(\gamma,{\bar \phi})\;\;.
\end{equation}
According to (6.1), the coefficient of $u^2$ in this expression is the angle
distribution of the vacuum radiation.

The integration over the directions of radiation, i.e. over the angles $\phi$
at ${\cal I}^+\;$ \footnote{
Not to be confused with the angles $\phi$ at ${i}^+$ on which the function
$W(\gamma,\phi)$ in (6.42) depends. A summary table of the integrals
over the 2-sphere used in the paper is given in Appendix B.},
can be done explicitly:
\begin{equation}
\int d^2{\cal S}(\phi)\Bigl(1 - \gamma\cos\omega(\phi,{\bar \phi})\Bigr)^{-3}
= \frac{4\pi}{(1-\gamma^2)^2}
\end{equation}
whence for the total radiation energy (6.1) one obtains the result
\begin{equation}
M(-\infty) - M(\infty) = - \frac{1}{4\pi^2}\int\limits_0^1 
d\gamma \gamma^2 \int d^2{\cal S}(\phi) \frac{W(\gamma,\phi)}{
(1 - \gamma^2)^{5/2}}
\end{equation}
in terms of the coefficient in (6.42). Eq. (6.31) for the case (6.42)
is of the form
\begin{equation}
W(\gamma,\phi) = (1-\gamma)^{5/2}w(\gamma,\phi)
\end{equation}
which ensures the convergence of the integral (6.45).
The coefficient $W(\gamma,\phi)$ in (6.42) will be calculated and shown to be
negative definite.

It will be  recalled that condition (6.46) emerges from the correspondence
between the limits $i^+$ and ${\cal I}^+$ and is a  consequence of the
convergence condition (4.26) at ${\cal I}^+$. However,
this correspondence concerns only the future of ${\cal I}^+$. Therefore,
for the validity of (6.46) and hence for the validity of the result (6.45)
it is only important that the convergence condition at ${\cal I}^+$ hold
at late time. There should be a value $u'$ of retarded time, no mattar how
late, such that
\begin{equation}
\Bigl(I\biggl|_{{\cal I}^+[u,\phi,r\to\infty]}\Bigr)_{u>u'} = 
O\Bigl(\frac{1}{r^3}\Bigr)\;\;.
\end{equation}

\subsection*{\bf The vertices and trees in the future of ${\cal I}^+$ .}

$$ $$

The analysis above of the behaviour of $D_{\1}(u,\phi|X)$ as $u\to\infty$
will now be applied to the case $X=J$ where $J$ is any of the physical
sources (1.8). Since the support of $J$ is confined to a tube, of the four
contributions (6.34)-(6.37) there remains only the contribution of
${\cal T}^+$, Eq. (6.37). Since, at ${\cal T}^+$ , $J$ is stationary, 
we obtain immediately
\begin{equation}
D_{\1}(u,\phi|J)\biggl|_{u\to\infty} = O(u^0)\;\;.
\end{equation}
This result can be made more precise. Let $u'$ be the value of
retarded time such that for $u>u'$ the intersection of the hyperplane
$U_{\phi}({\bar u},{\bar \phi},{\bar r}) = u$ with
the support tube of $J$ is entirely in the future Killing domain
${\bar u}>u^+.$ Then for $u>u'$ one may repeat the calculation in 
(5.37)-(5.38) to obtain by (5.2)
\begin{equation}
\frac{\partial}{\partial u}D_{\1}(u,\phi|J)\biggl|_{u>u'} = 0\;\;.
\end{equation}

Eq. (6.49) determines the behaviours of the vertex functions in the future
of ${\cal I}^+$
 since it provides a cut off for  the time integrals. Thus from (4.29)
 we obtain
 \begin{eqnarray}
 \Bigl(F(m,n)J_1J_2\biggl|_{{\cal I}^+[u,\phi,r\to\infty]}
 \Bigr)_{u\to\infty} = \hspace{7cm}\\
 = 2(-1)^{m+n+1}\frac{u^{m+n}}{r^3}
 \int\limits_{-\infty}^{\infty} d\tau
 \Bigl[\Bigl(\frac{d}{d\tau}\Bigr)^{m+1}D_{\1}(\tau,\phi|J_1)\Bigr]
 \Bigl[\Bigl(\frac{d}{d\tau}\Bigr)^{n+1}D_{\1}(\tau,\phi|J_2)\Bigr]\;.
 \nonumber
 \end{eqnarray}
 Specifically, the behaviour of the vertex function 
 ${\hat V}_{\mbox{\scriptsize scalar}}$ in (2.43) is
 \begin{equation}
\Bigl({\hat V}_{\mbox{\scriptsize scalar}
}\biggl|_{{\cal I}^+[u,\phi,r\to\infty]}\Bigr)_{u\to\infty} = 
- \frac{u^2}{r^3}
 \int\limits_{-\infty}^{\infty} d\tau
 \Bigl[\frac{d^2}{d\tau^2}{\hat D}^{\;Q}_{\1}(\tau,\phi)\Bigr]
 \Bigl[\frac{d^2}{d\tau^2}{\hat D}^{\;Q}_{\1}(\tau,\phi)\Bigr]
 \end{equation}
where the abbreviation (4.36) is used. In the same way, for 
${\hat V}_{\mbox{\scriptsize cross}}$ we obtain from (4.64)
 \begin{equation}
\Bigl(\nabla_{\alpha}\nabla_{\beta}
{\hat V}^{\alpha\beta}_{\mbox{\scriptsize cross}
}\biggl|_{{\cal I}^+[u,\phi,r\to\infty]}\Bigr)_{u\to\infty} = 
- \frac{1}{2}\frac{u^2}{r^3}
 \int\limits_{-\infty}^{\infty} d\tau
 \Bigl[\frac{d^2}{d\tau^2}D^{\;R}_{\1}(\tau,\phi)\Bigr]
 \Bigl[\frac{d^2}{d\tau^2}{\hat D}^{\;Q}_{\1}(\tau,\phi)\Bigr]\;.
 \end{equation}

More important are, however, the behaviours of the vector and tensor
vertices since they are $O(1/r^2)$ at ${\cal I}^+$ .
For ${\hat V}_{\mbox{\scriptsize vect}}$ we have from (4.65)
 \begin{equation}
\Bigl(\nabla_{\alpha}\nabla_{\beta}
{\hat V}^{\alpha\beta}_{\mbox{\scriptsize vect}
}\biggl|_{{\cal I}^+[u,\phi,r\to\infty]}\Bigr)_{u\to\infty} = 
 \frac{1}{6}\frac{u}{r^2}
 \int\limits_{-\infty}^{\infty} d\tau g_{\alpha\beta}
 \Bigl[\frac{d}{d\tau}{\hat D}^{\;\alpha}_{\1}(\tau,\phi)\Bigr]
 \Bigl[\frac{d}{d\tau}{\hat D}^{\;\beta}_{\1}(\tau,\phi)\Bigr]
 + O\Bigl(\frac{1}{r^3}\Bigr)
 \end{equation}
 but the result for 
 ${\hat V}_{\mbox{\scriptsize grav}}$
 is completely different:
 \begin{equation}
\Bigl(\nabla_{\alpha}\nabla_{\beta}\nabla_{\mu}\nabla_{\nu}
{\hat V}^{\alpha\beta\mu\nu}_{\mbox{\scriptsize grav}
}\biggl|_{{\cal I}^+[u,\phi,r\to\infty]}\Bigr)_{u>u'} = 
O\Bigl(\frac{1}{r^3}\Bigr)
\end{equation}
owing to (4.81) and (6.49). The latter result applies also
to the only tree that is $O(1/r^2)$ at ${\cal I}^+,$
Eq. (4.101).

Inspecting Eqs. (2.37)-(2.39) one infers that the behaviours 
obtained here and in Sec.4 can be summarized as the following 
result for the scalar $I$:
\begin{equation}
\Bigl(I\biggl|_{{\cal I}^+[u,\phi,r\to\infty]}\Bigr)_{u>u'} = 
\frac{1}{6r^2}\int\limits_{-\infty}^{u}d\tau (u-\tau)
\tr\Bigl[g_{\alpha\beta}
\Bigl(\frac{d}{d\tau}{\hat D}_{\1}^{\;\alpha}\Bigr)
\Bigl(\frac{d}{d\tau}{\hat D}_{\1}^{\;\beta}\Bigr)\Bigr] + 
O\Bigl(\frac{1}{r^3}\Bigr)
\end{equation}
where the only $1/r^2$ term at late time comes from the vector source.
Since, in (6.55), $\tau\le u$ and
\begin{equation}
\tr\Bigl[g_{\alpha\beta}
\Bigl(\frac{d}{d\tau}{\hat D}_{\1}^{\;\alpha}\Bigr)
\Bigl(\frac{d}{d\tau}{\hat D}_{\1}^{\;\beta}\Bigr)\Bigr] \le 0
\end{equation}
(see Sec.8), the integrand is negative definite. The only possibility
for vanishing of the $1/r^2$ term in (6.55) is, therefore,
\begin{equation}
\tr\Bigl[g_{\alpha\beta}
\Bigl(\frac{d}{d\tau}{\hat D}_{\1}^{\;\alpha}\Bigr)
\Bigl(\frac{d}{d\tau}{\hat D}_{\1}^{\;\beta}\Bigr)\Bigr] \equiv 0\;\;.
\end{equation}
By imposing this limitation on the vector source we force condition
(6.47) to hold.

\subsection*{\bf The trees at $i^+$ and ${\cal T}^+$ .}

$$ $$

The inference from the analysis above is that the real vacuum energy
is determined by the behaviour of the scalar $I$ at $i^+$ and possibly
at ${\cal T}^+$. We begin the study of these behaviours with the trees.

The only nonlocal object that figures in the trees is $(1/\Box)J(x)$
with $1/\Box$ in (3.13) and $J(x)$ in (1.8). The past light cone
of $x$ with $x$ moving to $i^+$ or ${\cal T}^+$ is similar to the null 
hyperplane shifted to the future in a sense that there exists a value
$u'$ of retarded time such that for $u(x)>u'$ the intersection of the past
 light cone of $x$ with the support tube of $J$ is entirely in the 
 future Killing domain. Then also the null geodesics connecting
 the points of this intersection with the vertex $x$ of the cone are entirely
 in the Killing domain.
 Therefore, at $u(x)>u'$ one may use Eqs. (5.32) and (5.2) to obtain
 \begin{equation}
 \LXI\frac{1}{\Box}J(x) = 0\;\;,\;\;u(x)>u'\;\;.
 \end{equation}
 For the behaviour at ${\cal T}^+$ this gives straight away
 \begin{equation}
 \frac{1}{\Box}J\biggl|_{{\cal T}^+[r,\phi,u\to\infty]} = O(u^0)\;\;.
\end{equation}
As far as the behaviour at $i^+$ is concerned, it is the same as the 
behaviour at ${\cal I}^+$ for $u(x)>u'$ since, in the static case,
there is no dependence on the direction at infinity. In this way we obtain
\begin{equation}
\frac{1}{\Box}J\biggl|_{{i}^+[\gamma,\phi,s\to\infty]} = 
O\Bigl(\frac{1}{s}\Bigr)\;\;.
\end{equation}

It follows that at ${\cal T}^+$
all the trees (4.82),(4.84) and (4.96) are $O(u^0).$
At $i^+$, the trees (4.82) vanish identically, and the trees
(4.84) and (4.96) are $O(1/s^2).$ In terms of Eqs. (6.29) and
(6.30) the maximum exponents of the trees are, therefore,
$k=0$ and $p=-2$ whereas for making a nonvanishing contribution
to the real vacuum energy these exponents should be $k\ge2$ , $p\ge-1$ .
Thus the contribution of the  trees vanishes.

Having excluded the trees we may go over to the chief thing:
the behaviours of the vertex functions at $i^+$.
}

\newpage

{\renewcommand{\theequation}{7.\arabic{equation}}  

\begin{center}
\section{\bf    The late-time behaviours (continued)}
\end{center}

\subsection*{\bf Spacelike hyperplanes.}

$$ $$

The behaviour of the world function as one of its points tends to $i^+$
and the other one stays in a compact domain can be obtained
as follows. Since
$\sqrt{-2\sigma(x,{\bar x})}$ is the geodetic
distance between $x$ and ${\bar x}$, in the principal approximation
as $x\to i^+$ there should be
\begin{equation}
\sqrt{-2\sigma(x,{\bar x})}\biggl|_{x\to i^+} =
s(x)\Bigl(1 + {\cal O}\Bigr)\;\; , \;\; {\cal O}\biggl|_{i^+} = 0
\end{equation}
where $s(x) \to \infty$ is the proper time of the point $x$
moving to $i^+$. In a curved spacetime there is one more growing term,
proportional to $\log s(x)$. Therefore, generally we have an
expansion of the form
\begin{eqnarray}
\sqrt{-2\sigma(x,{\bar x})}\biggl|_{x \in i^+[\gamma,\phi,s\to\infty]}
 = c(s) - \frac{T_{\gamma\phi}({\bar x})}{\sqrt{1 - \gamma^2}} +
 O\Bigl(\frac{1}{s}\Bigr)\;\;,\\
 c(s) = s + O(\log s) \hspace{6cm}
 \end{eqnarray}
where $c(s)$ is restricted by an additional condition that it doesn't
depend on ${\bar x}$, and a notation is introduced for the
$O(s^0)$ term of the expansion.
In this term, $T_{\gamma\phi}({\bar x})$ is some function of the point
${\bar x}$ depending also on the coordinates $\gamma,\phi$ of the point
$x$, and the normalization factor
$\sqrt{1 - \gamma^2}$ is introduced for further convenience.
The $T_{\gamma\phi}({\bar x})$ is defined up to an addition of an
arbitrary finite function of $x$ i.e. a function of $\gamma,\phi$.

The insertion of the asymptotic behaviour (7.2) in the equation (3.2)
for $\sigma$ with respect to the point ${\bar x}$ yields
\begin{equation}
\Bigl({\bar \nabla}T({\bar x})\Bigr)^2 = - (1 - \gamma^2)\quad ,
\quad T({\bar x}) \equiv T_{\gamma\phi}({\bar x})\;\;,
\end{equation}
and, since ${\bar \nabla}\sigma(x,{\bar x})$ is past directed, so
is ${\bar \nabla}T({\bar x}).$ The function $T_{\gamma\phi}(x)$ with fixed
$\gamma$ and $\phi$ defines the family of spacelike hypersurfaces
\begin{equation}
T_{\gamma\phi}(x) = \tau = \mbox{const.}
\end{equation}
with time $\tau$ growing towards the
future. The vector field orthogonal
to these hypersurfaces
\begin{equation}
N^{\alpha}(x) = \nabla^{\alpha}\frac{T_{\gamma\phi}(x)}{\sqrt{1 - \gamma^2}}
\quad , \quad N^2(x) = -1
\end{equation}
is a gradient and at the same time has a unit norm. A combination
of these two properties signifies that the integral curves of this
vector field
\begin{equation}
\frac{dx^{\alpha}}{ds} = - N^{\alpha}(x)
\end{equation}
are geodesics. The specification of these geodesics can be read off from 
Eq.(7.2):
\begin{equation}
- N^{\alpha}({\bar x}) = {\bar \nabla}^{\alpha} \sqrt{-2\sigma(x,{\bar x})}
\biggl|_{x\in i^+[\gamma,\phi,s\to\infty]}\;\;.
\end{equation}
At every point ${\bar x}$ the vector field $N^{\alpha}({\bar x})$
is tangent to a timelike geodesic that, when traced to the future,
appears at the asymptotically flat infinity with the energy
$E = (1-\gamma^2)^{-1/2}$ at the point $\phi$ of the celestial 
sphere. In a compact domain the geodesics having one and the same
$\gamma$ and $\phi$ differ by "translations" and make a 3-parameter
congruence. It follows from (7.2) that the geodesic congruence thus
defined is hypersurface-orthogonal, and the orthogonal hypersurfaces
are just (7.5). The hypersurfaces (7.5) will be called hyperplanes,
and there are different families of spacelike hyperplanes for different
values of $\gamma$ and $\phi$.\footnote{
The Bondi and ADM masses refer to the Lorentz frame at infinity in which
the asymptotically flat spacetime rests as a whole, i.e. the center-of-mass
frame. The parameter $\gamma$ refers to the same frame. Therefore, the dependence 
on $\gamma$ cannot be boosted away and is real.}

By using Eq. (6.5), the expansion  (7.2) can be rewritten in terms of 
$r(s)\to \infty:$
\begin{equation}
\sqrt{-2\sigma(x,{\bar x})}\biggl|_{x\in i^+[\gamma,\phi,s\to\infty]}
=
\frac{\sqrt{1 - \gamma^2}}{\gamma}\Bigl( r + O(\log r)\Bigr) 
- \frac{T_{\gamma\phi}({\bar x})}{\sqrt{1 - \gamma^2}} + 
O\Bigl(\frac{1}{r}\Bigr)
\end{equation}
whence, going over to the limit $\gamma \to 1$, we obtain
\begin{equation}
\Bigl(\sigma(x,{\bar x})\biggl|_{x\in i^+[\gamma,\phi,s\to\infty]}
\Bigr)_{\gamma\to 1} 
=
r\Bigl( T_{\gamma\phi}({\bar x})\biggl|_{\gamma=1} + {\cal O}\Bigr)\;\;, 
\;\;{\cal O}\biggl|_{i^+} = 0\;\;.
\end{equation}
By the correspondence (1.29), this sequence of limits should coincide with
the future of ${\cal I}^+$. On the other hand, at ${\cal I}^+$ we have
the expansion (4.5) in which the function $U_{\phi}({\bar x})$ is independent
of the time along ${\cal I}^+$. It follows that
\begin{equation}
T_{\gamma\phi}({\bar x})\biggl|_{\gamma=1} = 
U_{\phi}({\bar x}) + \mbox{const.}
\end{equation}
where the const. is independent of ${\bar x}$. Thus we infer that the 
function $T_{\gamma\phi}(x)$ admits the limit $\gamma\to 1$ and, at 
this limit, the spacelike hyperplanes (7.5) turn into the null hyperplanes
(4.7). The geodesics orthogonal to the spacelike hyperplanes
also become null and turn into the generators of the null hyperplanes.

The insertion of expansion (7.2) in (3.4) yields 
\begin{equation}
{\bar \nabla}_{\bar \mu}\frac{T_{\gamma\phi}({\bar x})}{\sqrt{1 - \gamma^2}}
=
g_{\bar \mu}^{\;\;\mu}({\bar x}, x)\nabla_{\mu}s(x)
\biggl|_{x\to i^+[\gamma,\phi]}
\end{equation}
which is the law of parallel transport of the tangent vector along a 
timelike geodesic orthogonal to the hyperplanes. By using the asymptotic
form (6.5) of the geodesic, one may expand $\nabla s$ over the vector
basis at $i^+:$
\begin{equation}
\nabla_{\mu}s(x) = \frac{1}{\sqrt{1 - \gamma^2}}
\Bigl(\nabla_{\mu}t(x) - \gamma \nabla_{\mu}r(x)\Bigr)\;\;,\;\;
x \to i^+[\gamma,\phi]\;\;.
\end{equation}
Then the law of parallel transport takes the form
\begin{equation}
{\bar \nabla}_{\bar \mu}T_{\gamma\phi}({\bar x}) = 
g_{\bar \mu}^{\;\;\mu}({\bar x}, x)
\Bigl(\nabla_{\mu}t(x) - \gamma \nabla_{\mu}r(x)\Bigr)\biggl|_{
x \to i^+[\gamma,\phi]}
\end{equation}
and at $\gamma = 1$ goes over into (4.11).

In presence of a timelike Killing vector the spacelike hyperplanes possess
the properties similar to (5.20) and (5.21):
\begin{eqnarray}
&&{\bar \xi}^{\mu}{\bar \nabla}_{\mu}T_{\gamma\phi}({\bar x}) = 1\;\;,\\
&&{\cal L}_{\bar \xi}
{\bar \nabla}_{\alpha_1}
\ldots {\bar \nabla}_{\alpha_n}T_{\gamma\phi}({\bar x}) = 0\;\;,
\;\;n\ge 1\;\;.
\end{eqnarray}

Finally, in the case where ${\bar x}$ is at the future asymptotically
flat infinity ${\cal I}^+$ or $i^+$, one can use the flat-spacetime
formula for $T_{\gamma\phi}({\bar x})$:
\begin{eqnarray}
T_{\gamma\phi}({\bar x})\biggl|_{{\bar x}\to {\cal I}^+,i^+} & = &
{\bar u} + {\bar r}\Bigl(1 - \gamma\cos\omega(\phi,{\bar \phi})\Bigr)
\nonumber\\
&=& {\bar t} - \gamma n_i(\phi){\bar \x}^i
\end{eqnarray}
with the same notation as in (4.15).

\subsection*{\bf The radiation moments and conserved charges.}

$$ $$

The function $T_{\gamma\phi}(x)$ may formally be regarded as a two-point
function with one point at $i^+$:
\begin{equation}
T_{\gamma\phi}({\bar x}) \equiv
T(x,{\bar x})\biggl|_{x\to i^+[\gamma,\phi]} \;\;. 
\end{equation}
With the aid of this notation and with the sources in (1.8), define 
the following {\it radiation moments}:
\begin{eqnarray}
{\hat D}(x)\biggl|_{i^+} &=& 
\frac{1}{4\pi}\int d{\bar x}{\bar g}^{1/2}
\delta\Bigl(T(x,{\bar x}) - \tau\Bigr)
{\hat P}({\bar x})\biggl|_{x\to i^+}\;\;,\\
{\hat D}^{\alpha}(x)\biggl|_{i^+} &=& 
\frac{1}{4\pi}\int d{\bar x}{\bar g}^{1/2}
\delta\Bigl(T(x,{\bar x}) - \tau\Bigr)
g^{\alpha}_{\;\;{\bar \alpha}}(x,{\bar x})
{\hat J}^{\bar \alpha}({\bar x})\biggl|_{x\to i^+}\;\;,\\
{D}^{\alpha\beta}(x)\biggl|_{i^+} &=& 
\frac{1}{4\pi}\int d{\bar x}{\bar g}^{1/2}
\delta\Bigl(T(x,{\bar x}) - \tau\Bigr)
g^{\alpha}_{\;\;{\bar \alpha}}(x,{\bar x})
g^{\beta}_{\;\;{\bar \beta}}(x,{\bar x})
{J}^{{\bar \alpha}{\bar \beta}}({\bar x})\biggl|_{x\to i^+}\;\;.
\end{eqnarray}
These objects are respectively a matrix, a matrix vector, and a tensor defined,
however, only for $x$ at $i^+.$ The specification $i^+$ will, therefore,
be omitted and we shall write simply ${\hat D},{\hat D}^{\alpha},
D^{\alpha\beta}.$ The  similarly defined moments for the Ricci scalar $R$
and the matrix ${\hat Q}$ will be denoted $D^{R}$ and ${\hat D}^Q$.
In addition to the parameters $\gamma$ and $\phi$ of the point at $i^+$,
the radiation moments depend on the parameter $\tau$ of the hyperplane.
As follows from (7.11), the previously introduced moments $D_{\1}$ are the
special cases of the moments $D$ corresponding to $\gamma=1$:
\begin{equation}
{\hat D}_{\1} = {\hat D}\biggl|_{\gamma = 1}\;\;,\;\;
{\hat D}_{\1}^{\;\alpha} = {\hat D}^{\alpha}\biggl|_{\gamma = 1}\;\;,\;\;
{D}_{\1}^{\;\alpha\beta} = {D}^{\alpha\beta}\biggl|_{\gamma = 1}\;\;,
\;\;\mbox{etc.}
\end{equation}

Owing to the law of parallel transport (7.14) and the conservation laws
(1.9), the  vector and tensor moments satisfy the relations
\begin{eqnarray}
(\nabla_{\alpha}t - \gamma\nabla_{\alpha}r)\frac{d}{d\tau}
{\hat D}^{\alpha} &=& 0\;\;,\\
(\nabla_{\alpha}t - \gamma\nabla_{\alpha}r)\frac{d}{d\tau}
{D}^{\alpha\beta} &=& 0\;\;
\end{eqnarray}
which generalize (4.42) and (4.48), and are proved similarly. The conserved
quantities in these relations
\begin{equation}
{\hat e} = (\nabla_{\mu}t - \gamma\nabla_{\mu}r){\hat D}^{\mu} = 
\frac{1}{4\pi} \int d{\bar x}{\bar g}^{1/2}\delta
\Bigl(T(x,{\bar x}) - \tau\Bigr){\bar \nabla}_{\bar \mu}T
{\hat J}^{\bar \mu}({\bar x})\biggl|_{x\to i^+}
\end{equation}
and
\begin{equation}
{p^{\nu}} = \frac{1}{2}(\nabla_{\mu}t - \gamma\nabla_{\mu}r){D}^{\mu\nu} = 
\frac{1}{8\pi} \int d{\bar x}{\bar g}^{1/2}\delta
\Bigl(T(x,{\bar x}) - \tau\Bigr)g^{\nu}_{\;\;{\bar \nu}}(x,{\bar x})
{\bar \nabla}_{\bar \mu}T
{J}^{{\bar \mu}{\bar \nu}}({\bar x})\biggl|_{x\to i^+}
\end{equation}
are the full charges (4.44) and (4.50) since the latter are independent
of the choice of the hypersurface $\Sigma = 0$ crossing the support tube 
of $J$.

By using (3.6) and the conservation laws, the vector and tensor moments
can be brought to the forms      
\begin{equation}
{\hat D}^{\alpha} = \frac{d}{d\tau} {\hat {\cal D}}^{\alpha}\;\;,\;\;
{D}^{\alpha\beta} = \frac{d^2}{d\tau^2} {{\cal D}}^{\alpha\beta}
\end{equation}
with
\begin{equation}
{\hat {\cal D}}^{\alpha}(x)\biggl|_{i^+}  = 
- \frac{1}{4\pi} \int d{\bar x} {\bar g}^{1/2}\delta\Bigl(
T(x,{\bar x}) - \tau\Bigr)\sigma^{\alpha}(x,{\bar x})
{\bar \nabla}_{\bar \mu}T {\hat J}^{\bar \mu}({\bar x})\biggl|_{x\to i^+}
\;\;,
\end{equation}
\begin{equation}
{{\cal D}}^{\alpha\beta}(x)\biggl|_{i^+}  = 
 \frac{1}{8\pi} \int d{\bar x} {\bar g}^{1/2}\delta\Bigl(
T(x,{\bar x}) - \tau\Bigr)
\sigma^{\alpha}(x,{\bar x})\sigma^{\beta}(x,{\bar x})
{\bar \nabla}_{\bar \mu}T {\bar \nabla}_{\bar \nu}T 
{J}^{{\bar \mu}{\bar \nu}}({\bar x})\biggl|_{x\to i^+}
\;\;.
\end{equation}
Although we called moments the $D$ 's, this name is more appropriate for the 
${\cal D}$ 's. Their relation to the textbook multipole moments is considered
in Sec.8. The ${\cal D}$ 's always appear differentiated with respect
to $\tau$. Undifferentiated, they are infinite at the limit $i^+$ as 
distinct from the $D$ 's. The consideration below is carried out in terms 
of the $D$ 's in (7.19)-(7.21).

\subsection*{\bf The vertex operators at $i^+$ .}

$$ $$

An essential point concerning the vertex functions is that the hyperboloid
$\sigma(x,{\bar x})=q$ with $q$ ranging from $0$ to $-\infty$ sweeps the whole
causal past of $x$. As $x\to i^+$, the hyperboloid of $x$ with any fixed
$q$ will go out of the support of the nonstationary sources but there will
always be a range of $q$ for which the hyperboloid crosses this support.
For $q$ to stay in this range, it should be shifted to $-\infty$
simultaneously with $x\to i^+$. This suggests the following replacement
of the integration variable in (3.23), (3.32) and (3.33):
\begin{equation}
\sqrt{-2q} = c(s) - \frac{\tau}{\sqrt{1 - \gamma^2}}
\end{equation}
where $\tau$ is the new integration variable,
$c(s)$ is the function in (7.3), and $\gamma,s$ are the parameters of 
$x\in i^+[\gamma,\phi,s\to \infty].$ With $q$ replaced as in (7.30)
and $\tau$ fixed, we obtain from (7.2)
\begin{equation}
\delta\Bigl(\sigma(x,{\bar x}) - q\Bigr)\biggl|_{x\in i^+[\gamma,\phi,
s\to \infty]} = 
\frac{\sqrt{1 - \gamma^2}}{s} \delta\Bigl(T_{\gamma\phi}({\bar x}) - \tau
\Bigr)\;\;.
\end{equation}
Hence
\begin{equation}
{\cal H}_q X(x)\biggl|_{x \in i^+[\gamma,\phi,s \to \infty]} = 
\frac{\sqrt{1 - \gamma^2}}{s}D_{\gamma}(\tau,\phi|X)
\end{equation}
where
\begin{equation}
D_{\gamma}(\tau,\phi|X) = \frac{1}{4\pi} \int d{\bar x}{\bar g}^{1/2}
\delta\Bigl(T_{\gamma\phi}({\bar x}) - \tau\Bigr){\bar X}
\end{equation}
is the full ($\gamma$- dependent) radiation moment of the test source $X$.
The moment in (4.19) is its limiting case
\begin{equation}
D_{\1}(\tau,\phi|X) = D_{\gamma}(\tau,\phi|X)\biggl|_{\gamma=1}\;\;.
\end{equation}

Another essential point concerns the spacelike hyperplane (7.5)
at late time $\tau$. This case is analogous to 
the case of the null hyperplane. As the hyperplane (7.5) shifts to the future,
its intersection with the support tube of $J$ turns out to be in the 
future Killing domain as well as the timelike geodesics emanating from
this intersection towards $i^+$. Eq. (7.15) can then be used to obtain
for the spacelike hyperplane the result analogous to (6.49). Namely,
there exists an instant $\tau'$ of time $\tau$ such that
\begin{equation}
\frac{d}{d\tau}D_{\gamma}(\tau,\phi|J)\biggl|_{\tau > \tau'} = 0\;\;.
\end{equation}

One is now ready for obtaining the vertex functions at $i^+$.
Making the replacement (7.30) in the integral (3.23), and using (7.32)
and (7.35) one arrives at the following result:
\begin{eqnarray}
F(m,n)J_1J_2\biggl|_{i^+[\gamma,\phi,s\to \infty]} = \hspace{9cm}\\
= s^{m+n-3}\Bigl(-\frac{1}{2}\Bigr)^{m+n-1}
\Bigl(\sqrt{1 - \gamma^2}\Bigr)^{m+n+3}
\int\limits_{-\infty}^{\infty}
d\tau 
\Bigl[\Bigl(\frac{d}{d\tau}\Bigr)^{m+1}D_{\gamma}
(\tau,\phi|J_1)\Bigr]
\Bigl[\Bigl(\frac{d}{d\tau}\Bigr)^{n+1}D_{\gamma}
(\tau,\phi|J_2)\Bigr]\;\;.                   \nonumber
\end{eqnarray}
{\it This is the sought for growth in time.}
Comparing this result with Eq. (6.50) one can check the fulfillment
of the correspondence (1.29) between the limits $i^+$ and ${\cal I}^+$.
A similar calculation for the operators (3.32) and (3.33) yields
\begin{eqnarray}
\frac{1}{\Box_2}
F(m,n)J_1J_2\biggl|_{i^+[\gamma,\phi,s\to \infty]} = \hspace{9cm}\\
= - s^{m+n-3}\Bigl(-\frac{1}{2}\Bigr)^{m+n-1}
\Bigl(\sqrt{1 - \gamma^2}\Bigr)^{m+n+1}
\int\limits_{-\infty}^{\infty}
d\tau 
\Bigl[\Bigl(\frac{d}{d\tau}\Bigr)^{m+1}D_{\gamma}
(\tau,\phi|J_1)\Bigr]
\Bigl[\Bigl(\frac{d}{d\tau}\Bigr)^{n-1}D_{\gamma}
(\tau,\phi|J_2)\Bigr]\;\;,                   
\nonumber\\
\frac{1}{\Box_1\Box_2}
F(m,n)J_1J_2\biggl|_{i^+[\gamma,\phi,s\to \infty]} = \hspace{9cm}\\
=  s^{m+n-3}\Bigl(-\frac{1}{2}\Bigr)^{m+n-1}
\Bigl(\sqrt{1 - \gamma^2}\Bigr)^{m+n-1}
\int\limits_{-\infty}^{\infty}
d\tau 
\Bigl[\Bigl(\frac{d}{d\tau}\Bigr)^{m-1}D_{\gamma}
(\tau,\phi|J_1)\Bigr]
\Bigl[\Bigl(\frac{d}{d\tau}\Bigr)^{n-1}D_{\gamma}
(\tau,\phi|J_2)\Bigr]\;\;.                   \nonumber
\end{eqnarray}

Specializing to the vertex functions in (2.40)-(2.43) one obtains
\begin{eqnarray}
{\hat V}_{\mbox{\scriptsize 
scalar}}\biggl|_{i^+[\gamma,\phi,s\to\infty]} &=& 
- \frac{1}{4s} (1 - \gamma^2)^{5/2}
\int\limits_{-\infty}^{\infty} d\tau 
\Bigl(\frac{d^2}{d\tau^2}{\hat D}^{
Q}\Bigr)
\Bigl(\frac{d^2}{d\tau^2}{\hat D}^{
Q}\Bigr)\;\;, \\
{\hat V}^{\alpha\beta}_{\mbox{\scriptsize 
cross}}\biggl|_{i^+[\gamma,\phi,s\to\infty]} &=& 
 \frac{s}{48} (1 - \gamma^2)^{5/2}
\int\limits_{-\infty}^{\infty} d\tau 
\Bigl(\frac{d^2}{d\tau^2}{D}^{
\alpha\beta}\Bigr)
\Bigl(\frac{d^2}{d\tau^2}{\hat D}^{
Q}\Bigr)\;\;, \\
{\hat V}^{\alpha\beta}_{\mbox{\scriptsize 
vect}}\biggl|_{i^+[\gamma,\phi,s\to\infty]} &=& 
 \frac{s}{48} (1 - \gamma^2)^{3/2}
\int\limits_{-\infty}^{\infty} d\tau 
\Bigl(\frac{d}{d\tau}{\hat D}^{
\alpha}\Bigr)
\Bigl(\frac{d}{d\tau}{\hat D}^{
\beta}\Bigr)\;\;, \\
{\hat V}^{\alpha\beta\mu\nu}_{\mbox{\scriptsize 
grav}}\biggl|_{i^+[\gamma,\phi,s\to\infty]} &=& 
- {\hat 1} \frac{s^3}{180\times 32} (1 - \gamma^2)^{5/2}
\int\limits_{-\infty}^{\infty} d\tau 
\Bigl(\frac{d^2}{d\tau^2}{D}^{
\alpha\beta}\Bigr)
\Bigl(\frac{d^2}{d\tau^2}{D}^{
\mu\nu}\Bigr)\;\;. 
\end{eqnarray}
In each case, only the highest exponents $m$ and $n$ of the vertex operators
$F(m,n)$ work.

As seen from the expressions above, the function 
${\hat V}_{\mbox{\scriptsize scalar}}$
has precisely the needed power of growth at $i^+$
(cf. Eq. (6.42)) but the  remaining functions grow apparently
too fast. On the other hand, in Eq. (2.39) these latter functions
appear differentiated. Unlike in the case of ${\cal I}^+$,
here the derivatives alone help as we show below. However, the conservation
laws simplify the result greately.

\subsection*{\bf The non-scalar vertices at $i^+$ .}

$$ $$

In view of (7.13), the conservation laws (7.23) and (7.24) for the
moments can be written in the form
\begin{equation}
\nabla_{\alpha}s \,\frac{d}{d\tau}{\hat D}^{\alpha} = 0 \quad ,\quad
\nabla_{\alpha}s \,\frac{d}{d\tau}{D}^{\alpha\beta} = 0\;\;.
\end{equation}
Therefore, when projected on $\nabla s$, the senior asymptotic terms
of the vertex functions (7.40)-(7.42) vanish. For further use we shall
record this fact in the form
\begin{equation}
\nabla_{\alpha}s \, V^{\alpha\ldots}\biggl|_{i^+} = 
O\Bigl(\frac{1}{s} V^{\alpha\ldots}\Bigr)
\end{equation}
where $V^{\alpha\ldots}$ is any of the vertex functions bearing 
indices \footnote{
When referring to tensors at infinity we always mean their components
in the null-tetrad basis or, equivalently, in the Minkowski frame. The
components of radiation moments in this frame are finite.}.

For tensors $V^{\alpha\beta}$ and $V^{\alpha\beta\mu\nu}$ possessing the 
property (7.44) we are to calculate the behaviours at $i^+$ of the quantities
\begin{equation}
\nabla_{\alpha}\nabla_{\beta}\,V^{\alpha\beta}\;\;\;,\;\;\;
\nabla_{\alpha}\nabla_{\beta}\nabla_{\mu}\nabla_{\nu}
\,V^{\alpha\beta\mu\nu}
\end{equation}
appearing in (2.39). For this purpose it is convenient to go over to new
coordinates of the point at $i^+$:
\begin{equation}
x\biggl|_{i^+} = (s,\gamma,\phi) = (s,\gamma^i)\;\;,
\;\;\gamma^i = \gamma n^i(\phi)
\end{equation}
where $s,\gamma^i$ are the new coordinates, and $n^i(\phi)$ is defined
in (1.27). The coordinates $s,\gamma^i$ can be related also to the Minkowski
coordinates at $i^+$. In the Minkowski frame (1.25) the geodesic (6.5) is
of the form
\begin{equation}
t = \frac{s}{\sqrt{1 - \gamma^2}}\;\;,\;\;
\x^i = \frac{\gamma^i s}{\sqrt{1 - \gamma^2}}\;\;,\;\; s\to \infty
\end{equation}
which gives the transformation law.

The transformation law (7.47) makes it easy to obtain
the metric at $i^+$ in the new frame. We have
\begin{eqnarray}
g^{00}\equiv (\nabla s,\nabla s)= -1 \;\;&,&\;\;
g^{0i}\equiv (\nabla s,\nabla \gamma^i)= 0 \;\;,\\ \;\;
g^{ik}\equiv (\nabla \gamma^i,\nabla \gamma^k)= 
\frac{1}{s^2}{\tilde g}^{ik} \;\;&,&\;\;
{\tilde g}^{ik}{\tilde g}_{kp} = \delta^i_p \nonumber
\end{eqnarray}
where ${\tilde g}_{ik}$ is the 3-dimensional metric depending
only on $\gamma^i$:
\begin{eqnarray}
{\tilde g}_{ik} = \frac{1}{1-\gamma^2}\Bigl(\delta_{ik} +
\frac{\gamma_i\gamma_k}{1 - \gamma^2}\Bigr)\;\;&,&
\;\; \gamma_i = \delta_{ik}\gamma^k \;\; ,   \\
{\tilde g}^{ik} = (1-\gamma^2)(\delta^{ik} - \gamma^i\gamma^k)\;\;&,&
\;\;\sqrt{\tilde g} \equiv \sqrt{\det {\tilde g}_{ik}} = 
\frac{1}{(1 - \gamma^2)^2}\;\;.\nonumber
\end{eqnarray}
Covariant derivatives with respect to the 3-metric ${\tilde g}_{ik}$
will be denoted ${\tilde \nabla}_i$ , and the following relation
will be quoted:
\begin{equation}
{\tilde \nabla}_i{\tilde \nabla}_k \frac{1}{\sqrt{1 - \gamma^2}}
= 
\frac{1}{\sqrt{1 - \gamma^2}}\; {\tilde g}_{ik}\;\;.
\end{equation}
It will also be noted that, in  terms of $\gamma^i$, the integral
over the parameters at $i^+$ that figures in (6.45) is of the form
\begin{equation}
\int\limits_0^1 d\gamma \gamma^2 \int d^2 {\cal S}(\phi)
X\biggl|_{i^+[\gamma,\phi,s\to\infty]} = 
\int d^3 \gamma\; \theta(1 - \gamma) X\biggl|_{i^+[\gamma^i,s\to\infty]}\;\;.
\end{equation}

By a direct calculation in the metric (7.48) one obtains
\begin{equation}
\nabla_{\alpha}\nabla_{\beta}V^{\alpha\beta} = 
{\tilde \nabla}_i{\tilde \nabla}_k V^{ik} + 
\Bigl(s \frac{\partial}{\partial s} + 4\Bigr){\tilde g}_{ik}V^{ik} + 
\Bigl(2\frac{\partial}{\partial s} + \frac{8}{s}\Bigr){\tilde \nabla}_i
V^{0i} + 
\Bigl(\frac{\partial^2}{\partial s^2} + \frac{6}{s}
\frac{\partial}{\partial s} + \frac{6}{s^2}\Bigr)V^{00}
\end{equation}
which is valid identically for any symmetric $V^{\alpha\beta}$.
That every index "$0$" on $V^{\cdots}$ is accompanied here by an
extra power of $1/s$ (or $\partial/\partial s$) is a general rule.
A similar calculation for a symmetrized fourth-rank tensor yields
\begin{eqnarray}
\nabla_{\alpha}\nabla_{\beta}\nabla_{\mu}\nabla_{\nu}
V^{(\alpha\beta\mu\nu)} & = &
{\tilde \nabla}_i{\tilde \nabla}_k{\tilde \nabla}_m{\tilde \nabla}_n 
V^{(ikmn)} + 
\Bigl(6s \frac{\partial}{\partial s} + 32\Bigr)
{\tilde g}_{ik}{\tilde \nabla}_m{\tilde \nabla}_n V^{(ikmn)} 
\hspace{2cm}\nonumber \\ & + &
\Bigl(3s^2\frac{\partial^2}{\partial s^2} + 
33{s}\frac{\partial}{\partial s} + 72\Bigr)
{\tilde g}_{ik}{\tilde g}_{mn}V^{(ikmn)} \nonumber \\&+&
\mbox{terms in}\; \frac{1}{s}V^{0kmn},\;\frac{1}{s^2}V^{00mn},\;
\frac{1}{s^3}V^{000n},\;\frac{1}{s^4}V^{0000}\;\;.
\end{eqnarray}

In Eqs. (7.52) and (7.53) there figure the coordinate components
of $V^{\alpha\ldots}$ i.e. the projections
\begin{equation}
V^{0\ldots} = \nabla_{\alpha} s \, V^{\alpha\ldots}\;\;,\;\;
V^{i\ldots} = \nabla_{\alpha}\gamma^i \, V^{\alpha\ldots}\;\;.
\end{equation}
The virtue of the introduced coordinate frame is in the fact
that the set $\nabla_{\alpha}s$ , $\nabla_{\alpha}\gamma^i$
is not a finite vector basis at $i^+$ as distinct from the
coordinate basis of the Minkowski frame. From the transformation
law (7.47) it follows that 
\begin{equation}
\nabla_{\alpha}s\biggl|_{i^+} = O(1) \;\;,\;\;\nabla_{\alpha}\gamma^i
\biggl|_{i^+} = O\Bigl(\frac{1}{s}\Bigr)\;\;.
\end{equation}
Therefore, the projection of $V^{\alpha\ldots}$ in (7.54)
that has $n$ indices "$i$" has also the extra $1/s^n$ as 
compared to $V^{\alpha\ldots}$ in the Minkowski frame. With
regard for this fact, all terms of an expression like
(7.52) or (7.53) are, generally, of one and the same order at $i^+$.
We have
\begin{equation}
\nabla_{\alpha}\nabla_{\beta}V^{\alpha\beta}\biggl|_{i^+} =
O\Bigl(\frac{1}{s^2}V\Bigr)\;\;,\;\;\nabla_{\alpha}\nabla_{\beta}\nabla_{\mu}
\nabla_{\nu}V^{\alpha\beta\mu\nu}\biggl|_{i^+} = 
O\Bigl(\frac{1}{s^4}V\Bigr)\;\;.
\end{equation}
This is precisely what is needed for the vertex functions (7.40)-(7.42)
since inserted in (7.56) they all become $O(1/s)$. Specifically,
the spatial components of all vertex functions in the frame (7.54) are
\begin{equation}
{\hat V}^{ik}_{\mbox{\scriptsize cross}}\biggl|_{i^+} = 
O\Bigl(\frac{1}{s}\Bigr)\;\;,\;\;
{\hat V}^{ik}_{\mbox{\scriptsize vect}}\biggl|_{i^+} = 
O\Bigl(\frac{1}{s}\Bigr)\;\;,\;\;
{\hat V}^{ikmn}_{\mbox{\scriptsize grav}}\biggl|_{i^+} = 
O\Bigl(\frac{1}{s}\Bigr)\;\;.
\end{equation}

The inference above obtains without using the conservation equation (7.44).
Using this equation one can assert more, namely that all terms in (7.52)
and (7.53) with at least one index "0" on $V^{\ldots}$ can be discarded
since they decrease by one power of $1/s$ faster than the "i" terms.
Using also (7.57), one obtains for both 
${\hat V}_{\mbox{\scriptsize cross}}^{\alpha\beta}$ and
${\hat V}_{\mbox{\scriptsize vect}}^{\alpha\beta}$ the expression
\begin{equation}
\nabla_{\alpha}\nabla_{\beta}V^{\alpha\beta}\biggl|_{i^+} = 
\Bigl({\tilde \nabla}_i{\tilde \nabla}_k + 3{\tilde g}_{ik}\Bigr)V^{ik} +
O\Bigl(\frac{1}{s^2}\Bigr)
\end{equation}
and for 
${\hat V}_{\mbox{\scriptsize grav}}^{\alpha\beta\mu\nu}$ the expression
\begin{equation}
\nabla_{\alpha}\nabla_{\beta}\nabla_{\mu}\nabla_{\nu}
V^{(\alpha\beta\mu\nu)}\biggl|_{i^+} = 
\Bigl({\tilde \nabla}_i{\tilde \nabla}_k{\tilde \nabla}_m{\tilde \nabla}_n + 
26{\tilde g}_{ik}{\tilde \nabla}_m{\tilde \nabla}_n +
45{\tilde g}_{ik}{\tilde g}_{mn}\Bigr)V^{(ikmn)} +
O\Bigl(\frac{1}{s^2}\Bigr).
\end{equation}

These expressions are next subject to the integration over the parameters 
at $i^+$ with the measure in (6.45):
\begin{equation}
\int\limits_0^1 d\gamma \gamma^2 \int d^2{\cal S}(\phi)\frac{1}{(1-\gamma^2)^{5/2}}
(\ldots) = 
\int d^3\gamma \sqrt{\tilde g}\; \theta(1 - \gamma)\;
\frac{1}{\sqrt{1-\gamma^2}}(\ldots)\;\;.
\end{equation}
Therefore, one needs them only up to terms vanishing
in this integral. Relation (7.50) makes it easy to integrate
by parts in (7.60) or, equivalently, to bring
expressions (7.58) and (7.59) to the following form:
\begin{eqnarray}
\nabla_{\alpha}\nabla_{\beta}V^{\alpha\beta}\biggl|_{i^+} &=&
4{\tilde g}_{ik}V^{ik}+ \sqrt{1 - \gamma^2}{\tilde \nabla}_i
Y^i\;\;,\\
\nabla_{\alpha}\nabla_{\beta}\nabla_{\mu}\nabla_{\nu}
V^{(\alpha\beta\mu\nu)}\biggl|_{i^+} &=&
72{\tilde g}_{ik}{\tilde g}_{mn}
V^{(ikmn)}+ \sqrt{1 - \gamma^2}{\tilde \nabla}_i
Z^i\;\;
\end{eqnarray}
with some quantities $Y^i$ and $Z^i$ in the total derivative
terms. It can be checked that these quantities vanish indeed at
$\gamma = 1$ which is the boundary of integration in (7.60).

One may now come back to the covariant form. With the metric in
(7.48) one has
\begin{equation}
s^2{\tilde g}_{ik}V^{ik..} = (g_{\alpha\beta} + \nabla_{\alpha}s
\nabla_{\beta}s)V^{\alpha\beta..}\;\;,
\end{equation}
and the projection $\nabla_{\alpha}sV^{\alpha\ldots}$ vanishes at $i^+$
by virtue of the conservation equation (7.44). Thus we obtain
the following final results:
\begin{eqnarray}
\nabla_{\alpha}\nabla_{\beta}V^{\alpha\beta}\biggl|_{i^+} &=&
\frac{4}{s^2}g_{\alpha\beta}V^{\alpha\beta}
+ \;\;\mbox{total derivatives}
\;\;,\\
\nabla_{\alpha}\nabla_{\beta}\nabla_{\mu}\nabla_{\nu}
V^{(\alpha\beta\mu\nu)}\biggl|_{i^+} &=&
\frac{72}{s^4}{g}_{\alpha\beta}{g}_{\mu\nu}
V^{(\alpha\beta\mu\nu)}+ \;\;\mbox{total derivatives}\;\; .
\end{eqnarray}

Substituting for $V^{\alpha\beta}$ and $V^{\alpha\beta\mu\nu}$
the expressions (7.40)-(7.42) we obtain the behaviours at
$i^+$ of all non-scalar vertices (the total derivatives are
omitted):
\begin{eqnarray}
\nabla_{\alpha}\nabla_{\beta}{\hat V}^{\alpha\beta}_{
\mbox{\scriptsize cross}}\biggl|_{i^+[\gamma,\phi,s\to\infty]} &=&
- \frac{1}{12s}(1 - \gamma^2)^{5/2}\int\limits_{-\infty}^{\infty}
d\tau \Bigl(\frac{d^2}{d\tau^2}D^R\Bigr)
 \Bigl(\frac{d^2}{d\tau^2}{\hat D}^Q\Bigr)\;\;,       \\
\nabla_{\alpha}\nabla_{\beta}{\hat V}^{\alpha\beta}_{
\mbox{\scriptsize vect}}\biggl|_{i^+[\gamma,\phi,s\to\infty]} &=&
 \frac{1}{12s}(1 - \gamma^2)^{3/2}\int\limits_{-\infty}^{\infty}
d\tau g_{\alpha\beta}\Bigl(\frac{d}{d\tau}{\hat D}^{\alpha}\Bigr)
 \Bigl(\frac{d}{d\tau}{\hat D}^{\beta}\Bigr)\;\;,       \\
\nabla_{\alpha}\nabla_{\beta}\nabla_{\mu}\nabla_{\nu}
{\hat V}^{\alpha\beta\mu\nu}_{
\mbox{\scriptsize grav}}\biggl|_{i^+[\gamma,\phi,s\to\infty]} &=&
- \frac{\hat 1}{120s}(1 - \gamma^2)^{5/2}\int\limits_{-\infty}^{\infty}
d\tau (g_{\alpha\mu}g_{\beta\nu} + \frac{1}{2}g_{\alpha\beta}g_{\mu\nu})
\nonumber \\
& &\times\;\, \Bigl(\frac{d^2}{d\tau^2}D^{\alpha\beta}\Bigr)
 \Bigl(\frac{d^2}{d\tau^2}{D}^{\mu\nu}\Bigr)\;\;.
\end{eqnarray}
With the behaviour of
${\hat V}_{\mbox{\scriptsize scalar}}$ in (7.39) included, we
have the complete list. All the behaviours are of the form (6.42)
i.e. proportional to $1/s$, and for all but one the proportionality
coefficient contains the factor $(1 - \gamma^2)^{5/2}$ as
required in Eq. (6.46).
The exception is the vector vertex (7.67) in which instead of 
$(1 - \gamma^2)^{5/2}$  we have
$(1 - \gamma^2)^{3/2}$. This is a consequence of the lack of convergence
at ${\cal I}^+$. That the results are correct can be checked by comparing
the asymptotic expressions above with the behaviours in the future of
${\cal I}^+$, Eqs. (6.51)-(6.54).
The correspondence (1.29) between the limits $i^+$ and ${\cal I}^+$
holds in all cases.

An alternative way of handling the non-scalar vertices is considered
in Appendix B.

\subsection*{\bf The vertices at $\T^+$ .}

$$ $$

Since we assume that all timelike geodesics are infinitely extendable
 to the future, the behaviour (7.1) holds at $\T^+$ as well. The
 parameter $\sqrt{-2q}$ is then to be shifted by $s$ like in Eq. (7.30),
 and, as a result, there will appear integrals over the hypersurfaces
 orthogonal to the geodesics coming to $\T^+$. These integrals analogous to
 the moments will be stationary at late time like in Eq. (7.35) since their
 only ingredients will be the metric and the stationary source. It
 follows that, in terms of the proper time along the world lines
 filling the tube, the behaviours of the vertex operators at $\T^+$
 are the same as at $i^+$:
 \begin{equation}
 F(m,n)J_1J_2\biggl|_{\T^+}\propto s^{m+n-3}\;\;, \;\; s \to \infty\;\;.
 \end{equation}
 Since we assume that there is no future horizon, the proper time of
  an observer in the tube is analytically related to the external
  time $u$ (normalized in (1.14)). Hence
 \begin{equation}
 F(m,n)J_1J_2\biggl|_{\T^+}\propto u^{m+n-3}\;\;, \;\; u \to \infty\;\;.
 \end{equation}

Eq. (7.70) signifies that the vertex terms in the scalar $I$ have one
and the same power of growth at $\T^+$ and $i^+\;$ 
\footnote{
For ${\hat V}_{\mbox{\scriptsize scalar}}$ this follows directly
 from (7.70). For the non-scalar vertices the overall derivatives
 in the expressions like
$\nabla_{\alpha}\nabla_{\beta}{\hat V}^{\alpha\beta}_{
 \mbox{\scriptsize vect}}$ , etc. have at $\T^+$ the same effect as at $i^+$.
 This is clear from the treatment of these derivatives in Appendix B.}
 whereas, for making a contribution to the real vacuum energy,
 the growth at $\T^+$ should be by three powers of $u$ faster (Sec.6).
 Thus we infer that the only contribution to the total vacuum energy
 comes from the vertex functions at $i^+$.
  }

\newpage

{\renewcommand{\theequation}{8.\arabic{equation}}  

\begin{center}
\section{\bf    Creation of particles and radiation of waves}
\end{center}

\subsection*{\bf The energy of the vacuum particle production.}

$$ $$

The result of consideration in the previous section is that formula
(6.45) can indeed be used, and $W(\gamma,\phi)$ in this formula 
is the sum of expressions (7.39),(7.66),(7.67) and (7.68) with the
factors $1/s$ detached:
\begin{eqnarray}
W(\gamma,\phi) & = & -\frac{1}{4}(1 - \gamma^2)^{5/2}
\int\limits_{-\infty}^{\infty} d \tau
\;\tr\Bigl[
\Bigl(\frac{d^2}{d\tau^2}{\hat D}^Q\Bigr)
\Bigl(\frac{d^2}{d\tau^2}{\hat D}^Q\Bigr) + \frac{1}{3}
\Bigl(\frac{d^2}{d\tau^2}{D}^R\Bigr)
\Bigl(\frac{d^2}{d\tau^2}{\hat D}^Q\Bigr) \nonumber\\
&&{}-\frac{1}{3}\frac{1}{(1 - \gamma^2)}g_{\alpha\beta}
\Bigl(\frac{d}{d\tau}{\hat D}^{\alpha}\Bigr)
\Bigl(\frac{d}{d\tau}{\hat D}^{\beta}\Bigr)\nonumber\\ && {}+
\frac{\hat 1}{30}(g_{\alpha\mu}g_{\beta\nu} + \frac{1}{2}g_{\alpha\beta}
g_{\mu\nu})
\Bigl(\frac{d^2}{d\tau^2}{D}^{\alpha\beta}\Bigr)
\Bigl(\frac{d^2}{d\tau^2}{D}^{\mu\nu}\Bigr)\Bigr]\;.
\end{eqnarray}

By (1.12)
\begin{equation}
{\hat D}^Q = {\hat D} - \frac{\hat 1}{6}D^R
\end{equation}
where ${\hat D}$ is the radiation moment (7.19) of the potential matrix 
${\hat P}$. One may notice that the coefficients of the first two terms 
in (8.1) are precisely such
that the cross contribution ${\hat P}\times R$ cancels. 
This cancellation (although
nontrivial) is a consequence of adding $-\frac{1}{6}R{\hat 1}$
to the potential in Eq. (1.2) and is the only manifestation of the
conformal properties of the effective action [31]. Otherwise, these
properties play no role in the present calculation as the reader can see.
\footnote{
There is an unceasing controversy in connection with the old paper [32]
about the significance of the trace anomaly in dimensions higher than
two. The calculation in the present paper may serve as a commentary
to the following conversation that took place between two persons:\\
N: The anomaly is a window through which we can see ...\\
V: ... the back yard.}

Using also that 
\begin{equation}
D^R = - g_{\mu\nu}D^{\mu\nu}
\end{equation}
(cf. Eq. (4.37)) and inserting (8.1) in (6.45) we obtain the following
{\it final result} for the quantity (1.37) in terms of the radiation
moments (7.19)-(7.21):
\begin{eqnarray}
M(-\infty) - M(\infty) & = & \frac{1}{(4\pi)^2}
\int\limits_0^1 d\gamma \gamma^2
\int\limits_{-\infty}^{\infty} d\tau  \int d^2{\cal S}(\phi) 
\;\tr\Bigl[\Bigl(\frac{d^2}{d\tau^2}{\hat D}\Bigr)^2  \hspace{3cm}\nonumber\\
&&\qquad{}-\frac{1}{3}\frac{1}{(1-\gamma^2)}
g_{\alpha\beta}
\Bigl(\frac{d}{d\tau}{\hat D}^{\alpha}\Bigr)
\Bigl(\frac{d}{d\tau}{\hat D}^{\beta}\Bigr) \nonumber\\&&\qquad {}+
\frac{1}{30}{\hat 1}(g_{\mu\alpha}g_{\nu\beta} - 
\frac{1}{3}g_{\mu\nu}g_{\alpha\beta})
\Bigl(\frac{d^2}{d\tau^2}{D}^{\alpha\beta}\Bigr)
\Bigl(\frac{d^2}{d\tau^2}{D}^{\mu\nu}\Bigr)\Bigr]\;\;.
\end{eqnarray}
There is a pole at $\gamma = 1 $ in the term with the vector moment. However, 
under the limitation (6.57):
\begin{equation}
\tr\Bigl[g_{\alpha\beta}
\Bigl(\frac{d}{d\tau}{\hat D}^{\alpha}\Bigr)
\Bigl(\frac{d}{d\tau}{\hat D}^{\beta}\Bigr)\Bigr]\biggl|_{\gamma = 1} = 0
\end{equation}
the pole cancels. As shown below, this limitation is a condition that the 
vector connection field contains no outgoing wave.

Expression (8.4) is the total energy of the particles produced from
the  vacuum by external fields.
It is useful to realize that the integral over $\gamma$ in (8.4)
is none other than the integral over the energies of the outgoing
particles. The vacuum radiation is thus obtained along with its spectrum.

\subsection*{\bf Positivity.}

$$ $$

Owing to the conservation laws for the moments, all contractions with the
metric in (8.4) are positive definite. In particular, by (1.23) and (7.24),
\begin{eqnarray}
(g_{\mu\alpha}g_{\nu\beta} - \frac{1}{3}
g_{\mu\nu}g_{\alpha\beta})
\Bigl(\frac{d^2}{d\tau^2}{D}^{\alpha\beta}\Bigr)
\Bigl(\frac{d^2}{d\tau^2}{D}^{\mu\nu}\Bigr) = 
\frac{1}{2}\Bigl|\frac{d^2}{d\tau^2}D^{\alpha\beta}m_{\alpha}m_{\beta}
\Bigr|^2  \hspace{35mm}
\\
{}+2(1 - \gamma^2)
\Bigl|\frac{d^2}{d\tau^2}D^{\alpha\beta}
\nabla_{\alpha}r \, m_{\beta}\Bigr|^2 + 
\frac{1}{6}\Bigl(\frac{d^2}{d\tau^2}D^{\alpha\beta}m_{\alpha}m^*_{\beta} - 
2(1 - \gamma^2)\frac{d^2}{d\tau^2}D^{\alpha\beta}
\nabla_{\alpha}r\nabla_{\beta}r\Bigr)^2 \nonumber
\end{eqnarray}
which proves the positivity of the gravitational-field contribution in
(8.4).

The positivity of the matrix contributions follows from assuming
the self-adjointness of the operator of small disturbances of the quantum
field \footnote{
For simplicity, let this field be boson and non-gauge. In the general case 
the calculation will include diagonalizing and squaring operators, and 
combining loops [24]. Eq. (8.4) gives the contribution of a generic loop.}
. There should exist a symmetric and nondegenerate matrix $\omega_{AB}$
such that the operator
\begin{equation}
\omega_{AC}H_B^C
\end{equation}
with $H^A_B$ in (1.1) is the Hessian of an action. The self-adjointness then
requires that i) the operator (8.7) be symmetric, and ii) the matrix
$\omega_{AB}$ be positive definite. The latter is a condition of the absence
of ghosts. The symmetry of the operator (8.7) implies that 
i) $\omega_{AB}$ behaves like a metric with respect to the covariant
differentiation:
\begin{equation}
\nabla_{\mu}\omega_{AB} = 0\;\;,
\end{equation}
and ii) $\omega_{AB}$ converts the potential matrix into a symmetric
form:
\begin{equation}
P^C_A\omega_{CB} - P^C_B\omega_{CA} = 0\;\;.
\end{equation}
From (8.8) it follows that
\begin{equation}
[\nabla_{\mu},\nabla_{\nu}]\omega_{AB} = 
- {\cal R}^C_{A\mu\nu}\omega_{CB} - 
 {\cal R}^C_{B\mu\nu}\omega_{CA} = 0 \;\;, 
\end{equation} 
i.e. $\omega_{AB}$ converts the commutator curvature into 
an antisymmetric form. 
Using (8.8) once again one obtains the antisymmetry relation for the source
of the commutator curvature in (1.7):
\begin{equation}
J_A^{C\alpha}\omega_{CB} + J_B^{C\alpha}\omega_{CA} = 0 \;\;,
\;\;J_B^{A\alpha} = {\hat J}^{\alpha}\;\;.
\end{equation}

At this point it is important that the integrals with matrices contain the
propagators of parallel transport for the matrix indices. The explicit forms
of Eqs. (7.19) and (7.20) are
\begin{eqnarray}
D_B^A &=& \frac{1}{4\pi} \int d{\bar x} {\bar g}^{1/2}
\delta\Bigl(T(x,{\bar x}) - \tau\Bigr)g^A_{\;\;{\bar A}}
g_B^{\;\;{\bar B}}{\bar P}^{\bar A}_{\bar B}\biggl|_{x \to i^+}\;\;,\\
D_B^{A\alpha} 
&=& \frac{1}{4\pi} \int d{\bar x} {\bar g}^{1/2}
\delta\Bigl(T(x,{\bar x}) - \tau\Bigr)g^A_{\;\;{\bar A}}
g_B^{\;\;{\bar B}}g^{\alpha}_{\;\;{\bar \alpha}}
{\bar J}^{{\bar A}{\bar \alpha}}_{\bar B}\biggl|_{x \to i^+}\;\;.
\end{eqnarray}
By (8.8), the law of parallel transport for $\omega_{AB}$ is
\begin{equation}
\omega_{AB}\,g^A_{\;\;{\bar A}}g^B_{\;\;{\bar B}} = 
{\bar \omega}_{{\bar A}{\bar B}}\;\;.
\end{equation}
Using this law it is not difficult to prove that the symmetries of the 
sources in (8.12) and (8.13) imply the symmetries of the moments:
\begin{eqnarray}
D_A^C\,\omega_{CB} = D^C_B\,\omega_{CA} \;,\\
D_A^{C\alpha}\,\omega_{CB} = - D^{C\alpha}_B\,\omega_{CA} \;.
\end{eqnarray}
The latter symmetries remain unchanged under a differentiation with respect
to $\tau$ and a contraction of (8.16) with any vector.

Let ${\hat \Gamma}^{+}$ and ${\hat \Gamma}^-$ be any matrices possessing
the properties
\begin{equation}
\Gamma^{\pm C}_{\;\;A}\,\omega_{CB} = 
\pm \Gamma^{\pm C}_{\;\;B}\,\omega_{CA}\;.
\end{equation}
The positive-definite metric $\omega_{AB}$ can be expanded over an
orthonormal basis:
\begin{equation}
\omega_{AB} = h_A(M)h_B(N)\delta(M,N)\;\;,\;\;
h_A(M)h^A(N) = \delta(M,N)\;\;,\;\;
h^A(M) = \omega^{-1AB}h_B(M)
\end{equation}
with $\delta(M,N)$ the Kronecker symbol. Denoting
\begin{equation}
\Gamma^{\pm}(M,N) = h^A(M)\Gamma^{\pm B}_{\;\;A}h_B(N)
\end{equation}
one has
\begin{equation}
\tr\Bigl({\hat \Gamma}^{\pm}\Bigr)^2 = 
\Gamma^{\pm A}_{\;\;B}\Gamma^{\pm B}_{\;\;A} = 
\pm \Gamma^{\pm A}_{\;\;B}\omega_{AC}\Gamma^{\pm C}_{\;\;D}\omega^{-1DB} = 
\pm \sum_{N,M}\Bigl(\Gamma^{\pm}(M,N)\Bigr)^2\;\;.
\end{equation}

Thus, identifying ${\hat \Gamma}^{\pm}$ with
\begin{equation}
{\hat \Gamma}^+ = \frac{d^2}{d\tau^2}{\hat D}\;\;,\;\; 
{\hat \Gamma}^- = \frac{d}{d\tau}{\hat D}^{\alpha}k_{\alpha}
\end{equation}
where $k_{\alpha}$ is an arbitrary vector, we obtain
\begin{equation}
\tr\Bigl(\frac{d^2}{d\tau^2}{\hat D}\Bigr)^2 \ge 0 \quad ,\quad
\tr\Bigl(\frac{d}{d\tau}{\hat D}^{\alpha}k_{\alpha}\Bigr)^2 \le 0\;\;.
\end{equation}
The first of these inequalities proves the positivity of the potential
contribution in (8.4). Finally, by (1.23) and (7.23),
\begin{equation}
g_{\alpha\beta}
\Bigl(\frac{d}{d\tau}{\hat D}^{\alpha}\Bigr)
\Bigl(\frac{d}{d\tau}{\hat D}^{\beta}\Bigr) = 
\Bigl(\frac{d}{d\tau}{\hat D}^{\alpha}m_{\alpha}\Bigr)
\Bigl(\frac{d}{d\tau}{\hat D}^{\beta}m^*_{\beta}\Bigr) +
(1 - \gamma^2)\Bigl(\frac{d}{d\tau}{\hat D}^{\alpha}\nabla_{\alpha}r\Bigr)^2
\;\;.
\end{equation}
Hence, by the second inequality in (8.22),
\begin{equation}
\tr\Bigl[g_{\alpha\beta}
\Bigl(\frac{d}{d\tau}{\hat D}^{\alpha}\Bigr)
\Bigl(\frac{d}{d\tau}{\hat D}^{\beta}\Bigr)\Bigr] \le 0 
\end{equation}
which proves the positivity of the commutator-curvature contribution
in (8.4).

The matrix $\omega_{AB}$ itself does not figure in any of the final
expressions (8.22)-(8.24). Only its existence is important.

\subsection*{\bf Radiation of waves.}

$$ $$

Generally, the flux of the non-coherent radiation caused by pair 
creation in the vacuum is only one term of the mass-loss formula (1.35).
To clear up the meaning of condition (8.5) and to complete the discussion of the 
radiation moments, we shall consider also the other terms which in the 
field-theoretic case stand for a radiation of waves ( not necessarily 
classical [23]).

By analogy with the electromagnetic field we may assume that the contribution
of the vector connection field to $T^{\mu\nu}_{\mbox{\scriptsize source}}$
in (1.32) is of the form
\footnote{
Outside the support of the source of $\R^{\mu\nu}$.}
\begin{equation}
T^{\mu\nu}_{\mbox{\scriptsize source}} = - \frac{1}{4\pi}
\tr \Bigl(g_{\alpha\beta}{\R}^{\mu\alpha}{\R}^{\nu\beta} - 
\frac{1}{4}g^{\mu\nu}{\R}^{\alpha\beta}{\R}_{\alpha\beta}\Bigr)
+ \mbox{a contribution of the potential}\;\;{\hat P}
\end{equation}
where the change of the overall sign as compared to the case where 
$\R^{\mu\nu}$ is literally replaced by the Maxwell tensor is owing to
the antisymmetry (8.10). The density of the flux of 
$T^{\mu\nu}_{\mbox{\scriptsize source}}$ through ${\cal I}^+$ is then
\begin{eqnarray}
\frac{1}{4}r^2T^{\mu\nu}_{\mbox{\scriptsize source}}
\nabla_{\mu}v \nabla_{\nu}v\Bigl|_{{\cal I}^+} &=&
- \frac{1}{4\pi}\tr \Bigl(\frac{1}{4}r^2\nabla_{\mu}v\,
{\R}^{\mu\alpha}g_{\alpha\beta}{\R}^{\nu\beta}\nabla_{\nu}v
\Bigr)\biggl|_{{\cal I}^+}    \\
&=& - \frac{1}{4\pi} \tr \Bigl[
\Bigl(\frac{1}{2}r\nabla_{\mu}v\,{\R}^{\mu\alpha}m_{\alpha}\Bigr)
\Bigl(\frac{1}{2}r\nabla_{\nu}v\,{\R}^{\nu\beta}m^*_{\beta}\Bigr)
\Bigr]_{{\cal I}^+}\nonumber
\end{eqnarray}
where the last form is obtained by inserting expression (1.21)
for $g_{\alpha\beta}$ and using the antisymmetry of 
${\R}^{\alpha\beta}$. If we define the complex news function
\footnote{
This term was introduced in [26,27] for the gravitational waves
 but the electromagnetic waves can be discussed along the same lines. Two
real  components of the news function are the initial data for the two
independent degrees of freedom of a radiation field counted per point of 
${\cal I}$.}
for the waves of the vector connection field
\begin{equation}
\frac{\partial}{\partial u}{\bf C}_{\mbox{\scriptsize vect}}(u,\phi) = 
- \frac{1}{2} \nabla_{\alpha}v \,m_{\beta}            
\Bigl(r{\R}^{\alpha\beta}\Bigr)\biggl|_{{\cal I}^+[u,\phi,r\to\infty]}\;\;,
\end{equation}
then, according to (8.26) and (1.35), the quantity
\begin{equation}
- \frac{1}{4\pi} \int d^2{\cal S}(\phi)\tr \Bigl|\frac{\partial}{\partial u}
{\bf C}_{\mbox{\scriptsize vect}}(u,\phi)\Bigr|^2
\end{equation}
is the energy flux of the outgoing radiation of this field.

Assuming that there is no incoming radiation, we may use the Jacobi
identities to express ${\R}^{\alpha\beta}$ through its source as in Eq.
(A.17) of Appendix A. Using also (4.17) we obtain to lowest order in $\Re$
\begin{equation}
\Bigl(r {\R}^{\alpha\beta}\Bigr)
\biggl|_{{\cal I}^+[\tau,\phi,r\to \infty]} = 
\nabla^{\alpha}u \,\frac{d}{d\tau}{\hat D}_{\1}^{\;\beta}(\tau,\phi) -
\nabla^{\beta}u\,\frac{d}{d\tau}{\hat D}_{\1}^{\;\alpha}(\tau,\phi)\;\;.
\end{equation}
Hence
\begin{equation}
\frac{\partial}{\partial \tau}{\bf C}_{\mbox{\scriptsize vect}}(\tau,\phi) = 
m_{\alpha}\frac{d}{d\tau}{\hat D}_{\1}^{\;\alpha}(\tau,\phi)\;\;,
\end{equation}
and
\begin{equation}
\Bigl|\frac{\partial}{\partial \tau}
{\bf C}_{\mbox{\scriptsize vect}}(\tau,\phi)\Bigl|^2 = 
\Bigl|\frac{d}{d\tau}{\hat D}_{\1}^{\;\alpha}m_{\alpha}\Bigr|^2
= (\delta_{ik} - n_in_k)\Bigl(\frac{d}{d\tau}{\hat D}_{\1}^{\;i}\Bigr)
\Bigl(\frac{d}{d\tau}{\hat D}_{\1}^{\;k}\Bigr)
\end{equation}
where, in the last form, ${\hat D}_{\1}^{\;i}$ are the spatial components of 
${\hat D}_{\1}^{\;\alpha}$ in the Minkowski frame, and use is made of
Eq. (1.28). Using the basis decomposition (1.21) for the metric and
the conservation law (4.42) for the moment, 
expression (8.31) can be written in the covariant form
\begin{equation}
\Bigl|\frac{\partial}{\partial \tau}
{\bf C}_{\mbox{\scriptsize vect}}(\tau,\phi)\Bigl|^2 = 
g_{\alpha\beta}\Bigl(\frac{d}{d\tau}{\hat D}_{\1}^{\;\alpha}\Bigr)
\Bigl(\frac{d}{d\tau}{\hat D}_{\1}^{\;\beta}\Bigr)\;\; .
\end{equation}
The total energy of the waves of the vector field emitted for the whole
history is, therefore,
\begin{equation}
- \frac{1}{4\pi} \int\limits_{-\infty}^{\infty}d\tau
\int d^2{\cal S}(\phi)\tr\Bigl[
g_{\alpha\beta}\Bigl(\frac{d}{d\tau}{\hat D}^{\alpha}\Bigr)
\Bigl(\frac{d}{d\tau}{\hat D}^{\beta}\Bigr)\Bigr]_{\gamma = 1}\;\;.
\end{equation}

It is now seen that condition (8.5) can be written in the form
\begin{equation}
\tr \Bigl|\frac{\partial}{\partial \tau}{\bf C}_{\mbox{\scriptsize vect}}
(\tau,\phi)\Bigr|^2 
\equiv 0 \;\;,
\end{equation}
and its meaning is that the source of the vector field radiates no waves.

Only the $\gamma = 1$ moments ${D}_{\1}$ appear in classical radiation
theory as is clear from Eq. (4.17).
However, even in this theory the concept of boosted ($\gamma$ - dependent)
moment $D$ is useful since it helps to understand the nature of the multipole
expansion. In the case of a nonrelativistic source, $\gamma$ is the
only parameter that contains the velocity of light 
since $\gamma$ is the velocity 
of a particle at $i^+$ per unit velocity of light. Therefore, in the 
nonrelativistic approximation, $D$ at $\gamma=1$ may be calculated as an
expansion at $\gamma = 0$:
\begin{equation}
{\hat D}_{\1}^{\;\alpha} = {\hat D}^{\alpha}\biggl|_{\gamma=0} +
\frac{d}{d\gamma}{\hat D}^{\alpha}\biggl|_{\gamma = 0} + \ldots \;\;.
\end{equation}
This expansion gives rise to the multipole moments. Since
at $\gamma = 0$ the dependence on the direction at $i^+$ should
disappear, the expansion in $\gamma$ is also an expansion
in the direction vector $n^i$. For the spatial components of 
${\hat D}^{\alpha}$ in the Minkowski frame one may write quite generally
\begin{eqnarray}
{\hat D}^i\biggl|_{\gamma=0} &=& \frac{d}{d\tau}\; {\hat d}^i\;\;,\\
\frac{d}{d\gamma}{\hat D}^i\biggl|_{\gamma=0} &=& n_k
\Bigl(\frac{1}{6}\frac{d^2}{d\tau^2}\;{\hat d}^{(ki)} + 
\frac{d}{d\tau}\;{\hat d}^{[ki]}\Bigr) + 
\mbox{a term} \propto n^i\;\;,\;\;\;{\hat d}^{(ki)}\delta_{ki} = 0
\end{eqnarray}
and so on. Here $d^i, d^{(ki)},d^{[ki]}$ are the dipole, quadrupole and
magnetic moments, and the term $\propto n^i$ drops out of (8.31).
The appearance of the time derivatives in expressions (8.36) and
(8.37) is owing to the fact that the multipole moments are coefficients in
the expansion of the ${\cal D}$ in Eq. (7.28) rather than the $D$. For the 
radiation of a nonrelativistic source one obtains the standard textbook 
result
\begin{equation}
\frac{1}{4\pi}\int d^2{\cal S}(\phi)\,\Bigl|\frac{\partial}{\partial \tau}
{\bf C}_{\mbox{\scriptsize vect}}(\tau,\phi)\Bigr|^2=
\frac{2}{3}\Bigl(\frac{d^2}{d\tau^2}\; d^i\Bigr)^2 + \frac{1}{180}
\Bigl(\frac{d^3}{d\tau^3}\; d^{(ik)}\Bigr)^2 +
\frac{1}{3}\Bigl(\frac{d^2}{d\tau^2}\; d^{[ik]}\Bigr)^2 + \ldots\;\;.
\end{equation}

The (differentiated) news function for the outgoing gravitational waves
is
\footnote{See, e.g., [23] and the original references [26,27].}
\begin{equation}
\frac{\partial^2}{\partial u^2}{\bf C}_{\mbox{\scriptsize grav}}(u,\phi) =
- \frac{1}{8} \nabla_{\alpha}v \nabla_{\mu}v \, m_{\beta}m_{\nu}
\Bigl(rR^{\alpha\beta\mu\nu}\Bigr)\biggl|_{{\cal I}^+[u,\phi,r \to \infty]}\;.
\end{equation}
Assuming again that there are no incoming waves and solving the Bianchi
 identities to express $R^{\alpha\beta\mu\nu}$ through its source [15,23]
 we obtain to lowest order in $\Re$
\begin{equation}
\Bigl(rR^{\alpha\beta\mu\nu}\Bigr)
\biggl|_{{\cal I}^+[\tau,\phi,r \to \infty]}=
-4 \nabla^{[\mu}u \nabla^{<\alpha}
u \;\frac{d^2}{d\tau^2}\;D_{\1}^{\;\nu]\beta>}(\tau,\phi)
\end{equation}
where both types of brackets $[]$ and $<>$ denote antisymmetrization of the
respective indices. Hence
\begin{equation}
\frac{\partial^2}{\partial \tau^2}{\bf C}_{\mbox{\scriptsize grav}}(\tau,\phi) =
\frac{1}{2}m_{\alpha}m_{\beta}
\frac{d^2}{d\tau^2}D_{\1}^{\;\alpha\beta}(\tau,\phi)\;\;.
\end{equation}
With the initial condition (5.41) one may  integrate this equation to
obtain the news function
\begin{equation}
\frac{\partial}{\partial \tau}{\bf C}_{\mbox{\scriptsize grav}}(\tau,\phi) =
\frac{1}{2}m_{\alpha}m_{\beta}
\frac{d}{d\tau}D_{\1}^{\;\alpha\beta}(\tau,\phi)\;\;.
\end{equation}
Thus
\begin{eqnarray}
& &\biggl|\frac{\partial}{\partial \tau}{\bf C}_{\mbox{\scriptsize grav}}
(\tau,\phi)\biggr|^2 =
\frac{1}{4}\biggl|
\frac{d}{d\tau}D_{\1}^{\;\alpha\beta}
m_{\alpha}m_{\beta}\biggr|^2 = \\
& &= \frac{1}{2}\Bigr[
(\delta_{im} - n_i n_m)(\delta_{kn} - n_k n_n) -
\frac{1}{2}(\delta_{ik} - n_i n_k)(\delta_{mn} - n_m n_n)\Bigr]
\Bigl(\frac{d}{d\tau}D_{\1}^{\;ik}\Bigr)
\Bigl(\frac{d}{d\tau}D_{\1}^{\;mn}\Bigr) \nonumber
\end{eqnarray}
where, in the last form, $D_{\1}^{\;ik}$ are the spatial
components of $D_{\1}^{\;\alpha\beta}$ in the Minkowski frame.
Bringing this expression to the covariant form yields the result
\begin{equation}
\biggl|\frac{\partial}{\partial \tau}{\bf C}_{\mbox{\scriptsize grav}}
(\tau,\phi)\biggr|^2 =
\frac{1}{2}
(g_{\alpha\mu}g_{\beta\nu} - \frac{1}{2}g_{\alpha\beta}g_{\mu\nu})
\Bigl(\frac{d}{d\tau}D_{\1}^{\;\alpha\beta}\Bigr)
\Bigl(\frac{d}{d\tau}D_{\1}^{\;\mu\nu}\Bigr)     \;\;.
\end{equation}
Therefore, according to Eq. (1.35), the total energy of the gravitational
waves emitted for the whole history is
\begin{equation}
\frac{1}{4\pi} \int\limits_{-\infty}^{\infty} d\tau \int d^2{\cal S}(\phi)
\frac{1}{2}
(g_{\alpha\mu}g_{\beta\nu} - \frac{1}{2}g_{\alpha\beta}g_{\mu\nu})
\Bigl(\frac{d}{d\tau}D^{\alpha\beta}\Bigr)
\Bigl(\frac{d}{d\tau}D^{\mu\nu}\Bigr)\biggl|_{\gamma=1}\;\;. 
\end{equation}

For a nonrelativistic source,
\begin{equation}
D^{\alpha\beta} = D^{\alpha\beta}\biggl|_{\gamma=0} + O(\gamma)\;\;,
\end{equation}
and
\begin{equation}
D^{ik}\biggl|_{\gamma=0} = 
\frac{1}{3}\frac{d^2}{d\tau^2}\;d^{ik} + \mbox{a term}\;\propto
\delta^{ik}\;\;,\;\;\; d^{ik}\delta_{ik} = 0
\end{equation}
where $d^{ik}$ is the quadrupole moment, and the term $\propto \delta^{ik}$
drops out of (8.43). Here again, $d^{ik}$ is the coefficient in the expansion
of the ${\cal D}$ in Eq. (7.29) rather than the $D$. Therefore, in (8.47)
there appears the second time derivative. For the radiation of the
gravitational waves by a nonrelativistic source Eqs. (8.43) and (8.47)
 yield the textbook result
\begin{equation}
\frac{1}{4\pi}\int d^2{\cal S}(\phi)\,
\biggl|\frac{\partial}{\partial \tau}{\bf C}_{\mbox{\scriptsize grav}}
(\tau,\phi)\biggr|^2 = \frac{1}{45}\Bigl(\frac{d^3}{d\tau^3}\;d^{ik}\Bigr)^2 +
\ldots \;\;.
\end{equation}

Compare Eqs. (8.33) and (8.45) with (8.4). The similarity is striking.
As a result of the derivation above, the quantum problem of particle
creation becomes almost the same thing as the classical problem of
radiation of waves. The whole difference is that, in the quantum problem,
 there figure the boosted moments $D$ and, instead of setting $\gamma=1$,
 one is {\it to integrate over} $\gamma$. In the nonrelativistic
 case even this difference disappears since all moments will be expanded
 at $\gamma=0$ , and the integral over $\gamma$ will be removed trivially. Note
 also that in the case of the vector field both the classical and quantum
 radiation effects are determined by the second-order time derivative of the
 moment (in terms of the ${\cal D}$ , Eq. (7.27)). In the case of the
 gravitational field there appears one more distinction between the classical
 and quantum contributions to the radiation energy. Namely, the classical
 radiation is determined by the third-order and quantum by the fourth-order
 time derivative of the respective moment ${\cal D}$ . This is a consequence
 of the dimension of the coupling constant in the quantum loop.
 } 

\newpage

{\renewcommand{\theequation}{9.\arabic{equation}}  

\begin{center}
\section{\bf    Specializations and examples}
\end{center}

\subsection*{\bf Spherical symmetry.}

$$ $$

The fact that the energy of vacuum radiation is expressed through
the moments with $\gamma$ different from unity has an important
consequence. Inspecting Eqs. (8.6), (8.23) and (8.43) one can see
that at $\gamma = 1$ there survive only the projections of the moments 
on the 2-sphere ${\cal S}$ (the transverse projections) whereas at
$\gamma \ne 1$ there are also projections on $\nabla r$ i.e. on the
direction of motion of the partice at $i^+$ (the longitudinal 
projections). For a spherically symmetric source, the vector moment
can have no projection on the sphere, and the projection of the
tensor moment on the sphere can only be proportional to the  metric on the 
sphere in which case the orthogonality $(m,m)=0$ comes into effect. It
follows that spherically symmetric sources cannot emit waves but can produce
particles from the vacuum. This makes spherical symmetry an interesting
case for studying the effect of vacuum radiation. In addition, the
limitation (8.5) is in this case fulfilled  automatically and so we can study
the particle production by vector fields. 

For spherically symmetric sources we have from (8.23) and (8.6)
\begin{equation}
g_{\alpha\beta}\Bigl(\frac{d}{d\tau}{\hat D}^{\alpha}\Bigr)
\Bigl(\frac{d}{d\tau}{\hat D}^{\beta}\Bigr) = 
(1 - \gamma^2)
\Bigl(\frac{d}{d\tau}{\hat D}^{\alpha}\nabla_{\alpha}r\Bigr)^2\;\;,
\end{equation}
\begin{equation}
(g_{\mu\alpha}g_{\nu\beta} - \frac{1}{3}g_{\mu\nu}g_{\alpha\beta})
\Bigl(\frac{d^2}{d\tau^2}{D}^{\alpha\beta}\Bigr)
\Bigl(\frac{d^2}{d\tau^2}{D}^{\mu\nu}\Bigr) = 
\frac{1}{6}\Bigl(\frac{d^2}{d\tau^2}D^R 
 + 3(1 - \gamma^2)
\frac{d^2}{d\tau^2}D^{\alpha\beta}\nabla_{\alpha}r\nabla_{\beta}r\Bigr)^2\;\;.
\end{equation}
Since these quantities are spherically symmetric scalars, they do not depend
on the angles at $i^+$. Therefore, the integral over ${\phi}$ in (8.4)
removes trivially, and we obtain
\begin{eqnarray}
M(-\infty) - M(\infty) &=& \frac{1}{4\pi}\int\limits_0^1 d\gamma \gamma^2
\int\limits_{-\infty}^{\infty}d\tau\; \tr\Bigl[
\Bigl(\frac{d^2}{d\tau^2}{\hat D}\Bigr)^2 - \frac{1}{3}
\Bigl(\frac{d}{d\tau}{\hat D}^{\alpha}\nabla_{\alpha}r\Bigr)^2  
\hspace{2cm}\nonumber\\
& &\qquad{}+ \frac{1}{180}{\hat 1}\Bigl(\frac{d^2}{d\tau^2}D^R + 
3(1 - \gamma^2)\frac{d^2}{d\tau^2}D^{\alpha\beta}\nabla_{\alpha}r
\nabla_{\beta}r \Bigr)^2\Bigr]
\end{eqnarray}
where there is no more pole in $\gamma$ including in the term with the 
vector moment.

The longitudinal projections of the moments are conveniently obtained as 
follows. One differentiates the law of parallel transport (7.14) with respect 
to $\gamma$. This yields
\begin{equation}
\nabla_{\mu}r(x)\,
g^{\mu}_{\;\;{\bar \mu}}(x,{\bar x})\biggl|_{x\to i^+[\gamma,\phi]}
= - {\bar \nabla}_{\bar \mu}\frac{\partial}{\partial \gamma}
T_{\gamma\phi}({\bar x}) + (\nabla_{\mu}t(x) - \gamma\nabla_{\mu}r(x))
\frac{\partial}{\partial \gamma}\,g^{\mu}_{\;\;{\bar \mu}}(x,{\bar x})
\biggl|_{x\in i^+[\gamma,\phi]}\;\;.
\end{equation}
The derivative
\begin{equation}
\frac{\partial}{\partial \gamma}
\Bigl(\, g^{\mu}_{\;\;{\bar \mu}}(x,{\bar x})
\biggl|_{x\in i^+[\gamma,\phi]}\,\Bigr) = O[\Re]
\end{equation}
can be calculated as a result of parallel transport of a vector along 
the closed contour consisting of two geodesics which emanate from 
${\bar x}$ and come to $i^+$ with boosts $\gamma$ and $\gamma + \delta 
\gamma$. With the contribution of (9.5) neglected , one obtains from 
(7.20) and (7.21)
\begin{equation}
\nabla_{\alpha}r {\hat D}^{\alpha} =
-\frac{1}{4\pi}\int d{\bar x} {\bar g}^{1/2} \delta(T_{\gamma\phi}({\bar x})
- \tau)\Bigl({\bar \nabla}_{\bar \mu}\frac{\partial}{\partial \gamma}
{\bar T}_{\gamma\phi}\Bigr){\hat J}^{\bar \mu}({\bar x}) + O[\Re^2]\;\;,
\end{equation}
\begin{equation}
\nabla_{\alpha}r\nabla_{\beta}r  {D}^{\alpha\beta} =
\frac{1}{4\pi}\int d{\bar x} {\bar g}^{1/2} \delta(T_{\gamma\phi}({\bar x})
- \tau)
\Bigl({\bar \nabla}_{\bar \mu}\frac{\partial}{\partial \gamma}
{\bar T}_{\gamma\phi}\Bigr)
\Bigl({\bar \nabla}_{\bar \nu}\frac{\partial}{\partial \gamma}
{\bar T}_{\gamma\phi}\Bigr)
{J}^{{\bar \mu}{\bar \nu}}({\bar x}) 
+ O[\Re^2]\;\;.
\end{equation}

The metric of a spherically symmetric spacetime is generally of the form
\begin{equation}
ds^2 = d\Gamma^2 + r^2\biggl|_{\Gamma}d{\cal S}^2\;\;,
\;\;d\Gamma^2 = g_{AB}(y)dy^Ady^B\;\;,\;\;
d{\cal S}^2 = g_{ab}(\phi)d\phi^a d\phi^b
\end{equation}
where $\Gamma$ is some 2-dimensional Lorentzian space, $\;{\cal S}$ is the unit 
2-sphere, and $r=r(y)$ is a function on $\Gamma$. The line $r=0$ is a 
boundary in $\Gamma$. We shall denote $g^{\Gamma}$ the determinant of 
$g_{AB}(y)$ and $\nabla_A$  the derivative in $\Gamma$. The world function
 on a spherically symmetric spacetime is a regular function of the following
 arguments:
 \begin{equation}
 \sigma(x,{\bar x}) = \sigma(y,{\bar y},\cos\omega)\;\;,
 \;\;\; \omega = \omega(\phi,{\bar \phi})
 \end{equation}
 where $\frac{1}{2}{\omega}^2(\phi,{\bar \phi})$ is the 
world function on the unit
 2-sphere, and an expression for $\cos\omega$ is given in (4.16).
 Hence
 \begin{equation}
 T_{\gamma\phi}({\bar x}) = T_{\gamma}({\bar y},\cos\omega)\;\;,\;\;\;
 \omega = \omega(\phi,{\bar \phi})
 \end{equation}
 and Eq. (7.4) takes the form 
 \begin{equation}
 {\bar g}^{AB}{\bar \nabla}_{A}T_{\gamma}{\bar \nabla}_{B}T_{\gamma}
 + \frac{1}{{\bar r}^2}(1 - \cos^2\omega)\Bigl(
 \frac{d}{d\cos\omega}T_{\gamma}\Bigr)^2 = - (1 - \gamma^2)\;\;.
 \end{equation}

 Denote
 \begin{equation}
 T_{\gamma}^{\;\pm} = T_{\gamma}(y,\pm 1)\;\;.
 \end{equation}
 By (9.11), the integral curves of each of the vector fields
 \begin{equation}
 N^{\;A}_+(y) = \frac{1}{\sqrt{1 - \gamma^2}}\nabla^AT_{\gamma}^{\;+}\quad ,
 \quad N^{\;A}_-(y) = \frac{1}{\sqrt{1 - \gamma^2}}\nabla^AT_{\gamma}^{\;-}
 \end{equation}
 are timelike geodesics in $\Gamma$. With fixed $\gamma$, the geodesics
 generated by $N_+$ make a 1-parameter congruence foliating $\Gamma$.
 The geodesics generated by $N_-$ make another such congruence. The geodesics
 from the two congruences may be joined pairwise, a pair consisting of
 the two geodesics that hit one and the same point of the boundary $r=0.$
 A world line in $\Gamma$ consisting of two geodesics joined at $r=0$
 is the mapping on $\Gamma$ of a 4-dimensional radial geodesic (the broken
 line in Fig.7). This line is generated by $N_+$ in the future from $r=0$ and
 by $N_-$ in the past from $r=0$.

 The spacelike lines orthogonal to the geodesics generated by $N_+$ also
 make a 1-parameter congruence foliating $\Gamma$. The spacelike lines
 orthogonal to the geodesics generated by $N_-$ make one more such
 congruence. These congruences are respectively
\begin{equation}
T_{\gamma}^{\;+} = \tau \quad ,\quad
T_{\gamma}^{\;-} = \tau
\end{equation}
 with $\tau$ the parameter of the congruence. By (9.11),
 \begin{equation}
 \frac{d}{d\cos \omega}\Bigl(T_{\gamma}({\bar y},\cos\omega)\biggl|_{
 r({\bar y})=0}\Bigr) = 0\;\;.
 \end{equation}
 It follows that the spacelike lines (9.14) join pairwise at $r=0$ as shown
 in Fig.7. The lines $T_{\gamma}^{\;+} = \tau$ and $T_{\gamma}^{\;-} = \tau$
 with one and the same $\tau$ hit one and the same point of the boundary.
 {\it The hyperplane} $T_{\gamma\phi}(x) = \tau$ {\it maps on the
 interior domain between these two lines.} However, the boundary of this
 domain, i.e. the line
 \begin{equation}
 \Bigl(T_{\gamma}^{\;+} = \tau\Bigr)\cup\Bigl(
 T_{\gamma}^{\;-} = \tau\Bigr)
 \end{equation}
 itself corresponds to no 4-dimensional original. Only in flat spacetime
 is this line a mapping of a spacelike geodesic.

 In two limiting cases, $\gamma=1$ and $\gamma=0$ , there come about
 coincidences. At $\gamma=1$ , the line $T_{\gamma}^{\;+} = \tau$
 coincides with a geodesic generated by $N_+$ , and the line 
 $T_{\gamma}^{\;-} = \tau$ coincides with a geodesic generated by $N_-$ .
 Both geodesics are then null, and their union is the mapping on $\Gamma$
 of a  4-dimensional radial light ray. At $\gamma=0$ , the line 
 $T_{\gamma}^{\;+} = \tau$ coincides with the line 
 $T_{\gamma}^{\;-} = \tau$ . In this case the mapping of the hyperplane
 on $\Gamma$ degenerates into a line.

 When all external fields are spherically symmetric, there appear only
 moments of the  sources ${\bar X}$ that either do not depend on the
 angles ${\bar \phi}$ or depend on them through $\omega(\phi,{\bar \phi})$
 where $\phi$  are the angles at $i^+$. The moment of any such source
 can be calculated as follows:
 \begin{equation}
 D = \frac{1}{4\pi}\int d{\bar x}{\bar g}^{1/2}\delta\Bigl(
 T_{\gamma\phi}({\bar x}) - \tau\Bigr){\bar X} = 
 \frac{1}{2}\int d^2{\bar y}\sqrt{-{\bar g}^{\Gamma}}\,{\bar r}^2
 \int\limits_{-1}^1 d(\cos\omega)
 \delta\Bigl(T_{\gamma}({\bar y},\cos\omega) - \tau\Bigr){\bar X}\;\;.
 \end{equation}
 Here the equation of the integration hyperplane is to be solved with
 respect to $\cos\omega$. Let $\cos\omega = f$ be the solution. Then the moment
 is obtained as the integral in $\Gamma$
 \begin{equation}
 D = \frac{1}{2}\int d^2{\bar y}\sqrt{-{\bar g}^{\Gamma}}
\, \theta\Bigl(T_{\gamma}^{\;-} - \tau\Bigr)
 \theta\Bigr(\tau - T_{\gamma}^{\;+}\Bigr)\,{\bar r}^2\biggl|
 \frac{dT_{\gamma}}{d\cos\omega}\biggr|^{-1}{\bar X}\biggl|_{\cos\omega = f}
 \end{equation}
 in which the integration domain is bounded by the lines 
 $T_{\gamma}^{\;-} = \tau$ and $T_{\gamma}^{\;+} = \tau$. This calculation
 will be carried out below for flat spacetime, and here it will only be
 noted that the Jacobian appearing in (9.18) is
 \begin{equation}
 \Bigl|\frac{dT_{\gamma}}{d\cos\omega}\Bigr| = 
 \frac{\sqrt{1 - \gamma^2}}{\sin\omega}\Bigl|
 L({\bar y},\omega,\gamma)\Bigr|
 \;\;,\;\;\; \omega = \omega(\phi,{\bar \phi})
 \end{equation}
 where $L({\bar y},\omega,\gamma)$ is the angular momentum of the 
 geodesic that, when traced backwards in time, comes from 
 $i^+[\gamma,\phi]$ to the point $({\bar y},{\bar \phi})$ of a  compact
 domain. This follows from the geodesic equation (7.7) whose angle 
 component takes the form
 \begin{equation}
 \frac{d\omega}{ds} = - \frac{1}{r^2}\frac{1}{\sqrt{1 - \gamma^2}} \frac{d}{d\omega}
 T_{\gamma}(y,\cos\omega)\;\;.
 \end{equation}

 \subsection*{\bf Electrically charged shell expanding in the self field.}

$$ $$

It is not the purpose of the present paper to discuss applications of the 
result (8.4) but to give a simple example we shall consider the particle
production by an electrically charged spherical shell expanding in the
self field. Below,$\;e,m,$ and ${\cal E}$ are respectively the shell's charge,
rest mass, and energy in excess of the rest energy.

In terms of time $t$ orthogonal to $r$ , $(\nabla t,\nabla r)\equiv 0$
, a spherically symmetric electromagnetic 
field is  completely determined by a 
single arbitrary function $e(r,t)$ which is the charge contained on the 
hypersurface $t=\mbox{const.}$ inside the 2-sphere of area $4\pi r^2$:
\begin{equation}
e(r,t) = \int d{\bar x}{\bar g}^{1/2}\delta({\bar t} - t)\theta(r - {\bar r})
{\bar \nabla}_{\bar \mu}{\bar t}\:{\bar j}^{\bar \mu}\;\;,
\end{equation}
\begin{equation}
 \nabla_{\nu}F^{\mu\nu} = 4\pi j^{\mu}\;\;.
\end{equation}
Here $F^{\mu\nu}$ is the Maxwell tensor, and $j^{\mu}$ is the electromagnetic
current. The choice of $e(r,t)$ is limited only by the condition of
regularity of the field at $r=0$ , $e(0,t)=0$ , and the  condition that the 
support of $j^{\mu}$ belongs to a spacetime tube. Let $r=\rho(t)$ be the 
equation of the boundary of the tube. Then
\begin{equation}
e(r,t)\biggl|_{r>\rho(t)} = e = \mbox{const.}
\end{equation}
where $e$ is the full charge. In flat spacetime, with the normalization
of time $(\nabla t)^2 = -1,$ the general solution of the conservation
equation is
\begin{equation}
4\pi j^{\mu} = - (\nabla^{\mu}t)\frac{1}{r^2}\frac{\partial e(r,t)}{
\partial r} - 
(\nabla^{\mu}r)\frac{1}{r^2}\frac{\partial e(r,t)}{\partial t}\;\;,
\end{equation}
and the general solution of the Maxwell equations is
\begin{equation}
F^{\mu\nu} = E^{\mu}\nabla^{\nu}t - E^{\nu}\nabla^{\mu}t\;\;,\;\;
E^{\mu} = \frac{e(r,t)}{r^2}\nabla^{\mu}r 
\end{equation}
where only an electric field is present.

In the case where the support of the charge is a thin shell, we have
\begin{equation}
e(r,t) = e\:\theta(r - \rho(t))
\end{equation}
and, by (9.24),
\begin{eqnarray}
4\pi(\nabla_{\mu}t)j^{\mu} &=& \frac{e}{\rho^2(t)}\delta(r - \rho(t))\;\;,
\\
4\pi(\nabla_{\mu}r)j^{\mu} &=& 
- \Bigl(\frac{d}{dt}\frac{e}{\rho(t)}\Bigr)\delta(r - \rho(t))\;\;
\end{eqnarray}
where $r=\rho(t)$ is a law of motion of the shell. In this case there is
only a static Coulomb field outside the shell; inside the field vanishes.
However, this simplicity is only apparent. Because the shell should
expand in the self field, this field is nonstationary, and the law of expansion 
is not very simple \footnote{
The appearance of $1/2$ in the Coulomb energy is owing to the fact that the
force exerted on the surface charge is determined by one half of the
sum of the electric fields on both sides of the surface [33].}:
\begin{equation}
\frac{mc^2}{\displaystyle\sqrt{1 
- \frac{1}{c^2}\Bigl(\frac{d\rho}{dt}\Bigr)^2}} +
\frac{1}{2}\frac{e^2}{\rho} = mc^2 + {\cal E}\;\;.
\end{equation}

The shell can be contracted as much as one wants by raising its energy
but for given energy its minimum radius is
\begin{equation}
r_{\mbox{\scriptsize min}} = \frac{e^2}{2{\cal E}}\;\;.
\end{equation}
Set free, it expands monotonically from $r = r_{\mbox{\scriptsize min}}$ to
$r=\infty$ with increasing speed.

We shall consider two limiting cases, the nonrelativistic shell
\begin{equation}
\frac{d\rho}{dt} = \sqrt{\frac{1}{m}\Bigl(2{\cal E} - \frac{e^2}{\rho}\Bigr)}
\;\;,\;\;{\cal E}<< mc^2
\end{equation}
and the ultrarelativistic shell
\begin{equation}
\rho = \frac{e^2}{2{\cal E}}\Bigl(1 - \frac{mc^2}{\cal E}\Bigr) +
\sqrt{c^2(t - t_{\mbox{\scriptsize start}})^2 + \Bigl(\frac{e^2}{2{\cal E}}   
\frac{mc^2}{\cal E}\Bigr)^2}\;\;,
\;\; t_{\mbox{\scriptsize start}} = \mbox{const.}\;\;,\;\;{\cal E}>> mc^2\;.
\end{equation}
It is important that in the latter case the shell will remain timelike. 
The timelike ultrarelativistic shell  is always slow near 
$r = r_{\mbox{\scriptsize min}}$ and only at $r\to\infty$ it approaches the
speed of light.

\subsection*{\bf Particle production by a spherically symmetric 
electromagnetic field.}

$$ $$

The expanding spherical shell emits no electromagnetic waves but,
as we show below, it excites the vacuum of charged fields, and
the vacuum emits the quanta of these fields.

The coupling of an external electromagnetic field $A_{\mu}$ to the
charged vacuum fields will be introduced through the following
form of the covariant derivative in (1.2):
\begin{equation}
\nabla_{\mu} = \partial_{\mu}{\hat 1} + qA_{\mu}{\hat \Omega}\quad ,
\quad \tr\:{\hat \Omega}^2 < 0 
\end{equation}
where $q$ is the charge of the vacuum particles, $\;{\hat \Omega}$
is a numerical matrix, and the negativity of the trace of 
${\hat \Omega}^2$ is a corollary of the  self-adjointness of the
equation of the quantum field (see Sec.8).

For the commutator curvature and its source, Eq. (9.33) yields the
expressions
\begin{equation}
\R^{\mu\nu} = q\,{\hat \Omega}F^{\mu\nu}\;\;,\;\;
{\hat J}^{\mu} = 4\pi q\,{\hat \Omega}\, j^{\mu}
\end{equation}
in terms of the Maxwell tensor and the electromagnetic current. It is  
convenient to factor the coupling parameters out of the  
longitudinal projection of the vector moment:
\begin{equation}
\nabla_{\alpha}r \,{\hat D}^{\alpha} \equiv q\,{\hat \Omega}D_{||}\;\;.
\end{equation}
Then for $D_{||}$ we obtain from (9.6)
\begin{equation}
D_{||} = - \int d{\bar x}{\bar g}^{1/2}\delta\Bigl(
T_{\gamma\phi}({\bar x}) - \tau\Bigr)\Bigl(
{\bar \nabla}_{\bar \mu}\frac{\partial}{\partial \gamma}
{\bar T}_{\gamma\phi}\Bigr){\bar j}^{\bar \mu}\;\;.
\end{equation}

With the metric of the Lorentzian section in (9.8)
\begin{equation}
d\Gamma^2 = - dt^2 + dr^2
\end{equation}
one may use expression (7.17) for the hyperplane:
\begin{equation}
T_{\gamma\phi}({\bar x}) = {\bar t} - {\bar r}\gamma\cos\omega
(\phi,{\bar \phi})\;\;.
\end{equation}
Hence
\begin{equation}
T_{\gamma}^{\;\pm} = {\bar t}  \mp \gamma{\bar r}\;\;,  
\end{equation}
and the lines (9.14) are spacelike geodesics (Fig. 7). With a spherically
symmetric current $j^{\mu}$ we obtain
\begin{equation}
\Bigl({\bar \nabla}_{\bar \mu}\frac{\partial}{\partial \gamma}
{\bar T}_{\gamma\phi}\Bigr){\bar j}^{\bar \mu}  = 
- \cos\omega ({\bar \nabla}_{\bar \mu}{\bar r}){\bar j}^{\bar \mu}\;\;,
\end{equation}
and Eq. (9.36) takes the form
\begin{equation}
D_{||} = \int d^2{\cal S}({\bar \phi})\int\limits_0^{\infty} d{\bar r}\,
{\bar r}^2\int\limits_{-\infty}^{\infty}d{\bar t}\;
\delta({\bar t} - \tau - {\bar r}\gamma\cos\omega)\cos\omega 
({\bar \nabla}_{\bar \mu}{\bar r}){\bar j}^{\bar \mu}\;\;.
\end{equation}
The integration over the angles in this expression brings the longitudinal
moment to its final form
\begin{equation}
D_{||} = \frac{2\pi}{\gamma^2}\int\limits_{-\infty}^{\infty} dt
\int\limits_0^{\infty}dr\; \theta(t - \tau + r\gamma)\theta(\tau - t +r\gamma)
\, (t - \tau)(\nabla_{\mu}r)j^{\mu}
\end{equation}
which is a specialization of (9.18). Here the longitudinal projection of 
a spherically symmetric current is to be inserted from (9.24):
\begin{equation}
(\nabla_{\mu}r)j^{\mu} = - \frac{1}{4\pi r^2}
\frac{\partial e(r,t)}{\partial t}\;\;.
\end{equation}

In terms  of $D_{||}$ the vacuum energy released by a spherically
symmetric electromagnetic field is, by (9.3) and (9.35),
\begin{equation}
M(-\infty) - M(\infty) = \frac{q^2}{12\pi}\Bigl(-\tr\;{\hat \Omega}^2\Bigr)
\int\limits_0^1 d\gamma \gamma^2\int\limits_{-\infty}^{\infty}d\tau
\Bigl(\frac{d}{d\tau}D_{||}\Bigr)^2\;\;.
\end{equation}

\subsection*{\bf Radiation of the nonrelativistic shell.}

$$ $$

The longitudinal moment of the charged spherical shell is obtained 
by inserting expression (9.28) in (9.42):
\begin{equation}
D_{||} = - \frac{e}{2\gamma^2}\int\limits_{t_-}^{t_+}dt \;(t - \tau)
\:\frac{d}{dt} \frac{1}{\rho(t)}\;\;.
\end{equation}
Here $t_+$ and $t_-$ are solutions of the following equations:
\begin{equation}
t_+ - \gamma\rho(t_+) = \tau \quad , \quad t_- + \gamma\rho(t_-) = \tau\;\;.
\end{equation}
These are the time instants at which the world line of the shell intersects
the lines (9.14). The moment (9.45) is an integral along the world line 
of the shell between the points $t_-$ and $t_+$ as shown in Fig.7.
Integrating by parts in (9.45), and using Eq. (9.46) we obtain
\begin{equation}
D_{||} = \frac{e}{2\gamma^2}\biggl(\;
\int\limits_{t_-}^{t_+}
dt \frac{1}{\rho(t)} - 2 \gamma\;\biggr)\;\;.
\end{equation}

For the nonrelativistic shell, the moment can be calculated by expanding
it at $\gamma = 0$. Since, by (9.46),
\begin{equation}
t_+ = \tau + O(\gamma)\quad , \quad t_- = \tau + O(\gamma)\;\;,
\end{equation}
this amounts to expanding $\rho(t)$ at $t = \tau$.
A straightforward calculation using only Eq. (9.46) yields
\begin{equation}
\int\limits_{t_-}^{t_+}
dt \frac{1}{\rho(t)} =   2 \gamma + 
\frac{1}{3}\gamma^3\frac{d^2}{d\tau^2}
\rho^2(\tau) + O(\gamma^4)
\end{equation}
whence
\begin{equation}
D_{||} = 
\frac{e\gamma}{6}\frac{d^2}{d\tau^2}
\rho^2(\tau) + O(\gamma^2)\;\;.
\end{equation}
Inserting this expression in (9.44) one obtains the result
\begin{equation}
M(-\infty) - M(\infty) = 
\frac{q^2e^2}{180\times 12\pi}(-\tr\;{\hat \Omega}^2)
\int\limits_{-\infty}^{\infty}d\tau 
\Bigl(\frac{d^3}{d\tau^3}\rho^2(\tau)\Bigr)^2
\end{equation}
which holds for the nonrelativistic shell irrespectively of the specific form
of the law $\rho(\tau)$.

It is now important to know the entire history of the shell. We shall assume
that before some time instant $t = t_{\mbox{\scriptsize start}}$ the shell
was kept at the point of maximum contraction $r=r_{\mbox{\scriptsize min}}$
and next was let go. Beginning with $t = t_{\mbox{\scriptsize start}}$ 
it was expanding unboundedly in the self field. This world line is shown
in Fig.7.

Since
\begin{equation}
\frac{d\rho(t)}{dt}\biggl|_{t< t_{\mbox{\scriptsize start}}} = 0\;\;, 
\end{equation}
we have
\begin{equation}
\int\limits_{-\infty}^{\infty} dt \Bigl(\frac{d^3}{dt^3}
\rho^2(t)\Bigr)^2 = 
\int\limits_{t_{\mbox{\scriptsize start}}}^{\infty} dt
\Bigl[(\rho^2)'''\Bigr]^2 = 
\int\limits_{r_{\mbox{\scriptsize min}}}^{\infty} 
\frac{d\rho}{\rho'}\Bigl[(\rho^2)'''\Bigr]^2  
\end{equation}
where the primes denote the derivatives of $\rho(t)$ ,
$r_{\mbox{\scriptsize min}}$ is given in (9.30), and the replacement
of the integration variable $t \to \rho(t)$ is to be made with 
the law $\rho(t)$ in (9.31). It is convenient to write
\begin{equation}
(\rho^2)''' = 6\rho'\rho'' + 2\rho\rho'\frac{d}{d\rho}(\rho'')
\end{equation}
and use the law of motion (9.31):
\begin{equation}
\rho'' = \frac{e^2}{2m}\frac{1}{\rho^2}\;\;,\;\;
\rho' = \sqrt{\frac{2{\cal E}}{m} - \frac{e^2}{m\rho} }\;\;.
\end{equation}
One obtains
\begin{equation}
\int\limits_{r_{\mbox{\scriptsize min}}}^{\infty} 
\frac{d\rho}{\rho'}\Bigl[(\rho^2)'''\Bigr]^2  =
\frac{(2{\cal E})^{7/2}}{e^2m^{5/2}}
\int\limits_1^{\infty} dx
\frac{\sqrt{x-1}}{x^{9/2}}\;\;,
\end{equation}
and
\begin{equation}
\int\limits_1^{\infty} dx
\frac{\sqrt{x-1}}{x^{9/2}}= 2\: \frac{2}{3}\:\frac{2}{5}\:\frac{2}{7}\;\;.
\end{equation}

Finally, restoring $\hbar$ and $c$, we infer that for the whole time of expansion
the nonrelativistic shell looses the energy
\begin{equation}
M(-\infty) - M(\infty) = \frac{(-\tr\;{\hat \Omega}^2)}{
81\times 25 \times 7\pi} \;\frac{q^2}{\hbar c}\Bigl(\frac{2{\cal E}}{mc^2}
\Bigr)^{5/2} 2 {\cal E}\;\;,\;\;\;
{\cal E} << mc^2\;\;.
\end{equation}

\subsection*{\bf Radiation of the ultrarelativistic shell.}

$$ $$

Coming back to the exact expression (9.47) for the longitudinal moment, and
using (9.46) we obtain
\begin{equation}
\frac{dt_+}{d\tau} = \frac{1}{\displaystyle 1 
- \gamma\frac{d\rho}{dt}\biggl|_{t_+}}\;\;,
\;\;
\frac{dt_-}{d\tau} = \frac{1}{\displaystyle 1 
+ \gamma\frac{d\rho}{dt}\biggl|_{t_-}}
\end{equation}
and
\begin{equation}
\frac{d}{d\tau}D_{||} = 
\frac{e}{2\gamma^2}\Biggl(
\frac{1}{\displaystyle \rho\Bigl(1 
- \gamma\frac{d\rho}{dt}\Bigr)\biggl|_{t_+}} - 
\frac{1}{\displaystyle \rho\Bigl(1 
+ \gamma\frac{d\rho}{dt}\Bigr)\biggl|_{t_-}}
\Biggr)
\end{equation}
where the notation assumes that the substitution $t = t_+$ or
$t = t_-$ is to be made.

Since the law $\rho(t)$ is different for $t<t_{\mbox{\scriptsize start}}$
and  $t>t_{\mbox{\scriptsize start}}$ , the range of integration over
$\tau$ should be divided into three intervals,
\begin{equation}
\int\limits_{-\infty}^{\infty}d\tau \Bigl(\frac{d}{d\tau}D_{||}\Bigr)^2 = 
\Bigl(\int\limits_{-\infty}^{\tau_1}d\tau +
\int\limits_{\tau_1}^{\tau_2}d\tau + 
\int\limits^{\infty}_{\tau_2}d\tau\Bigr)
\Bigl(\frac{d}{d\tau}D_{||}\Bigr)^2\;\;,
\end{equation}
according to the location of the points $t_+$ and $t_-$:
\begin{eqnarray}
-\infty < \tau < \tau_1 \;\;\;&:&\;\;\; t_- < t_{\mbox{\scriptsize start}}\;\;\;,
\;\;t_+ < t_{\mbox{\scriptsize start}} \nonumber \\
\tau_1 < \tau < \tau_2 \;\;\;&:&\;\;\; t_- < t_{\mbox{\scriptsize start}}\;\;\;,
\;\;t_+ > t_{\mbox{\scriptsize start}} \nonumber \\
\tau_2 < \tau < \infty \;\;\;&:&\;\;\; t_- > t_{\mbox{\scriptsize start}}\;\;\;,
\;\;t_+ > t_{\mbox{\scriptsize start}}\;\;. \nonumber 
\end{eqnarray}
The motion of the points $t_+$ and $t_-$ along the world line of the shell 
as $\tau$ increases can be traces in Fig. 7. We have
\begin{eqnarray}
\int\limits_{-\infty}^{\tau_1} d\tau \Bigl(\frac{d}{d\tau}D_{||}\Bigr)^2 &=& 
0\;\;,\\
\int\limits_{\tau_1}^{\tau_2} 
d\tau \Bigl(\frac{d}{d\tau}D_{||}\Bigr)^2 &=& 
\frac{e^2}{4\gamma^4}\int\limits_{\tau_1}^{\tau_2} d\tau
\Biggl[
\frac{1}{\displaystyle \rho\Bigl(1 
- \gamma\frac{d\rho}{dt}\Bigr)\biggl|_{t_+}} - 
\frac{1}{r_{\mbox{\scriptsize min}}}
\Biggr]^2\;\;,\\
\int\limits^{\infty}_{\tau_2} 
d\tau \Bigl(\frac{d}{d\tau}D_{||}\Bigr)^2 &=& 
\frac{e^2}{4\gamma^4}\int\limits_{\tau_2}^{\infty} d\tau
\Biggl[
\frac{1}{\displaystyle \rho\Bigl(1 
- \gamma\frac{d\rho}{dt}\Bigr)\biggl|_{t_+}} - 
\frac{1}{\displaystyle \rho\Bigl(1 
+ \gamma\frac{d\rho}{dt}\Bigr)\biggl|_{t_-}}
\Biggr]^2\;\;,
\end{eqnarray}
and, from (9.46),
\begin{equation}
\tau_1 = t_{\mbox{\scriptsize start}} - \gamma r_{\mbox{\scriptsize min}}
\;\;,\;\;
\tau_2 = t_{\mbox{\scriptsize start}} + \gamma r_{\mbox{\scriptsize min}}
\;\;.
\end{equation}
In (9.63) and (9.64) the law $\rho(t)$ pertains already to the 
expansion stage.

Consider a shell that moves all the time with the speed of light. For such
a shell, assuming that it passes through $r = r_{\mbox{\scriptsize min}}$
 at  $t = t_{\mbox{\scriptsize start}}$ , the solutions of Eqs. (9.46) are 
\begin{equation}
\rho(t_{\pm}) = \frac{\tau - t_{\mbox{\scriptsize start}} + 
r_{\mbox{\scriptsize min}}}{1 \mp \gamma}\;\;,
\;\;
\frac{d\rho(t)}{dt}\equiv 1 \;\;,
\end{equation}
and expression (9.60) vanishes identically. Hence we infer that
{\it a null shell creates no real energy from the vacuum}. \footnote{
This is not the case for a gravitationally charged shell 
contracting in the self field.}
Now consider a shell that before 
$t = t_{\mbox{\scriptsize start}}$ is at rest and then suddenly starts
expanding with the speed of light. This shell creates an infinite amount
of energy owing to the jump of its velocity at
$t = t_{\mbox{\scriptsize start}}$ . Indeed, in this case the term (9.64) 
vanishes
\begin{equation}
\int\limits_{\tau_2}^{\infty} d \tau \Bigl(\frac{d}{d\tau}D_{||}\Bigr)^2
\equiv 0
\end{equation}
but the term (9.63) doesn't:
\begin{equation}
\int\limits_{\tau_1}^{\tau_2} d \tau \Bigl(\frac{d}{d\tau}D_{||}\Bigr)^2
= \frac{{\cal E}}{2\gamma^4}\Bigl(\frac{1}{1 - \gamma} - \frac{1}{1 + \gamma} +
2 \log\frac{1 - \gamma}{1 + \gamma} + 2 \gamma\Bigr)\;\;.
\end{equation}
The latter expression behaves like $O(1/\gamma^2)$ at $\gamma = 0$ which in 
view of the presence of $\gamma^2$ in the measure in (9.44) is a regular 
behaviour. However, at $\gamma = 1$ expression (9.68) has a pole, and 
the total energy (9.44) diverges:
\begin{equation}
\Bigl( M(-\infty) - M(\infty)\Bigr) \propto - {\cal E} 
\log(1 - \gamma)\biggl|_{\gamma = 1}\;\;.
\end{equation}
The pole comes from the lower limit in (9.68) at which 
the regime changes from static to null. 
These considerations suggest that, in the case of the timelike 
ultrarelativistic shell, the effect may come only from a neighbourhood of
$r = r_{\mbox{\scriptsize min}}$ and it should be finite owing to the
continuity of the shell's velocity. It is clear in  advance that the divergent 
term (9.69) will be regularized  with the regularization parameter
$mc^2/{\cal E}$ , and the regularized term will be of order 
${\cal E}\log({\cal E}/mc^2)$ . Since this term grows as $mc^2/{\cal E}\to 0$ ,
it will make a dominant contribution to the energy of the particle production.

 Indeed, for the timelike shell, Eq. (9.67) gets replaced with
\begin{equation}
\int\limits_{\tau_2}^{\infty} d \tau \Bigl(\frac{d}{d\tau}D_{||}\Bigr)^2
\to 0\;\;, \;\;\;\frac{mc^2}{\cal E} \to 0
\end{equation}
while the term (9.63) can be expressed directly through the law of motion
(9.32) by making the change $\tau \to t_+(\tau)$ of the integration variable
and inserting the Jacobian from (9.59):
\begin{eqnarray}
\int\limits^{\tau_2}_{\tau_1} d \tau \Bigl(\frac{d}{d\tau}D_{||}\Bigr)^2
&=&
\frac{e^2}{4\gamma^4}\int\limits_{t_+(\tau_1)}^{t_+(\tau_2)}
dt \Bigl( 1 - \gamma\frac{d\rho(t)}{dt}\Bigr)
\Biggl[\frac{1}{\displaystyle \rho(t)\Bigl(1 
- \gamma\frac{d\rho(t)}{dt}\Bigr)} - 
\frac{1}{r_{\mbox{\scriptsize min}}}\Biggr]^2  \\
&=&
\frac{e^2}{4\gamma^4}\int\limits_{r_{\mbox{\scriptsize min}}}^{
\rho(t_+(\tau_2))}
\frac{d\rho}{\rho'} 
\Bigl[\frac{1}{\rho^2(1 - \gamma\rho')} - 
\frac{2}{\rho r_{\mbox{\scriptsize min}}}
+ \frac{(1 - \gamma\rho')}{r^2_{\mbox{\scriptsize min}}}
\Bigr]\;\;.\nonumber
\end{eqnarray}
Here the approximation $\rho'= 1$ can be used in all terms except
in the denominator $(1 - \gamma\rho')$ where the deceleration of the
shell near $r = r_{\mbox{\scriptsize min}}$  cannot be neglected.
Specifically, the upper limit in (9.71) can be calculated in the
approximation $\rho' = 1$ using (9.66) and (9.65):
\begin{equation}
\rho(t_+(\tau_2)) = \frac{1 + \gamma}{1 - \gamma}r_{\mbox{\scriptsize min}}
\;\;,\;\;\;\frac{mc^2}{\cal E} \to 0 \;\;.
\end{equation}
However, for the insertion in the denominator $(1 - \gamma\rho')$ ,
 the law of motion (9.32) should be approximated better:
\begin{equation}
\rho' = \frac{\rho - r_{\mbox{\scriptsize min}}}{\displaystyle
\sqrt{(\rho - r_{\mbox{\scriptsize min}})^2 + 
(r_{\mbox{\scriptsize min}}\frac{mc^2}{\cal E})^2}}\;\;.
\end{equation}
In this way, introducing the integration variable
\begin{equation}
x = \frac{\rho - r_{\mbox{\scriptsize min}}}{r_{\mbox{\scriptsize min}}}\;\;,
\end{equation}
we obtain for $mc^2/{\cal E}\to 0$
\begin{eqnarray}
\int\limits_{-\infty}^{\infty} d \tau \Bigl(\frac{d}{d\tau}D_{||}\Bigr)^2 &=& 
\frac{{\cal E}}{{2\gamma^4}}
\biggl\{ \frac{1}{1 - \gamma} - \frac{1}{1 + \gamma} + 2\log
\frac{1 - \gamma}{1 + \gamma} + 2\gamma \\
& &- \frac{\gamma}{1 - \gamma}
\Bigl(\frac{mc^2}{\cal E}\Bigr)^2
\int\limits_0^{2\gamma/(1 - \gamma)} dx \frac{1}{(1 + x)^2
[(1 - \gamma^2)x^2 + (mc^2/{\cal E})^2]}\biggr\}\;\;.\nonumber
\end{eqnarray}
This expression has already a regular behaviour at both $\gamma = 0$ and 
$\gamma = 1$ (cf. Eq. (9.68)).

Upon doing the integral in (9.75) and next the integral over $\gamma$ one is 
to retain the senior terms at the limit $mc^2/{\cal E}\to 0$ .
These are seen to be the senior terms of (9.75) at the limit
\begin{equation}
(1 - \gamma) \to 0 \;\;,\;\; \frac{mc^2}{\cal E} \to 0\;\;,
\;\; (1 - \gamma)\Bigl(\frac{{\cal E}}{{mc^2}}\Bigr)^2 = \mbox{finite}\;\;.
\end{equation}
At this limit expression (9.75) becomes
\begin{equation}
\int\limits_{-\infty}^{\infty} d\tau \Bigl(\frac{d}{d\tau}D_{||}\Bigr)^2 = 
{\cal E}\int\limits_0^{\infty} dx\frac{x^2}{(1 + x)^2
[2(1 - \gamma)x^2 + (mc^2/{\cal E})^2]}\;\;.
\end{equation}
Hence, interchanging the integrations over $x$ and $\gamma$ , we obtain
\begin{equation}
\int\limits_0^1 d\gamma \gamma^2 \int\limits_{-\infty}^{\infty}
d\tau \Bigl(\frac{d}{d\tau}D_{||}\Bigr)^2 = 
{\cal E}\Bigl(\log\frac{{\cal E}}{{mc^2}}\Bigr)\int\limits_0^{\infty}
dx\frac{1}{(1 + x)^2} + O(1)\;\;,\;\;\; \frac{mc^2}{\cal E} \to 0
\end{equation}
where the contribution comes from the end point $\gamma = 1$.

Thus we infer that for the whole time of expansion the ultrarelativistic shell
looses the energy
\begin{equation}
M(-\infty) - M(\infty) = 
\frac{(-\tr\;{\hat \Omega}^2)}{12\pi}\frac{q^2}{\hbar c}\Bigl(\log
\frac{{\cal E}}{{mc^2}}\Bigr){\cal E}\;\;,\;\;\;
{\cal E}>> mc^2\;\;.
\end{equation}
This is already interesting since, at a sufficiently large ${\cal E}$ ,
the shell radiates more energy than it has. Accounting for the backreaction
of radiation becomes necessary.

Both quantities (9.58) and (9.79) are independent of the charge $e$
of the shell and remain nonvanishing at the limit $e = 0$ . This paradox
is a consequence of the  property of classical theory that the
charge cannot be switched off gradually. The 
backreaction of the vacuum may change this result since the expectation-value
equations contain a dimensionless parameter $e/q$ . Below 
we show that the range 
$e/q{<\atop\sim}1$ is within the limits of validity 
of the approximations made.

\subsection*{\bf Validity of the approximations.}

$$ $$

The calculational scheme presented uses two types of expansion. The effective
action is expanded in the number of loops, 
and within each loop order it is expanded 
over the basis of nonlocal invariants like in Eq. (2.2) \footnote{
In the axiomatic approach of Refs. [19-22] the expansion in loops is
avoided, and expansion (2.2) is considered as an ansatz for the full action.}
. The limits of validity of the latter expansion are presently to be considered. 
These limits are best seen from a comparison of the technique of nonlocal form 
factors with the Schwinger-DeWitt technique [30].

As shown in [34], the nonlocal form factors emerge as a result of a partial
summation of the local Schwinger-DeWitt series for the effective action.
At a given order in the dimensional coupling constants this series can 
symbolically be written down as follows:
\begin{equation}
S_{\mbox{\scriptsize vac}} = \mbox{const.} \int dx g^{1/2}
\sum_p\frac{1}{\mu^{2p}} \sum_{n=1}^{p}(\nabla)^{2p - 2n}\Bigl(\Re\Bigr)^n
\end{equation}
where $\mu$ is the mass of the vacuum particles. The nonlocal expansion
of the effective action is obtained by summing all terms in (9.80) with 
a given power of $\Re$ and any number of derivatives. Within each order
in $\mu^2$ one neglects $\Re^{n+1}$ as compared to 
$\nabla\nabla\Re^n$ . Therefore, the condition of validity of the resultant
nonlocal expansion is generally of the form [34,19]
\begin{equation}
\Re^2 << \nabla\nabla\Re\;\;.
\end{equation}

The partially summed series admits the limit $\mu\to 0$ [34,14-17] and at 
this limit takes the form (2.2). There is no problem of principle in
generalizing the nonlocal expansion to the case of massive vacuum particles
but the neglect of the mass is correct if 
\begin{equation}
\mu^2\Re << \nabla\nabla\Re\;\;.
\end{equation}
The approximation in which Eqs. (9.81) and (9.82) hold may be qualified as 
the high-frequency approximation meaning the frequency of the external 
( or mean ) field.

Not all limitations implied in (9.81) and (9.82) should necessarily be
fulfilled since not all terms neglected in the partial summation of the 
series (9.80) may be important for the problem in question. Thus, the effect of
particle creation is caused by the external field's variability in time, and
its spatial inhomogeneity is unimportant. Let $l$ be the characteristic 
spatial scale of the source of the external field and $\nu$ be its 
characteristic frequency. Then, for the problem of particle creation, the
important limitation stemming from (9.82) is 
\begin{equation}
\mu c^2 << \hbar \nu\;\;.
\end{equation}
 Under this condition the results of the technique of nonlocal form factors 
 are valid for massive vacuum particles as well.

When specializing to the problem of particle creation by an external 
 electric field one is to consider only the mixed components of the
 commutator curvature ${\R}_{\mu\nu}$ . For these components condition (9.81)
 yields
 \begin{equation}
 \R_{0i} << \frac{\nu}{l}\;\;,
 \end{equation}
 and, by (9.34),
 \begin{equation}
 \R_{0i} = \mbox{const.}\; \frac{q}{\hbar}E
 \end{equation}
where $E$ is the characteristic strength of the external electric field:
\begin{equation}
E = \frac{e}{l^2}\;\;.
\end{equation}
Thus, for this problem, the condition of validity of the nonlocal expansion 
is
\begin{equation}
\frac{qe}{l} << {\hbar}\nu
\end{equation}
where the quantity on the left-hand side is the Coulomb energy of the vacuum
particle in the external electric field. Condition (9.87) is to be combined
with the condition of validity of the one-loop approximation:
\begin{equation}
\frac{q^2}{\hbar c}<<1\;\;.
\end{equation}

For the nonrelativistic shell, $l$ and $\nu$ are unrelated:
\begin{equation}
l = \frac{e^2}{\cal E}\;\;,\;\; \nu = \frac{{\cal E}^{3/2}}{m^{1/2}e^2}\;\;,
\;\;{\cal E}<< mc^2\;\;.
\end{equation}
In this case the energy  of the shell is clutched by the inequalities
(9.87) and (9.89) in the interval
\begin{equation}
\frac{q^2}{\hbar c}\;\frac{e}{q} << \sqrt{\frac{{\cal E}}{{mc^2}}} << 1\;\;.
\end{equation}
Owing to condition (9.88) this still leaves room. It is only important
that the ratio $e/q$ be not too large:
\begin{equation}
\frac{q^2}{\hbar c}\;\frac{e}{q} << 1\;\;.
\end{equation}
The weaker the coupling of the quantum field, the bigger charge of the 
external field can be considered. Thus, for the electron-positron vacuum,
$e = 10q$ is admissible but $e = 137q$ is not.

For the ultrarelativistic sources we have $\nu l = c$ , and  the inequality 
(9.87) takes again the form (9.91). Thus condition (9.91) is universal but the
relativistic case is better suited for the approximations made since no upper
bound on the energy ${\cal E}$ emerges.
}

\newpage

{\renewcommand{\theequation}{10.\arabic{equation}}

\begin{center}
\section*{\bf Concluding remarks}
\end{center}

$$ $$

This section will not be long since it may be called "reminiscences of the
future" with the exception of the following point.

As remarked in Sec.2, Eqs. (2.38)-(2.45) pertain to the case where
$\R_{\alpha\beta}$ and ${\hat P}$ are metric independent but {\it the final
result} (8.4) {\it is valid for arbitrary local} $\R_{\alpha\beta}$ {\it 
and} ${\hat P}$ . An inspection of the table of asymptotic behaviours in
[18] shows that, in the contributions of the variational derivatives
$\delta\R_{\alpha\beta}/\delta g_{\mu\nu}$ and $\delta{\hat P}/\delta 
g_{\mu\nu}$  to $T^{\mu\nu}_{\mbox{\scriptsize vac}}\Bigl|_{{\cal I}^+}$ , 
the highest
exponents $m$ and $n$  of the vertex operators $F(m,n)$ are by one smaller
than in (2.40)-(2.43). The contributions of these variational 
derivatives are, therefore,
{\it pure quantum noise}. This remarkable fact gives the result (8.4) the status
of a generating expression applicable to fields of any spin [24].

The remaining limitations imposed on the external fields require bigger
efforts for their removal.

The limitation (8.5) signals that the theory contains another effect:
the vacuum screening or amplification of the electromagnetic waves emitted by 
a source. This is equivalent to an {\it observable} renormalization 
of all multipole
moments. The emergence of the limitation on the vector field is connected
with the fact that, in the case of this field, the nonlocal vacuum polarization
and the local charge renormalization are of one and the same order
by dimension. The separation of the respective terms in the effective action is 
quite subtle (this issue is discussed for QED in [35]). Hence it is clear
why no limitation like (8.5) emerges in the case of the gravitational field.
It is also clear what should be done. Since there appear nonlocal sources
that behave at ${{\cal I}^+}$ like $O(1/r^2)$ rather than $O(1/r^3)$ , two
amendments are needed. First, terms $O(\Box^0)$ of the form factors can
no more be discarded. They should be extracted from the exact form factors
[17] and added to (2.36). Second, expression (3.14) for the kernel of
$\log(-\Box)$ is no more valid and should be replaced by the exact expression
[21,22]. It can be expected that, after introducing these amendments,
the difficulty with the convergence at ${{\cal I}^+}$ will be removed
along with the limitation (8.5), and the result will be the following
expression for the density of the radiation flux from the vector source:
\begin{eqnarray*}
-\frac{1}{4\pi} Z g_{\alpha\beta} \Bigl(\frac{d}{d\tau}{\hat D}^{\alpha}
\Bigr)\Bigl(\frac{d}{d\tau}{\hat D}^{\beta}\Bigr)\biggl|_{\gamma=1}
\hspace{9 cm} \\ \qquad
\qquad{}-\frac{1}{3{(4\pi)^2}}\int\limits_0^1 d\gamma\gamma^2
\frac{1}{(1-\gamma^2)}\Bigl[ g_{\alpha\beta}\Bigl(\frac{d}{d\tau}
{\hat D}^{\alpha}\Bigr)\Bigl(\frac{d}{d\tau}{\hat D}^{\beta}\Bigr)-
g_{\alpha\beta}\Bigl(\frac{d}{d\tau}{\hat D}^{\alpha}\Bigr)
\Bigl(\frac{d}{d\tau}{\hat D}^{\beta}\Bigr)\biggl|_{\gamma=1}\Bigr]
\end{eqnarray*}
(cf. Eqs. (8.4) and (8.33)). Here $Z$ is the missing renormalization constant.
This effect is analogous to the effect of the vacuum gravitational waves
[23]. The difference is only in the ways in which these effects appear
in the formalism and in the dimensions of the coupling constants. The
dimension of the coupling constant causes that the gravitational effect
does not boil down to a mere renormalization [23].

Removing the limitation that the metric contains no horizon is a separate 
problem since in this case the total radiation energy is infinite  but
the reason why it becomes infinite can be pointed out straight away.
What becomes wrong is the assumption of asymptotic stationarity of the
sources in the future which leads to Eq. (7.35) and hence to Eq. (7.36).
Since $\tau$ is an external time, as $\tau\to\infty$ the sources moving
in the tube hit the event horizon. Therefore, the asymptotic stationarity
gets replaced by the condition that at $\tau\to\infty$ the sources remain
finite along with their internal derivatives. The growth of the vertex
function at $i^+$ should then become by one power of $s$ faster.

Finally, the assumption that the support of the physical sources is
confined to a spacetime tube may seem technical and minor but is
in fact physical and inadmissibly restrictive. As pointed out in Sec.4,
the significance of this assumption is in the fact that it excludes
the radiation of charge. In the external-field problem this covers
interesting cases but in the self-consistent problem this is merely
wrong since what has been calculated above is exactly the radiation
of the gravitational charge. Obtaining the backreaction equations
is impossible without letting the sources out of the tube.
}

\newpage

\begin{center}
\section*{\bf Acknowledgments}
\end{center}

$$ $$

A work in some respect close to the present one can be found in [36].
The authors are grateful to B.S. DeWitt for pointing out this reference.
G.V. is grateful to V.D. Skarzhinsky, S.A. Surkov, and M.I. Zelnikov
for urgently initiating him in computer mysteries and thereby making
the appearance of this paper possible. The present work was supported 
in part by the Russian Foundation for Fundamental Research Grant 96-02-16295 
and INTAS Grant 93-493-ext.

\newpage
\appendix
\renewcommand{\thesection}{Appendix\enskip\Alph{section}.}

{\renewcommand{\theequation}{A.\arabic{equation}}

\section{\bf  Identities for the vertex operators.
Use of the Jacobi identities.}

$$ $$

The functions $A_i^{\; k} (\Box_m, \Box_n )$ and
$B_i^{\; k} (\Box_m, \Box_n)$ in the asymptotic expansions of the
form factors (2.33) are given in Ref.[18] as linear combinations
with rational coefficients of the vertex operators (2.44)
\begin{equation}
F(k_m, k_n) \equiv \Bigl(\frac{\partial}{\partial j_m}\Bigr)^{k_m}
\Bigl(\frac{\partial}{\partial j_n}\Bigr)^{k_n}
\frac{\log(j_m \Box_m / j_n \Box_n)}{j_m\Box_m  -j_n\Box_n}
\biggl |_{j=1} \quad , \quad m\ne n \quad .
\end{equation}
Expression (A.1) can also be written in the form
\begin{equation}
F(k_m,k_n) = \Box_m^{\;k_m}\Box_n^{\;k_n}\Bigl(\frac{\partial}{\partial \Box_m}
\Bigr)^{k_m}\Bigl(\frac{\partial}{\partial \Box_n}\Bigr)^{k_n}
\frac{\log(\Box_m/\Box_n)}{\Box_m-\Box_n}\quad .
\end{equation}

The results in [18] simplify drastically if one takes into account the
constraints that exist between the functions
(A.1) with different $k_m$ and $k_n$ . The constraint equation reads
\begin{equation}
F(k_m, k_n+1) +F(k_m+1,k_n)=-(k_m+k_n+1)F(k_m,k_n)
\end{equation}
and is obtained by acting with the operator
\begin{equation}
\Box_m^{\;k_m}\Box_n^{\;k_n}\Bigl(\frac{\partial}{\partial \Box_m}
\Bigr)^{k_m}\Bigl(\frac{\partial}{\partial \Box_n}\Bigr)^{k_n}
\end{equation}
on the easily verifiable identity
\begin{equation}
\Bigl(\Box_m\frac{\partial}{\partial\Box_m} + \Box_n
\frac{\partial}{\partial \Box_n}\Bigr) 
\frac{\log(\Box_m/\Box_n)}{\Box_m-\Box_n}\equiv
-\frac{\log(\Box_m/\Box_n)}{\Box_m-\Box_n}\quad .
\end{equation}

The first step in simplifying the functions $A_i^{\; k} (\Box_m,\Box_n)$
 and $B_i^{\; k} (\Box_m,\Box_n)$ is a removal of the dimensionless
coefficients $\Box_m/\Box_n$  that some of these functions have in front
of $F(k_m,k_n)$ [18]. This removal can be done so that there remain
linear combinations of $F(k_m,k_n)$ with numerical coefficients and
rational additions. Indeed, by writing
\begin{equation}
\frac{\Box_m}{\Box_n} F(k_m,k_n)\equiv \frac{\Box_m-\Box_n}{\Box_n}
F(k_m,k_n) + F(k_m,k_n)
\end{equation}
and commuting $(\Box_m-\Box_n)$ with the derivatives $\partial /\partial \Box$
 in (A.2), one obtains
\begin{eqnarray}
\frac{\Box_m-\Box_n}{\Box_n}F(0,k_n) &=& (-1)^{k_n}\frac{(k_n-1)!}{\Box_n}
+k_nF(0,k_n-1)\quad , \quad k_n\ne 0 \quad ,
\\
\frac{\Box_m-\Box_n}{\Box_n}F(k_m,k_n) &=& -k_n\frac{\Box_m-\Box_n}{\Box_n}
F(k_m-1,k_n)+(k_m+
k_n)F(k_m,k_n-1)
\\ &&
{}+k_m(k_m+k_n-1)F(k_m-1,k_n-1)\quad ,\quad k_m\ne 0\;,\; k_n\ne 0
\quad. \nonumber
\end{eqnarray}
The latter identity applied $k_m$ times with a subsequent use of (A.7)
and (A.3) brings to the following reduction formula:
\begin{equation}
\frac{\Box_m-\Box_n}{\Box_n}F(k_m,k_n)=(k_m+k_n)F(k_m,k_n-1)+
(-1)^{k_m+k_n}\frac{k_m!(k_n-1)!}{\Box_n}\quad,\quad k_m\ne 0\;,\; k_n\ne 0
\quad .
\end{equation}
This formula is useful also in the case where there is a dimensional
coefficient $1/\Box_n$ in front of $F(k_m,k_n)$ [18] since it enables one
to convert the $1/\Box_n$ into the $1/\Box_m$ and vice versa. For that,
it suffices to multiply Eqs. (A.7) and (A.9) by $1/\Box_m$ .

The relations above make it possible to bring the asymptotic form factors
in [18] to their final form in (2.39)-(2.45). The strategy of this calculation is
as follows. First, the relations (A.6)-(A.9) are used to get rid of the
coefficients $\Box_m/\Box_n$ in front of $F(k_m,k_n)$ and, if needed, to
replace the coefficients $1/\Box_n$ with $1/\Box_m$ . Next, the relation
(A.3) is used in the thus obtained linear combinations of $F(k_m,k_n)$ to
reduce the difference between $k_m$ and $k_n$ . For example,
\begin{equation}
F(0,4)+F(4,0)=2F(2,2)-48F(1,1)+24F(0,0)\quad .
\end{equation}
The equality of the exponents $k_m$ and $k_n$ ensures ultimately the
positive definiteness of the total radiation energy.

For illustration, consider two specific examples of the functions
$B_i^{\; k}$ in [18]:
\begin{equation}
B_{12}{}^1(\Box_2,\Box_3)=-\frac{2}{3\Box_2\Box_3}-\frac{1}{3\Box_2}F(1,0)
-\frac{1}{3\Box_3}F(0,1)
\end{equation}
and
\begin{eqnarray}
B_7{}^1(\Box_2,\Box_3)=\frac{1}{24}F(1,0)+\frac{1}{24}F(0,1)+\frac{1}{96}
F(2,0)+\frac{1}{96}F(0,2)+\frac{19}{48}F(1,1) \hspace{35 mm} \\ 
\quad{} -\frac{1}{24}F(3,1)
-\frac{1}{24}F(1,3) 
+\frac{\Box_3}{\Box_2}\Bigl(\frac{1}{96}F(2,0)-
\frac{1}{96}F(1,1)\Bigr)+\frac{\Box_2}{\Box_3}\Bigl(-\frac{1}{96}F(1,1)
+\frac{1}{96}F(0,0)\Bigr)\quad .\nonumber
\end{eqnarray}
In (A.11), make the factors $1/\Box$ in front of the $F$ 's uniform, and use
(A.3). As a result, $B_{12}{}^1$ proves to be a tree operator:
\begin{equation}
B_{12}{}^1(\Box_2,\Box_3)=-\frac{1}{3\Box_2\Box_3}\quad .
\end{equation}
In (A.12), first remove the factors $\Box_3/\Box_2$ , $\Box_2/\Box_3$ with
the aid of (A.7) and (A.9), and next use (A.3). The result is
\begin{equation}
B_7{}^1(\Box_2,\Box_3)=\frac{1}{12}F(2,2)-\frac{1}{6}F(1,1)\quad .
\end{equation}

Apart from the simplification of the form factors, the expression for the effective
action in [18] needs to be improved in the terms with the commutator curvature.
In the work [15-18], the basis of nonlocal gravitational invariants was
expressed from the outset through the source of the metric curvature by the
use of the Bianchi identities. It is an omission of this work that a similar
procedure has not been applied to the commutator curvature. By differentiating
and contracting the Jacobi identities (1.5) one obtains the equation
\begin{equation}
\Box \R_{\alpha\beta}=\nabla^{\gamma}\nabla_{\beta}\R_{\alpha\gamma}
-\nabla^{\gamma}\nabla_{\alpha}\R_{\beta\gamma}
\end{equation}
which can be solved iteratively to express the commutator curvature
$\R_{\alpha\beta}$ through its source ${\hat J}_{\alpha}=\nabla^{\gamma}
\R_{\alpha\gamma}$ and the initial data. The iteration procedure is started
by commuting the covariant derivatives on the right-hand side of (A.15)
with the aid of Eq. (1.10). Neglecting the commutators, one obtains to
lowest order in $\Re$
\begin{equation}
\Box \R_{\alpha\beta}=\nabla_{\beta}{\hat J}_{\alpha}-\nabla_{\alpha}
{\hat J}_{\beta}+O[\Re^2]\quad .
\end{equation}
The solution of this equation with the retarded Green function:
\begin{equation}
\R_{\alpha\beta}=\frac{1}{\Box}\Bigl(\nabla_{\beta}{\hat J}_{\alpha}-
\nabla_{\alpha}{\hat J}_{\beta}\Bigr) +O[\Re^2]
\end{equation}
is the solution with no incoming wave of the vector connection field.
(See Ref.[23] for the proof of a similar assertion in the case of
the gravitational field.)

From the results in [18] (with the vertex operators treated as above),
the contribution of the commutator curvature to the scalar $I_2$,
Eq. (2.39), obtains originally in the form
\begin{equation}
\frac{1}{2}f(\Box_1,\Box_2)\biggl\{\Bigl(\frac{1}{\Box_1}+\frac{1}{\Box_2}
\Bigr)\nabla_{\alpha}\R_1^{\;\alpha\lambda}\nabla_{\beta}
\R_2^{\;\beta}{}_{\lambda} +\R_1^{\;\alpha\beta}\R_{2\;\alpha\beta}\biggr\}
\end{equation}
where
\begin{equation}
f(\Box_1,\Box_2)=\frac{1}{6}F(2,2)-\frac{1}{3}F(1,1)\quad .
\end{equation}
The form (A.18) is redundant because the commutator curvature is
constrained by the Jacobi identities. By using (A.17), one finds
\begin{equation}
\R_1^{\;\alpha\beta}\R_{2\;\alpha\beta}=2\frac{1}{\Box_1\Box_2}
\nabla_{\beta}{\hat J}_1^{\;\alpha}\nabla^{\beta}{\hat J}_{2\;\alpha}-
2\frac{1}{\Box_1\Box_2}\nabla_{\beta}{\hat J}_1^{\;\alpha}\nabla_{\alpha}
{\hat J}_2^{\;\beta}
\end{equation}
where in the first term use can be made of the relation [15]
\begin{equation}
2\nabla_1\nabla_2=(\nabla_1+\nabla_2)^2-\Box_1-\Box_2\quad .
\end{equation}
The contribution of the operator $(\nabla_{1}+\nabla_{2})^2$ to (A.20) is
of the form
\begin{equation}
\frac{1}{\Box_1\Box_2}(\nabla_1+\nabla_2)^2{\hat J}_1^{\;\alpha}
{\hat J}_{2\;\alpha}\equiv \Box\Bigl[\Bigl(\frac{1}{\Box}
{\hat J}^{\alpha}\Bigr)\Bigl(\frac{1}{\Box}{\hat J}_{\alpha}\Bigr)\Bigr]
\quad .
\end{equation}
Since here appears an overall $\Box$ operator acting at the observation
point, the respective contribution to (A.18) can be discarded (see Sec.2).
As a result, expression (A.18) takes the form
\begin{equation}
-\frac{f(\Box_1,\Box_2)}{\Box_1\Box_2}\nabla_{\beta}{\hat J}_1^{\;\alpha}
\nabla_{\alpha}{\hat J}_2^{\;\beta}\quad .
\end{equation}

Finally, after the scalar $I_2$ in (2.39) has been expressed as a quadratic
combination of the conserved currents, the derivatives acting on the
individual currents can all be made overall owing to the conservation
laws (1.9). Thus, up to $O[\Re^3]$
\begin{eqnarray}
\Bigl(\nabla_{\beta}{\hat J}_1^{\;\alpha}\Bigr)\Bigl(\nabla_{\alpha}
{\hat J}_2^{\;\beta}\Bigr) &=& \nabla_{\beta}\nabla_{\alpha}
\Bigl({\hat J}_1^{\;\alpha}{\hat J}_2^{\;\beta}\Bigr)\quad ,
\\
\Bigl(\nabla_{\alpha}\nabla_{\beta} J_1^{\;\mu\nu}\Bigr)
\Bigl(\nabla_{\mu}\nabla_{\nu} J_2^{\;\alpha\beta}\Bigr) &=& 
\nabla_{\alpha}\nabla_{\beta}\nabla_{\mu}\nabla_{\nu}
\Bigl( J_1^{\;\mu\nu} J_2^{\;\alpha\beta}\Bigr)\quad .
\end{eqnarray}
In this way the final result (2.39)-(2.45) is obtained.
}

\newpage

{\renewcommand{\theequation}{B.\arabic{equation}}

\section{\bf  Alternative approach to the non-scalar vertices.}

$$ $$

There is an alternative to the special consideration of the non-scalar
vertices in Secs.4 and 7. It is based on the fact that up to terms
$O[\Re^3]$  the overall derivatives in (2.39) can be commuted with
the operator $\log(-\Box)$ . Retaining only the vertex terms we have
from (2.36) and (2.39)
\begin{eqnarray}
T^{\mu\nu}_{\mbox{\scriptsize vac}}\biggl |_{{\cal I}^+} &=& 
\frac{1}{(4\pi)^2}\;\tr\;\Biggl\{\nabla^{\mu}\nabla^{\nu}  \log(-\Box)
{\hat V}_{\mbox{\scriptsize scalar}}\nonumber
\\ && \quad\qquad\;{}
+\nabla^{\mu}\nabla^{\nu}\nabla_{\alpha}\nabla_{\beta} \log(-\Box)
\Bigl({\hat V}^{\alpha\beta}_{\mbox{\scriptsize cross}}+
{\hat V}^{\alpha\beta}_{\mbox{\scriptsize vect}}\Bigr)\nonumber
\\ && \quad\qquad\;{}
+\nabla^{\mu}\nabla^{\nu}\nabla_{\alpha}\nabla_{\beta}\nabla_{\gamma}
\nabla_{\sigma} \log(-\Box){\hat V}^{\alpha\beta\gamma\sigma}_{\mbox{
\scriptsize grav}}
+\mbox{trees} \Biggr\}\quad .
\end{eqnarray}
The derivatives act now at a point at ${\cal I}^+$ and, moreover, we
need only the senior terms of expansions like (4.59) and (4.76):
\begin{equation}
\nabla^{\mu}\nabla^{\nu}\;\to\;\nabla^{\mu}u\nabla^{\nu} u
\frac{\partial^2}{\partial u^2}\quad ,\quad \nabla^{\mu}\nabla^{\nu}
\nabla_{\alpha}\nabla_{\beta}\;\to\;\nabla^{\mu}u\nabla^{\nu}u
\nabla_{\alpha}u\nabla_{\beta}u \frac{\partial^4}{\partial u^4}
\quad ,\quad \mbox{etc.}
\end{equation}
Thus, making use of Eqs. (1.36) and (4.18), we obtain
\begin{eqnarray}
M(-\infty)-M(\infty) & = & -\lim_{u\to\infty}\frac{2}{(4\pi)^2}\;
\int d^2{\cal S}(\phi)\;
\tr\;\Biggl\{  \frac{\partial^2}{\partial u^2}D_{\1}(u,\phi|
{\hat V}_{\mbox{\scriptsize scalar}})\nonumber
\\  && {}
+\frac{\partial^4}{\partial u^4}D_{\1}(u,\phi|{\hat V}^{\alpha\beta}_{
\mbox{\scriptsize cross}})\nabla_{\alpha} u\nabla_{\beta} u
+\frac{\partial^4}{\partial u^4}D_{\1}(u,\phi|{\hat V}^{\alpha\beta}_{
\mbox{\scriptsize vect}})\nabla_{\alpha} u\nabla_{\beta} u\nonumber
\\ && {}
+\frac{\partial^6}{\partial u^6}D_{\1}(u,\phi|
{\hat V}^{\alpha\beta\gamma\sigma}_{\mbox{\scriptsize grav}})
\nabla_{\alpha} u\nabla_{\beta} u\nabla_{\gamma} u\nabla_{\sigma} u
\Biggr\}\quad .
\end{eqnarray}

The price for getting rid of the derivatives is that now we have to deal
with the moments $D_{\1}$ of the non-scalar vertices. Since these vertices
have different powers of growth at $i^+$, Eq. (6.41) is to be applied
in each case separately. Let us introduce a notation for the tensors
at $i^+$ that figure in the asymptotic expressions for the vertices,
Eqs. (7.40)-(7.42):
\begin{eqnarray}
\Delta^{\alpha\beta}_{\mbox{\scriptsize cross}}(x)\biggl |_{i^+} &=& 
\frac{1}{48}\int\limits_{-\infty}^{\infty} d\tau \Bigl(
\frac{d^2}{d\tau^2}D^{\alpha\beta}\Bigr)\Bigl(\frac{d^2}{d\tau^2}
{\hat D}^Q\Bigr)\quad ,
\\
\Delta^{\alpha\beta}_{\mbox{\scriptsize vect}}(x)\biggl |_{i^+} &=&
\frac{1}{48(1-{\gamma}^2)}\int\limits_{-\infty}^{\infty} d\tau \Bigl(
\frac{d}{d\tau}{\hat D}^{\alpha}\Bigr)\Bigl(\frac{d}{d\tau}
{\hat D}^{\beta}\Bigr)\quad ,
\\
\Delta^{\alpha\beta\gamma\sigma}_{\mbox{\scriptsize grav}}(x)
\biggl |_{i^+} &=& -\frac{{\hat 1}}{180\times 32}
\int\limits_{-\infty}^{\infty} d\tau \Bigl(\frac{d^2}{d\tau^2}
D^{\alpha\beta}\Bigr)\Bigl(\frac{d^2}{d\tau^2}D^{\gamma\sigma}\Bigr)\quad .
\end{eqnarray}
With the behaviours of the $V$ 's in (7.40)-(7.42), the algorithm (6.41)
yields
\begin{eqnarray}
D_{\1}(u,\phi|V^{\alpha\beta})\biggl |_{u\to\infty}=\frac{u^4}{4\pi}
\int\limits_0^1 d\gamma\,\gamma^2(1-\gamma^2)^3\int d^2 {\cal S}
({\bar\phi})\,(1-\gamma n{\bar n})^{-5}\nonumber
\\ 
\times \Bigl(g^{\alpha}_{\;\;{\bar \alpha}}(x,{\bar x})
g^{\beta}_{\;\;{\bar \beta}}(x,{\bar x})\Delta^{{\bar \alpha}{\bar\beta}}
({\bar x})\Bigr)\Biggr |_{\vbox{\hbox{$x\to {\cal I}^+[u,\phi]$}
\hbox{${\bar x}\to i^+[\gamma,{\bar \phi}]$}}}
\end{eqnarray}
for both ${\hat V}^{\alpha\beta}_{\mbox{\scriptsize cross}}$ and
${\hat V}^{\alpha\beta}_{\mbox{\scriptsize vect}}$ , and
\begin{eqnarray}
D_{\1}(u,\phi|V^{\alpha\beta\gamma\sigma})\biggl |_{u\to\infty}=
\frac{u^6}{4\pi}\int\limits_0^1 d\gamma\,\gamma^2(1-\gamma^2)^4
\int d^2 {\cal S}({\bar \phi})\,(1-\gamma n {\bar n})^{-7}\nonumber
\\
\times \Bigl(g^{\alpha}_{\;\;{\bar \alpha}}(x,{\bar x})
g^{\beta}_{\;\;{\bar\beta}}(x,{\bar x})
g^{\gamma}_{\;\;{\bar\gamma}}(x,{\bar x})
g^{\sigma}_{\;\;{\bar\sigma}}(x,{\bar x})
\Delta^{{\bar\alpha}{\bar\beta}{\bar\gamma}{\bar\sigma}}({\bar x})\Bigr)
\Biggl |_{\vbox{\hbox{$x\to{\cal I}^+[u,\phi]$}
\hbox{${\bar x}\to i^+[\gamma,{\bar\phi}]$}}}
\end{eqnarray}
for ${\hat V}^{\alpha\beta\gamma\sigma}_{\mbox{\scriptsize grav}}$ .
Here the appearance of 
\begin{equation}
n{\bar n}=n_i(\phi)n^i({\bar \phi})
\end{equation}
is owing to Eq. (4.16), and we didn't forget {\it the propagators of
parallel transport} connecting the point at ${\cal I}^+$ with the
point at $i^+$ .

The powers of $u$ in expressions (B.7) and (B.8) are just the ones
needed for the quantity (B.3) to be finite and nonvanishing but the
weight of calculation transfers now to carrying out the integration
over the directions of radiation i.e. over the angles $\phi$ at
${\cal I}^+$ . As seen from (B.3) and (B.7)-(B.8), one is to do
the integrals
\begin{eqnarray}
{\Pi}_{{\bar\alpha}{\bar\beta}}({\bar x})\biggl |_{i^+} & = &
\int d^2 {\cal S}(\phi)\, (1-\gamma n{\bar n})^{-5} 
\\
& \times &  \Bigl(\nabla_{\alpha}u(x)\nabla_{\beta}u(x)
g^{\alpha}_{\;\;{\bar\alpha}}(x,{\bar x})
g^{\beta}_{\;\;{\bar \beta}}(x,{\bar x})\Bigr)
\Biggl |_{\vbox{\hbox{$x\to {\cal I}^+[u,\phi]$}
\hbox{${\bar x}\to i^+[\gamma,{\bar \phi}]$}}}\nonumber \quad ,
\\
{\Pi}_{{\bar\alpha}{\bar\beta}{\bar\gamma}{\bar\sigma}}({\bar x})
\biggl |_{i^+} & = & \int d^2 {\cal S}(\phi)\, (1-\gamma n{\bar n})^{-7}
\\
& \times &  \Bigl( \nabla_{\alpha}u(x)\nabla_{\beta}u(x)\nabla_{\gamma}u(x)
\nabla_{\sigma}u(x)g^{\alpha}_{\;\;{\bar\alpha}}(x,{\bar x})
g^{\beta}_{\;\;{\bar\beta}}(x,{\bar x})
g^{\gamma}_{\;\;{\bar\gamma}}(x,{\bar x})
g^{\sigma}_{\;\;{\bar\sigma}}(x,{\bar x})
\Bigr)
\Biggl |_{\vbox{\hbox{$x\to {\cal I}^+[u,\phi]$}
\hbox{${\bar x}\to i^+[\gamma,{\bar \phi}]$}}}.\nonumber 
\end{eqnarray}
Since both points of the propagators
$g^{\alpha}_{\;\;{\bar\alpha}}(x,{\bar x})$ are at the future infinity,
one may use the flat-spacetime expressions for these propagators. The
integration is conveniently carried out in the Minkowski frame (1.25)
for both points $x$ and ${\bar x}$ . In this frame,
\begin{equation}
g^{\alpha}_{\;\;{\bar\alpha}}(x,{\bar x})={\delta}^{\alpha}_{\;\;{\bar\alpha}}
\quad , \quad {\nabla}_{\alpha}u(x)={\delta}^0_{\alpha}-{\delta}^i_{\alpha}
n_i\quad,
\end{equation}
and there emerge only integrals of the form
\begin{equation}
\int d^2 {\cal S}(\phi)\, (1-\gamma n{\bar n})^{-p}n_i \cdots  n_j
\end{equation}
summarized in the table below.

When contracting the $\Pi$ 's obtained in the Minkowski frame with the
$\Delta$ 's in Eqs. (B.4)-(B.6), it should be taken into account that
the Minkowski components of any of the $\Delta$ 's can be expressed
as follows:
\begin{eqnarray}
\Delta^{0\cdots}=\gamma \nabla_{\alpha}r\Delta^{\alpha\cdots}\quad ,
\quad n_i\Delta^{i\cdots}=\nabla_{\alpha}r\Delta^{\alpha\cdots}\quad ,
\nonumber \\
\delta_{ik}\Delta^{ik\cdots}=(g_{\alpha\beta}+\gamma^2\nabla_{\alpha}r
\nabla_{\beta}r)\Delta^{\alpha\beta\cdots}
\end{eqnarray}
by virtue of the conservation equations (7.23),(7.24).

In this way after many cancellations we obtain
\begin{equation}
\Pi_{\alpha\beta}\Delta^{\alpha\beta}(x)\biggl |_{i^+} =
\frac{4\pi}{3} \frac{1}{(1-\gamma^2)^3}g_{\alpha\beta}
\Delta^{\alpha\beta}(x)\biggl |_{i^+}
\end{equation}
for both $\Delta^{\alpha\beta}_{\mbox{\scriptsize cross}}$ and
$\Delta^{\alpha\beta}_{\mbox{\scriptsize vect}}$ , and
\begin{equation}
\Pi_{\alpha\beta\gamma\sigma}\Delta^{\alpha\beta\gamma\sigma}
(x)\biggl |_{i^+} =
\frac{4\pi}{5}\frac{1}{(1-\gamma^2)^4}g_{\alpha\beta}g_{\gamma\sigma}
\Delta^{(\alpha\beta\gamma\sigma)}(x)\biggl |_{i^+}
\end{equation}
for $\Delta^{\alpha\beta\gamma\sigma}_{\mbox{\scriptsize grav}}$
where the latter tensor appears symmetrized. The use of these
contractions in (B.7),(B.8) and (B.3) yields the same final result as in 
the main text, Eq. (8.4).

The integrals (B.13) are elementary but it is useful to have them.
Therefore, we present the ones needed for the calculation above.
Denoting
\begin{eqnarray*}
n\gamma=n_i(\phi)\gamma^i\quad , \quad \gamma_i=\delta_{ik}\gamma^k
\quad , \quad \gamma^2=\delta_{ik}\gamma^i\gamma^k 
\end{eqnarray*}
we have
\begin{eqnarray*}
\frac{1}{4\pi}\int d^2 {\cal S}(\phi)\, \frac{1}{(1-n\gamma)^3} &=&
\frac{1}{(1-\gamma^2)^2}\quad , \\
\frac{1}{4\pi}\int d^2 {\cal S}(\phi)\, \frac{1}{(1-n\gamma)^4} &=&
\frac{1}{3}\frac{(3+\gamma^2)}{(1-\gamma^2)^3}\quad , \\
\frac{1}{4\pi}\int d^2 {\cal S}(\phi)\, \frac{1}{(1-n\gamma)^5} &=&
\frac{(1+\gamma^2)}{(1-\gamma^2)^4}\quad , \\
\frac{1}{4\pi}\int d^2 {\cal S}(\phi)\, \frac{1}{(1-n\gamma)^6} &=&
\frac{1}{5}\frac{(5+10\gamma^2+\gamma^4)}{(1-\gamma^2)^5}\quad , \\
\frac{1}{4\pi}\int d^2 {\cal S}(\phi)\, \frac{1}{(1-n\gamma)^7} &=&
\frac{1}{3}\frac{(1+3\gamma^2)(3+\gamma^2)}{(1-\gamma^2)^6}\quad , \\
\frac{1}{4\pi}\int d^2 {\cal S}(\phi)\, \frac{n_i}{(1-n\gamma)^5} &=&
\frac{1}{3}\frac{(5+\gamma^2)}{(1-\gamma^2)^4}\gamma_i\quad , \\
\frac{1}{4\pi}\int d^2 {\cal S}(\phi)\, \frac{n_i n_k}{(1-n\gamma)^5} &=&
\frac{1}{3}\frac{1}{(1-\gamma^2)^3}\Bigl(\delta_{ik}
+6\frac{\gamma_i\gamma_k}{1-\gamma^2}\Bigr)\quad , \\
\frac{1}{4\pi}\int d^2 {\cal S}(\phi)\, \frac{n_i}{(1-n\gamma)^7} &=&
\frac{1}{15}\frac{(35+42\gamma^2+3\gamma^4)}{(1-\gamma^2)^6}
\gamma_i\quad , \\
\frac{1}{4\pi}\int d^2 {\cal S}(\phi)\, \frac{n_i n_k}{(1-n\gamma)^7} &=&
\frac{1}{15}\frac{(5+3\gamma^2)}{(1-\gamma^2)^5}\delta_{ik} 
 + \frac{8}{15}\frac{(7+3\gamma^2)}{(1-\gamma^2)^6}
\gamma_i\gamma_k\quad , \\
\frac{1}{4\pi}\int d^2 {\cal S}(\phi)\, \frac{n_i n_k n_p}{(1-n\gamma)^7} &=&
\frac{8}{15}\frac{(9+\gamma^2)}{(1-\gamma^2)^6}\gamma_i\gamma_k\gamma_p 
 +  \frac{1}{15}\frac{(7+\gamma^2)}{(1-\gamma^2)^5}(\delta_{ik}\gamma_p
+\delta_{ip}\gamma_k +\delta_{kp}\gamma_i)\quad , \\
\frac{1}{4\pi}\int d^2 {\cal S}(\phi)\, \frac{n_i n_j n_k n_p}{(1-n\gamma)^7}
  &=& \frac{16}{3}\frac{1}{(1-\gamma^2)^6}\gamma_i\gamma_j\gamma_k\gamma_p 
 +  \frac{1}{15}\frac{1}{(1-\gamma^2)^4}(\delta_{ij}\delta_{kp}
+\delta_{ik}\delta_{jp}+\delta_{ip}\delta_{jk}) \\
& + & \frac{8}{15}\frac{1}{(1-\gamma^2)^5}(\delta_{ij}\gamma_k\gamma_p
+\delta_{ik}\gamma_j\gamma_p +\delta_{ip}\gamma_j\gamma_k 
 +  \delta_{jk}\gamma_i\gamma_p +\delta_{jp}\gamma_i\gamma_k
+\delta_{kp}\gamma_i\gamma_j).
\end{eqnarray*}
}

\newpage

\begin{center}
\section*{\bf References}
\end{center}

$$ $$

\begin{enumerate}
\item B.S.DeWitt, Phys. Rep. 19 C (1975) 295.
\item A.A.Grib, S.G.Mamayev, and V.M.Mostepanenko, {\it Quantum effects
in intense external fields} (Atomizdat, Moscow, 1980), in Russian.
\item N.D.Birrell and P.C.W.Davies, {\it Quantum fields in curved
space} (Cambridge U.P., Cambridge, 1982).
\item A.G.Mirzabekian and G.A.Vilkovisky, Phys. Lett. B 414 (1997) 123.
\item V.P.Frolov and G.A.Vilkovisky, Phys. Lett. B 106 (1981) 307; in:
Proc. 2nd Seminar on Quantum Gravity (Moscow, 1981), eds. M.A.Markov
and P.C.West (Plenum, London, 1983) p. 267.
\item J.Schwinger, in: Proc. 1960 Brandeis Summer School (Brandeis U.,
Brandeis, 1960) p. 282; J. Math. Phys. 2 (1961) 407.
\item L.V.Keldysh, Zh. Eksp. Teor. Fiz. 47 (1964) 1515.
\item Yu.A.Golfand, Yad. Fiz. 8 (1968) 600.
\item P.Hajicek, in: Proc. 2nd Marcel Grossmann Meeting on General
Relativity (Trieste, 1979), ed. R.Ruffini (North Holland, Amsterdam,
1982) p. 483.
\item E.S.Fradkin and D.M.Gitman, Fortschr. der Phys. 29 (1981) 381.
\item J.L.Buchbinder, E.S.Fradkin, and D.M.Gitman, Fortschr. der Phys.
29 (1981) 187.
\item R.D.Jordan, Phys. Rev. D 33 (1986) 44.
\item E.Calzetta and B.L.Hu, Phys. Rev. D 35 (1987) 495.
\item A.O.Barvinsky and G.A.Vilkovisky, Nucl. Phys. B 282 (1987) 163.
\item A.O.Barvinsky and G.A.Vilkovisky, Nucl. Phys. B 333 (1990) 471;
 Nucl. Phys. B 333 (1990) 512.
\item G.A.Vilkovisky, CERN-TH.6392/92; Publ. Inst. Rech. Math. Avanc\'ee,
R.C.P. 25, Vol. 43 (Strasbourg, 1992) p. 203.
\item A.O.Barvinsky, Yu.V.Gusev, G.A.Vilkovisky, and V.V.Zhytnikov,
{\it Covariant perturbation theory} (IV). {\it Third order in the
curvature} (U. Manitoba, Winnipeg, 1993) pp. 1-192; J. Math. Phys.
35 (1994) 3543; Nucl. Phys. B 439 (1995) 561.
\item A.O.Barvinsky, Yu.V.Gusev, V.V.Zhytnikov, and G.A.Vilkovisky,
Class. Quantum Grav. 12 (1995) 2157.
\item G.A.Vilkovisky, Class. Quantum Grav. 9 (1992) 895.
\item A.O.Barvinsky, Yu.V.Gusev, G.A.Vilkovisky, and V.V.Zhytnikov,
J. Math. Phys. 35 (1994) 3525.
\item A.G.Mirzabekian and G.A.Vilkovisky, Phys. Lett. B 317 (1993) 517.
\item A.G.Mirzabekian, Zh. Eksp. Teor. Fiz. 106 (1994) 5 [Engl. trans.:
JETP 79 (1994) 1].
\item A.G.Mirzabekian and G.A.Vilkovisky, Phys. Rev. Lett. 75 (1995)
3974; Class. Quantum Grav. 12 (1995) 2173.
\item A.O.Barvinsky and G.A.Vilkovisky, Phys. Rep. 119 (1985) 1.
\item R.M.Wald, {\it General relativity} (Chicago U.P., Chicago, 1984).
\item H.Bondi, M.G.J. van der Burg, and A.W.K.Metzner, Proc. R. Soc.
A 269 (1962) 21.
\item R.Sachs, in: Relativity, Groups and Topology (1963 Les Houches
Lectures), eds. C.DeWitt and B.S.DeWitt (Gordon and Breach, New York,
1964).
\item A.M.Polyakov, Phys. Lett. B 103 (1981) 207.
\item J.L.Synge, {\it Relativity: the general theory} (North Holland,
Amsterdam, 1960).
\item B.S.DeWitt, {\it Dynamical theory of groups and fields} (Gordon
and Breach, New York, 1965).
\item A.G.Mirzabekian, G.A.Vilkovisky, and V.V.Zhytnikov, Phys. Lett. B
369 (1996) 215.
\item E.S.Fradkin and G.A.Vilkovisky, Phys. Lett. B 73 (1978) 209.
\item I.E.Tamm, {\it Foundations of electricity theory} (Gostekhizdat,
Moscow, 1956), in Russian.
\item G.A.Vilkovisky, in: Quantum Theory of Gravity, ed. S.M.Christensen
(Hilger, Bristol, 1984) p. 169.
\item A.A.Ostrovsky and G.A.Vilkovisky, J. Math. Phys. 29 (1988) 702.
\item N.M.J.Woodhouse, Phys. Rev. Lett. 36 (1976) 999.
\end{enumerate}

\newpage

\begin{center}
\section*{Captions to the figures}
\end{center}

$$ $$

\begin{itemize}
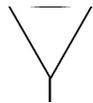
\item[Fig.1.] The Bondi-Sachs frame with the central geodesic
$r=0$ and the light cones of equal retarded time.
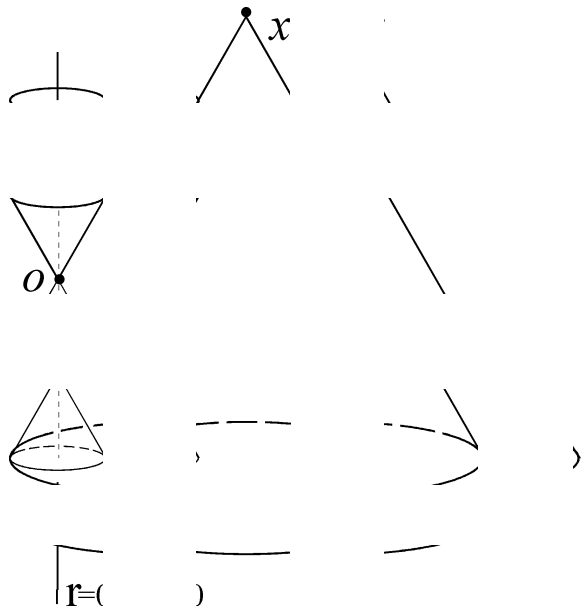
\item[Fig.2.] Shown are the past light cone of a point $x$,
an arbitrary point $o$ on its surface, and both sheets of the
light cone of $o$. (The line $r=0$ is the central geodesic.)
The past light cone of $x$ lies outside the light cone of $o$.
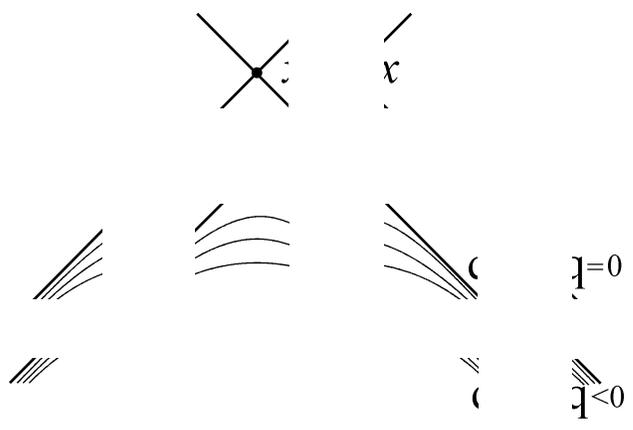
\item[Fig.3.] The past hyperboloids of $x$, $\sigma (x,{\bar x})=q$,
inscribed in the past light cone of $x$, $\sigma (x,{\bar x})=0$.
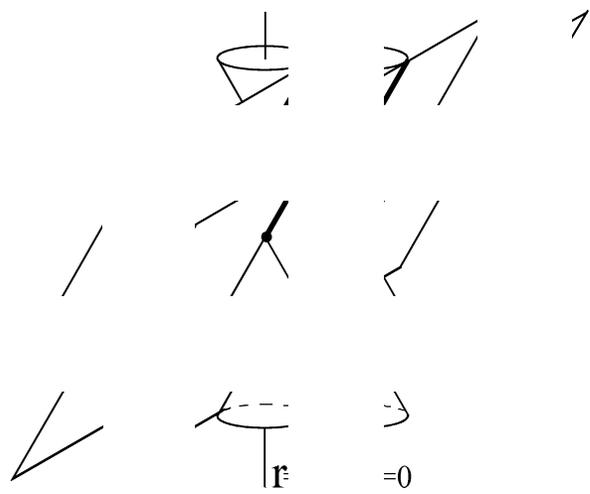
\item[Fig.4.] The past light cone of $x$ in Fig.2 after $x$ has
moved to ${\cal I}^+$ along the radial geodesic shown bold. The
resultant null hyperplane has the parameters $u$, $\phi$ of this
geodesic.
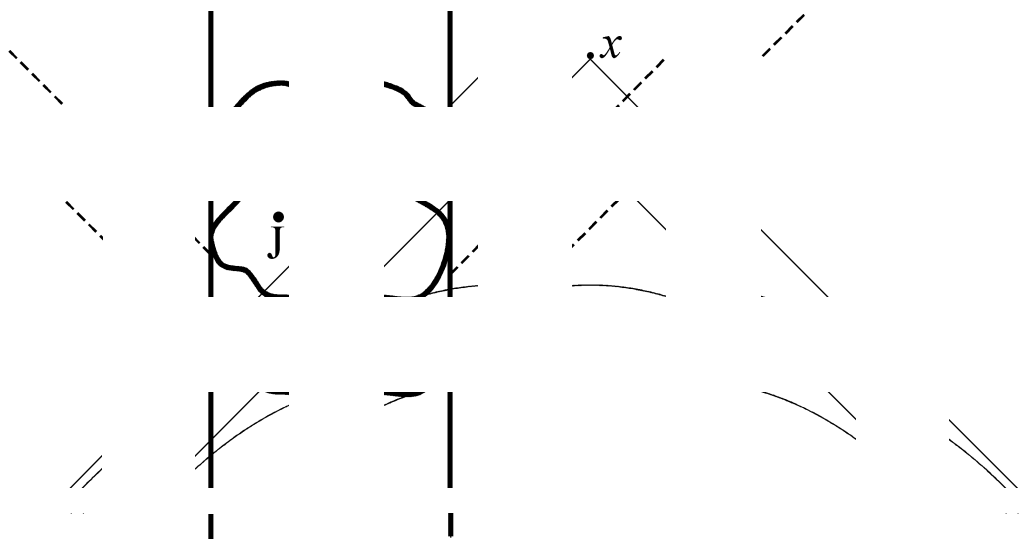
\item[Fig.5.] Shown are the support tube of a physical source $J$,
the compact support of $\J$, and the earliest hyperboloid of $x$
that still crosses the latter support. The broken lines bound the
causal future of the support of $\J$.
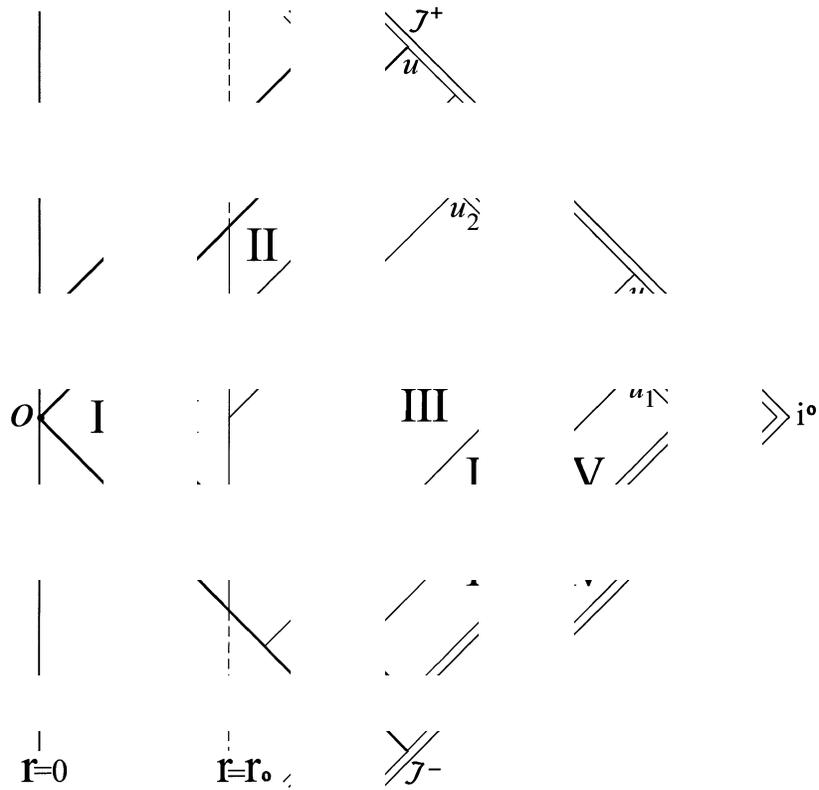
\item[Fig.6.] Penrose diagram for the section of fixed angles
in the Bondi-Sachs coordinates. The exterior of the light cone
of $o$ is divided into four subdomains bounded by two future
light cones $u=u_1$ and $u=u_2$, and the surface of the tube
$r<r_0$.
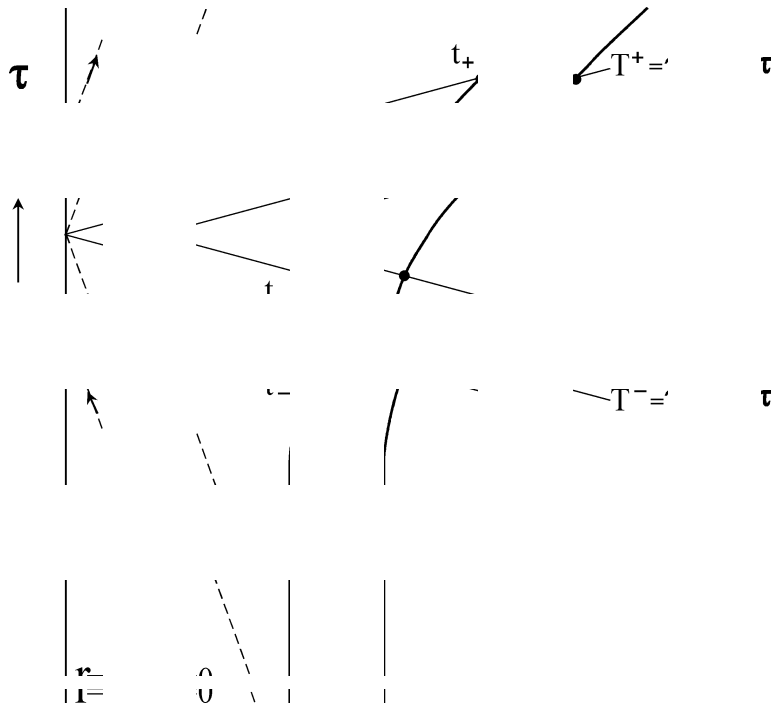
\item[Fig.7.] Lorentzian section $\Gamma$ of a spherically symmetric 
spacetime. The broken line is the mapping on $\Gamma$ of a timelike
radial geodesic. The lines $T^{\pm}=\tau$ bound the mapping on
$\Gamma$ of a spacelike hyperplane. The bold line is the world line
of the support shell (or the boundary of the support tube) of the source.

\end{itemize}

\newpage
\begin{figure}[p]
\begin{picture}(360,450)
\put(165,240){
\newbox\onebox
\newdimen\onew
\font\onea=onea at 72.27truept
\setbox\onebox=\vbox{\hbox{%
\onea\char0\char1}}
\onew=\wd\onebox
\setbox\onebox=\hbox{\vbox{\hsize=\onew
\parskip=0pt\offinterlineskip\parindent0pt
\hbox{\onea\char0\char1}
\hbox{\onea\char2\char3}
\hbox{\onea\char4\char5}
\hbox{\onea\char6\char7}}}
\ifx\parbox\undefined
    \def\setone{\box\onebox}
\else
    \def\setone{\parbox{\wd\onebox}{\box\onebox}}
\fi
\setone}
\end{picture}
\begin{center}
{\LARGE Fig.1}
\end{center}
\end{figure}
\newpage
\begin{figure}[p]
\begin{picture}(360,450)
\put(90,270){
\newbox\twobox
\newdimen\twow
\font\twoa=twoa at 72.27truept
\setbox\twobox=\vbox{\hbox{%
\twoa\char0\char1\char2\char3}}
\twow=\wd\twobox
\setbox\twobox=\hbox{\vbox{\hsize=\twow
\parskip=0pt\offinterlineskip\parindent0pt
\hbox{\twoa\char0\char1\char2\char3}
\hbox{\twoa\char4\char5\char6\char7}
\hbox{\twoa\char8\char9\char10\char11}
\hbox{\twoa\char12\char13\char14\char15}}}
\ifx\parbox\undefined
    \def\settwo{\box\twobox}
\else
    \def\settwo{\parbox{\wd\twobox}{\box\twobox}}
\fi
\settwo}
\end{picture}
\begin{center}
{\LARGE Fig.2}
\end{center}
\end{figure}
\newpage
\begin{figure}[p]
\begin{picture}(360,450)
\put(75,210){ 
\newbox\threebox
\newdimen\threew
\font\threea=threea at 72.27truept
\setbox\threebox=\vbox{\hbox{%
\threea\char0\char1\char2\char3}}
\threew=\wd\threebox
\setbox\threebox=\hbox{\vbox{\hsize=\threew
\parskip=0pt\offinterlineskip\parindent0pt
\hbox{\threea\char0\char1\char2\char3}
\hbox{\threea\char4\char5\char6\char7}
\hbox{\threea\char8\char9\char10\char11}}}
\ifx\parbox\undefined
    \def\setthree{\box\threebox}
\else
    \def\setthree{\parbox{\wd\threebox}{\box\threebox}}
\fi
\setthree}
\end{picture}
\begin{center}
{\LARGE Fig.3}
\end{center}
\end{figure}
\newpage
\begin{figure}[p]
\begin{picture}(360,450)
\put(84,255){
\newbox\fourbox
\newdimen\fourw
\font\foura=foura at 72.27truept
\setbox\fourbox=\vbox{\hbox{%
\foura\char0\char1\char2\char3}}
\fourw=\wd\fourbox
\setbox\fourbox=\hbox{\vbox{\hsize=\fourw
\parskip=0pt\offinterlineskip\parindent0pt
\hbox{\foura\char0\char1\char2\char3}
\hbox{\foura\char4\char5\char6\char7}
\hbox{\foura\char8\char9\char10\char11}
\hbox{\foura\char12\char13\char14\char15}}}
\ifx\parbox\undefined
    \def\setfour{\box\fourbox}
\else
    \def\setfour{\parbox{\wd\fourbox}{\box\fourbox}}
\fi
\setfour}
\end{picture}
\begin{center}
{\LARGE Fig.4}
\end{center}
\end{figure}
\newpage
\begin{figure}[p]
\begin{picture}(360,450)
\put(0,261){
\newbox\fivebox
\newdimen\fivew
\font\fivea=fivea at 72.27truept
\setbox\fivebox=\vbox{\hbox{%
\fivea\char0\char1\char2\char3\char4\char5}}
\fivew=\wd\fivebox
\setbox\fivebox=\hbox{\vbox{\hsize=\fivew
\parskip=0pt\offinterlineskip\parindent0pt
\hbox{\fivea\char0\char1\char2\char3\char4\char5}
\hbox{\fivea\char6\char7\char8\char9\char10\char11}
\hbox{\fivea\char12\char13\char14\char15\char16\char17}
\hbox{\fivea\char18\char19\char20\char21\char22\char23}}}
\ifx\parbox\undefined
    \def\setfive{\box\fivebox}
\else
    \def\setfive{\parbox{\wd\fivebox}{\box\fivebox}}
\fi
\setfive}
\end{picture}
\begin{center}
{\LARGE Fig.5}
\end{center}
\end{figure}
\newpage
\begin{figure}[p]
\begin{picture}(360,450)
\put(68,294){
\newbox\sixbox
\newdimen\sixw
\font\sixa=sixa at 72.27truept
\setbox\sixbox=\vbox{\hbox{%
\sixa\char0\char1\char2\char3\char4\char5}}
\sixw=\wd\sixbox
\setbox\sixbox=\hbox{\vbox{\hsize=\sixw
\parskip=0pt\offinterlineskip\parindent0pt
\hbox{\sixa\char0\char1\char2\char3\char4\char5}
\hbox{\sixa\char6\char7\char8\char9\char10\char11}
\hbox{\sixa\char12\char13\char14\char15\char16\char17}
\hbox{\sixa\char18\char19\char20\char21\char22\char23}
\hbox{\sixa\char24\char25\char26\char27\char28\char29}}}
\ifx\parbox\undefined
    \def\setsix{\box\sixbox}
\else
    \def\setsix{\parbox{\wd\sixbox}{\box\sixbox}}
\fi
\setsix}
\end{picture}
\begin{center}
{\LARGE Fig.6}
\end{center}
\end{figure}
\newpage
\begin{figure}[p]
\begin{picture}(360,450)
\put(69,310){
\newbox\sevenbox
\newdimen\sevenw
\font\sevena=sevena at 72.27truept
\setbox\sevenbox=\vbox{\hbox{%
\sevena\char0\char1\char2\char3\char4}}
\sevenw=\wd\sevenbox
\setbox\sevenbox=\hbox{\vbox{\hsize=\sevenw
\parskip=0pt\offinterlineskip\parindent0pt
\hbox{\sevena\char0\char1\char2\char3\char4}
\hbox{\sevena\char5\char6\char7\char8\char9}
\hbox{\sevena\char10\char11\char12\char13\char14}
\hbox{\sevena\char15\char16\char17\char18\char19}
\hbox{\sevena\char20\char21\char22\char23\char24}}}
\ifx\parbox\undefined
    \def\setseven{\box\sevenbox}
\else
    \def\setseven{\parbox{\wd\sevenbox}{\box\sevenbox}}
\fi
\setseven}
\end{picture}
\begin{center}
{\LARGE Fig.7}
\end{center}
\end{figure}

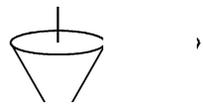
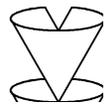
\end{document}